\documentclass[aps,aps,showpacs,onecolumn,preprintnumbers,
amsmath,amssymb,superscriptaddress,longbibliography,notitlepage,nofootinbib]{revtex4-1}
\usepackage{slashed}
\usepackage{amsmath}
\usepackage{cancel}
\usepackage{amsfonts}
\usepackage{amssymb}
\usepackage{xcolor,comment}
\usepackage{color}
\usepackage[colorlinks=True,linkcolor=blue,citecolor=blue,urlcolor=blue]{hyperref}
\usepackage{tcolorbox}
\usepackage{ulem}
\usepackage{collectbox}

\usepackage{epsfig}
\usepackage{graphicx}
\usepackage{dcolumn}
\usepackage{bm}

\newcommand{\beq}{\begin{equation}}
\newcommand{\eeq}{\end{equation}}
\newcommand{\eq}[1]{(\ref{#1})}
\newcommand{\beqn}{\begin{eqnarray}}
\newcommand{\eeqn}{\end{eqnarray}}
\newcommand{\beqa}{\begin{eqnarray}}
\newcommand{\eeqa}{\end{eqnarray}}
\newcommand{\avr}[1]{\left\langle #1 \right\rangle}
\newcommand{\bs}{\boldsymbol}

\def\gapp{\lower.35em\hbox{$\stackrel{\textstyle>}{\sim}$}}
\def\lapp{\lower.35em\hbox{$\stackrel{\textstyle<}{\sim}$}}

\definecolor{brickred}{rgb}{0.8, 0.25, 0.33}
\definecolor{klgreen}{rgb}{0.0, 0.5, 0.0}
\definecolor{mvblue}{rgb}{0.5, 0.0, 0.8}

\newcommand{\cL}{{\mathcal L}}
\newcommand{\cl}{{\mathrm{cl}}}

\begin{document}

\title{Thermal transport, geometry, and anomalies}
\author{Maxim N. Chernodub}
\affiliation{Institut Denis Poisson UMR 7013, Universit\'e de Tours, Tours, 37200, France}
\affiliation{Pacific Quantum Center, Far Eastern Federal University, 690950 Vladivostok, Russia}
\author{Yago Ferreiros}
\affiliation{IMDEA Nanociencia, Faraday 9, 28049 Madrid, Spain}
\author{Adolfo G. Grushin}
 \affiliation{Univ. Grenoble Alpes, CNRS, Grenoble INP, Institut N\'eel, 38000 Grenoble, France}
\author{Karl Landsteiner} 
\affiliation{Instituto de F\'isica Te\'orica UAM/CSIC, \\
Nicol\'as Cabrera 13-15, Cantoblanco, 28049 Madrid, Spain }
\author{Mar\'ia A. H. Vozmediano}
\affiliation{Instituto de Ciencia de Materiales de Madrid,\\
CSIC, Cantoblanco, E-28049 Madrid, Spain.}

\begin{abstract}
The relation between thermal transport and gravity was  highlighted in the seminal work by Luttinger in 1964, and has been extensively developed to understand thermal transport, most notably the thermal Hall effect. Here we review the novel concepts that relate thermal transport, the geometry of space-time and quantum field theory anomalies. We give emphasis to the cross-pollination between emergent ideas in condensed matter, notably Weyl and Dirac semimetals, and the understanding of gravitational and scale anomalies stemming from high-energy physics. We finish by relating to recent experimental advances and presenting a perspective of several open problems.
\end{abstract}
\maketitle

\tableofcontents

\clearpage

\begin{flushright}
\it In fact, if the gravitational field didn't exist, \\
one could invent one for the purposes of this paper.\footnote{From footnote 7 of J.M. Luttinger "Theory of thermal transport coefficients" \cite{Luttinger1964}.} \\[3mm]
\end{flushright}
\section{Introduction}
\label{sec_intro}
The main conceptual advances in physics have usually been prompted by almost simultaneous 
discoveries in its different disciplines. In the past century, statistical physics, quantum field theory and condensed matter
had their main developments in parallel with the best physicists contributing
to them all. After a long period of atomization and specialization of the different branches of physics,
recent experiments and conceptual developments in condensed matter physics are fostering a new era of
grand unification of low and high energy physics.  The new era started with the experimental realization of graphene,
a two-dimensional crystal made of carbon atoms  in 2004 \cite{Netal05,Zetal05}. As it is well known,
the dynamics of the low energy electronic excitations in graphene  is described by the 
massless  Dirac Hamiltonian in 2+1  dimensions and the interacting system shows parallelisms with 
quantum electrodynamics~\cite{Gonzalez:1993uz,RMP12}.
More recently, Dirac and Weyl semimetals, 3D materials described at low energies by the massless Dirac equation
\cite{JXH16,AMV18}, completed  the picture providing new perspectives on one of the most exciting aspects of QFT: Quantum anomalies \cite{Adler:1969,Bell:1969, Kimura:1969wj, Bert96, Fujikawa:2004cx} and
anomaly-induced transport phenomena \cite{Kharzeev:2013ffa,Landsteiner:2016led,Burkov_2015}.

Quantum anomalies were encountered and described in the earlier developments of QFT associated to
the physics of elementary particles. The  language and ideas at the time were based on
concepts such as current algebra or the analytic $S$ matrix,  which involved rather 
sophisticated mathematics and that have largely fallen into disuse. Despite the   successful reviews adapting the concepts to more modern language~\cite{Bilal:2008qx, Harvey:2005it},  the QFT literature seems to remain moderately inaccessible to a large fraction of
condensed matter researchers, both experimentalists and theoreticians. Equivalently, the condensed matter description of
topological matter and the literature involving the anomaly-related transport experiments, use a phenomenological language which is
often difficult to integrate in the QFT formalism. This is especially so in the case of the mixed axial--gravitational anomaly and 
its relation to the thermal transport phenomena observed in Dirac materials \cite{Gooth2017,Heremans2019}.

Quantum anomalies play an important role in high-energy physics~\cite{Shifman:1988zk}. The anomalies help to describe experimental data in phenomenology of fundamental particle interactions and constrain the space of physically self-consistent models in quantum field theory (QFT). One of the best known observational consequences of quantum anomalies is highlighted by the axial anomaly which shortens the lifetime of a neutral pion by opening the classically forbidden decay channel of this pseudoscalar particle into two photons, $\pi^0 \to 2 \gamma$~\cite{Adler:1969,Bell:1969}.

A different type of quantum anomaly appears in interacting field theories that possess classical invariance under appropriate global rescaling of coordinates and fields~\cite{Peskin:1995ev}. This scale symmetry implies the equivalence of classical processes that develop at different energies since the classical equations of motion contain no dimensionful parameters. The symmetry is often broken by the quantum scale anomaly which generates, due to quantum corrections, a dependence of the theory on the energy scale. The existence of the conformal anomaly was first identified via its signatures 
in correlation functions which signalled a nonzero expectation value of the trace of the energy-momentum tensor 
in a purely curved (gravitational) background~\cite{Capper:1973mv,Capper:1974ed}. Later it was recognised that the interactions of matter fields with classical electromagnetic fields also lead to a conformal anomaly in flat space~\cite{Deser:1976yx}. Perturbatively, the quantum scale anomaly gives rise to non-zero beta functions that describe how the couplings of the model change with the energy of the relevant process~\cite{Shifman:1988zk}. Nonperturbatively, the scale anomaly can generate a mass gap which sets a new energy or length scale in the theory~\cite{Weinberg:1975gm}. 
In the context of particle physics, both perturbative and non-perturbative features of the scale anomaly are manifest, for example, in Quantum Chromodynamics~\cite{Savvidy:1977as,Cornwall:1981zr}. As we will see in this review,  the perturbative scale anomaly  has also consequences for condensed matter.

On the purely theoretical side, the presence of quantum anomalies allows to narrow down the class of feasible gauge models of particle physics which are consistent with the experimentally established conservation laws. The anomalous contributions coming from all matter fields should mutually cancel each other thus making a tight constraint on the particle content of the model and the coupling of the particles with interaction carriers~\cite{Anomaly:free:1972}. For example, the existence of gauge invariance, which is important for the renormalizability of the theory, requires the cancellation of the triangular diagrams~\cite{Adler:1969,Bell:1969}. Similar self-consistency requirements imply the absence of the axial gauge fields in the standard model of fundamental interactions.

Here is the place where condensed matter realizations of Dirac and Weyl fermions go beyond the constraints of high energy physics. For example axial gauge fields can be realized in Weyl metals by applying strain on the crystal \cite{CFLV15,CKLV16}. This gives unique opportunities to experimentally probe certain types anomalies related to these gauge fields \cite{PCF16,GVetal16,Strain20}. Similarly, in the realm of geometry, it is commonly accepted that torsion as a dynamical field is not realized in nature. But in condensed matter systems certain crystal defects enter the electronics of Weyl and Dirac metals as a torsion field \cite{Kleinert89,KatVol92} whose coupling to the fermions  gives rise to new interesting physical phenomena \cite{YZV09}.

The main purpose of this review is to present the basic notions underlying  new developments in condensed matter in a language
equally accessible to both high energy and condensed matter communities. The more technical presentations will be
complemented with an intuitive description that will permit to skip the details keeping the physical ideas. The main concepts will be emphasized in boxes at the end of  each section.  We also include a general introductory section  to ease the jumps between the different disciplines involved in the topic. Thorough reviews are already available   on the material implementation of the chiral anomaly \cite{Burkov_2015,Landsteiner:2016led,Marsh_2017}, and on the condensed matter systems subjected to these anomalies (Dirac and Weyl semimetals)  \cite{JXH16,Felser17,AMV18}. 
In this review we will focus on the  gravitational and other geometric anomalies that emerged more recently, and on the new transport phenomena that they induce. 
Even though some of the topics included are still under debate in either condensed matter or high energy physics (or both), the basic facts around these geometrical anomalies are well established. Since these facts are scattered in a variety of  quite specialized contexts,  it is worthy to put them together. 
\\

The review is organized as follows.
We begin in Sec. \ref{sec_dic} by setting up a few well established facts about the similarities and differences
between the condensed matter description and the quantum field theory counterpart
encountered in the field of anomaly--induced transport. 

In  Sec. \ref{sec_TT} we fix the notation and some general notions of thermal transport. The section includes
the definition of the thermal and thermoelectric responses,
Wiedemann-Franz and Mott relations, and the Luttinger theory of thermal transport coefficients. 

In Sec. \ref{sec_magnet} we describe the
magnetization (electric and energy) currents that may arise in the absence of time reversal symmetry and see how they affect the interpretation of the Kubo formulas for the transport coefficients. We also show how they are framed within the Luttinger formalism.

Section \ref{sec_chiralanoms} is devoted to transport induced by triangle anomalies. We  
discuss  what is an anomaly in QFT (and what is not) and summarize the basic facts around the chiral and the mixed chiral--gravitational anomalies. Then we expose the main anomaly--related transport phenomena in general, and discuss in detail their relation with magnetotransport in Dirac and Weyl semimetals. We also show that the anomaly--induced thermoelectric transport coefficients obey the Wiedemann--Franz and Mott relations.

In section \ref{sec_axialfields} we introduce the axial gauge fields arising in condensed matter and describe the  new anomaly induced  phenomena associated to them. 

In Sec. \ref{sec_Scale}, we analyze the physical consequences of the scale anomaly. We introduce the concepts and the consequences of the anomaly to transport phenomena both in the bulk and in the boundary of the physical systems. We also discuss the thermomagnetic transport effects produced by the scale anomaly in Dirac and Weyl semimetals.

Sec. \ref{sec_torsion} is devoted to the potentially new anomaly--induced transport phenomena that arise when  considering the torsion degrees of freedom that occurs naturally in crystalline materials associated to dislocations of the lattice. We will describe the basic formalism and give an alternative perspective to some of the works encountered in the literature. In particular we review in what sense there are no mixed axial--torsional anomalies in the strict quantum field theory sense, and that there is no universal chiral transport as a response to geometric torsion.

In Sec. \ref{sec_exp} we give an overview of the experimental situation. We review the recent experimental efforts to measure the mixed axial-gravitational anomalies in thermo-electric and thermal transport in Weyl semimetals in a magnetic field. 

We include various appendices that give technical support to the material covered in the review. We have included boxes to highlight some of the main conclusions of each section. In the last Appendix we review our notation for different quantities.

Finally, let us comment on certain fascinating related issues left aside from this review. Since we address phenomena related to quantum field theory anomalies common to high energy and condensed matter physics, we have reduced our condensed matter systems to the topological materials whose low energy effective description matches the Dirac equation in three spatial dimensions describing, basically, Dirac and Weyl semimetals. This leaves aside other types of ``Dirac matter" as nodal~\cite{Nodal16} or multi-Weyl semimetals~\cite{Double16}, Kane fermions~\cite{Teppe16}, topological insulators and superconductors~\cite{RML12}. We have also omitted the very interesting analogs~\cite{FBB15,ZMetal18,HQetal18,NZetal19,PSetal19,HGetal19} and gapped quantum materials~\cite{Kapustin21}.


\section{Key concepts }
\label{sec_dic}
The anomaly related transport phenomena encountered in  modern condensed matter systems are sourced in different areas of physics which have their own terms to describe them. This broadness testifies to the richness of the field, but it also nourishes  misunderstandings. Same phenomena are called different names, and different phenomena can be mistaken as the same. An additional difficulty lies on the fact that the approximations assumed in the different areas of physics are often not explicitly established and their regime of validity can be incompatible with the experimental conditions. Since this review aims to provide a common basis for the understanding of how anomalies evolved from high energy to condensed matter, we begin with an attempt to clarify the main concepts participating in the field. 

In this section we will establish the similarities  and differences between the quantum field
theory  description of chiral fermions as relativistic particles  and the 
condensed matter  ``Dirac fermions''.  

The mere idea of the name ``Dirac fermions'' is somehow astonishing in particle physics;
what else can they be? Fermions $\equiv$ Dirac spinors. In condensed  matter the electrons are fermions in the
sense that they obey the Pauli exclusion principle, and the distribution function used in the statistical averages is the Fermi-Dirac distribution. In the standard model for metals, the Landau Fermi liquid \cite{L57}, the electronic dynamics is described by the Hamiltonian $H=p^2/2m$.

Dirac dynamics became popular in  
condensed matter with the advent of graphene  since the low-energy description of the electronic degrees 
of freedom (quasiparticles) around band crossings (point-like Fermi surface and linear dispersion relation) was the (2+1) Dirac equation describing massless fermions. In the graphene case, Dirac physics is not related to the physical spin of the electrons
but to a flavor degree of freedom coming from the structure of the underlying ionic lattice (pseudospin). So, even when
$H=v_F {\vec \sigma}\cdot{\vec p}$, still these fermions are spinless and the spin is added by doubling the degrees of freedom.
Also graphene lives in (2+1) dimensions where the chirality operator is not  defined, so the fermions are characterized by parity, but not chirality. The advent of Dirac and Weyl semimetals in (3+1) dimensions brought the condensed matter fermions closer to their QFT partners but the concepts remain sometimes mixed and obscure since, again, the "Diracness" there often arises associated to a pseudospin.

In what follows we will describe and define some of the most important concepts  involved in this review\footnote{A very interesting description of Dirac matter from a quantum field theory point of view is given by E. Witten in~\cite{Witten16}.}.
\begin{itemize}
\item {\bf The Noether theorem and quantum anomalies}

According to the first Noether theorem\footnote{ See ref. \cite{quigg2019} for a  historical review.}, continuum symmetries in a Lagrangian system generate conserved currents and charges.  Well known examples of  external symmetries are space (time) translations giving rise to momentum (energy) conservation or rotational invariance generating conserved angular momentum in classical and quantum mechanics. Perhaps more interestingly, internal symmetries as phase (gauge) rotations of the wave function in quantum mechanics give rise to  particle current and particle number conservation - electric charge conservation if the particles are charged -. In the process of quantization, classically conserved currents are promoted from vector fields  to composite local operators -  several quantum fields located at the same point - which often need to be regularized. An anomaly occurs  when you cannot find a regulator that retains all the symmetries of the classical action. 
The best example is the case of a  massless Dirac system which will be discussed in the next paragraph.  As we will see, the system has two classically conserved currents eqs. \eqref{eq:JVcurrent}, \eqref{eq:JAcurrent} associated to gauge invariance and chiral symmetry. Due to the quantum  (anti) commutation relation $\{\Psi_\alpha(x),\Psi^+_\beta(y)\}=\delta_{\alpha\beta}\delta^4 (x-y)$,
with two quantum fields  located at the same point the  currents need to be regularized. It so happens that no regularization procedure can keep both currents conserved.   In this simplest case of having two $U(1)_V$ and $U(1)_A$ currents, we can choose to keep the vector current conserved, in which case the divergence of the axial current  turns out to be \cite{Bert96}:
\begin{equation}
    \partial_\mu J_A^\mu=\frac{1}{4\pi^2}{\vec E}\cdot{\vec B},
\end{equation}
where ${\vec E}, {\vec B}$ are background electromagnetic fields. 
We absorb electric charge $e$ into the definition of the electromagnetic fields, $\vec E$ and $\vec B$, and restore it when it is convenient.

\item {\bf Chirality and Weyl fermions}

The  concept of chirality is clear in particle physics. Fermion fields transform as a spinor representations of the Lorentz group, they have an inherent spin 1/2 (or semi--integer in general, but we will restrict ourselves to spin 1/2), and their dynamics is described by 
\beq
{\cal L}=\bar\Psi(i \gamma^\mu\partial_\mu-m)\Psi,
\label{eq_DiracL}
\eeq
where $\partial_\mu \equiv \partial /\partial x^\mu$ is the coordinate derivative,
$\gamma^\mu$ are (3+1) dimensional Dirac matrices satisfying 
\beq
\left\{ \gamma^\mu, \gamma^\nu \right\} = - 2\eta^{\mu\nu} {\cal I}_4,
\label{eq_Cifford}
\eeq
and ${\cal I}_N$ is a $N \times N$ unit matrix in the spinor space.
The Dirac equation 
\beq
(i \gamma^\mu\partial_\mu - m)\Psi=0,
\label{eq_Dirac} 
\eeq
can be written as the Schrodinger equation $i\partial_t\Psi=H\Psi$ with $H=\beta m+{\vec\alpha}\cdot {\vec p}$, where $\beta=\gamma^0$ and $\vec \alpha = \beta.\vec\gamma$.

In odd space dimensions a $\gamma^5$ matrix exists that anti-commutes with all the $\gamma^\mu$ matrices and defines the chirality projection operator $P_{L,R}=(1\pm\gamma^5)/2$.

In the Weyl representation of the gamma matrices:
\begin{equation}
    \gamma^0=\left(\begin{array}{cc} 
    0&{\cal I}_2\\
    {\cal I}_2&0
    \end{array}\right)\,, \qquad
     \gamma^i=\left(\begin{array}{cc} 
    0&\sigma^i\\
    -\sigma^i&0
    \end{array}\right)\,, \qquad
     \gamma_5=\left(\begin{array}{cc} 
    {\cal I}_2&0\\
    0&-{\cal I}_2
    \end{array}\right)\,,
\end{equation}
and when $m=0$, the Dirac Hamiltonian splits into two bi-dimensional blocks
$H_{R,L}= \pm {\vec \sigma}\cdot{\vec p}$. The eigenfunctions are two dimensional spinors of well defined chirality: $\Psi_{L,R}=P_{L,R}\Psi$. These are the Weyl fermions.

One way to characterise the eigenstates is by noting that the positive energy states of right-handed fermions have their (pseudo)spin aligned with their momentum, whereas for left-handed fermions the (pseudo)spin of positive energy states is anti-aligned with the momentum. In high-energy physics positive energy states are identified as particles whereas negative energy states are re-interpreted as anti-particles. In condensed matter physics anti-particle states are referred to as holes. The Nielsen-Ninomiya theorem \cite{NN81,NN83} ensures that chiral states in a lattice always arise in pairs of opposite chiralities. 

As a consequence of chiral symmetry, the action of a massless 3+1 fermion is invariant under independent phase rotations of left and right fermions. We can define two currents that are conserved at the classical level: 
\beqn  \label{eq:JVcurrent}
J_V^\mu(x) & = & \bar\Psi(x)\gamma^\mu\Psi(x)=J_L+J_R\,, \\ \label{eq:JAcurrent}
J_5^\mu(x) & = & \bar\Psi(x)\gamma^\mu\gamma^5\Psi(x)=J_L-J_R\,.
\label{eq:J5current}
\eeqn
The vector $J_V$ current is associated to gauge invariance (conservation of the total electric charge), and the axial $J_5$ current
is associated to the chiral symmetry which leads, at the classical level,  to conservation of the difference between the numbers of right- and left-handed fermions. 
\\

\begin{figure}[!thb]
\begin{center}
\includegraphics[scale=0.25,clip=true]{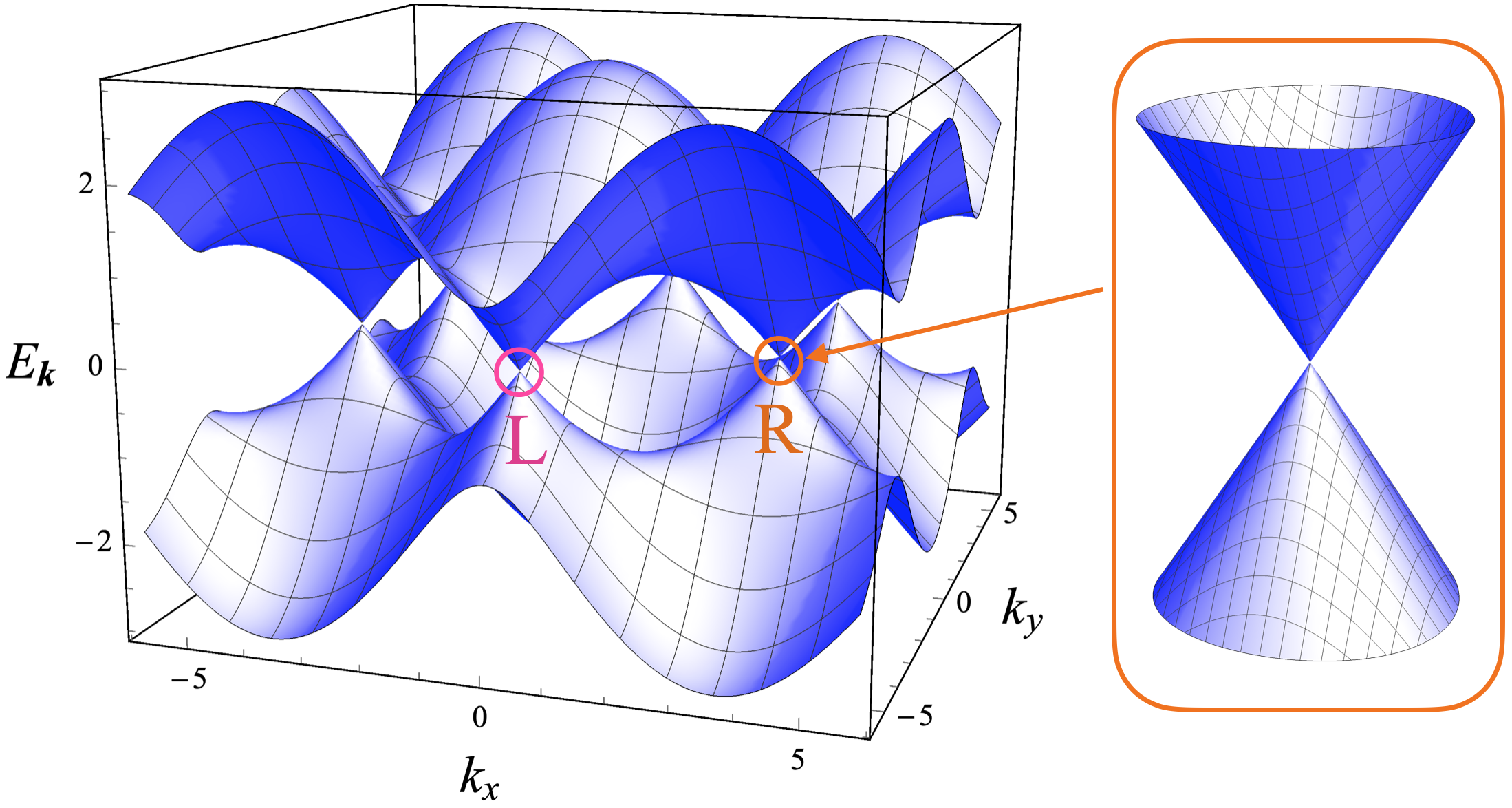}
\end{center}
\caption{Schematic dispersion relation of graphene. The Fermi level is located at the band crossings and the energy is bounded 
defining a compact Fermi sea.}
\label{fig:graphene}
\end{figure}

Massless Dirac fermions appear in condensed matter as the low energy effective description of electronic systems whose Fermi level sits near a two-band crossing. The Nielsen-Ninomiya theorem \cite{NN81} ensures that these crossings appear always in pairs of opposite chiralities. Real crystals typically host more than one pair of chiral fermions, usually close to high-symmetry points or planes. Restricting to a single pair, Dirac semi-metals are those three-dimensional crystals where the two chiralities sit at the same point in energy-momentum space (as massless Dirac particles). To realize a Weyl semimetal (isolated Weyl nodes) either inversion or time-reversal (or both) need to be absent, which allows the two chiralities to be separated in either energy or momentum (or both).  
A detailed description of these conditions and materials that realize them can be found in several recent reviews \cite{JXH16,Felser17,AMV18}.

Fig. \ref{fig:graphene} shows the band structure of graphene, the material which made ``Dirac fermions" popular in condensed matter. There are two inquivalent momenta, or "valleys", in the Brillouin zone where the dispersion is linear, marked as L, R in the figure. These are refered to as massless Dirac points. As mentioned in the introduction of this section, being a two dimensional material, the two valleys of graphene are not Weyl fermions. The dispersion relation shown is visually similar to the band structure of a (say, constant $k_z$)
cut of a three dimensional Weyl semimetal. As seen in Fig.~\ref{fig:graphene}, the bands of the two valleys are connected in the Brillouin zone. 

In condensed matter realizations, in addition to the two band crossings described by the Dirac or Weyl equation (linear dispersion relation),  the name of Dirac matter may be used to refer to other systems described by more exotic equations involving also Dirac matrices. Most noticeable are multi-Weyl fermions with different dispersion relations along different symmetry axis, higher-order polynomial dispersion relations, nodal Weyl semimetals which host a line of nodal points, and multifold fermions, which have point degeneracies of multiple bands~\cite{Hasan21} modelled by higher-spin generalizations of Weyl fermions. The phenomena described in this review do not necessarily apply to these systems, defining in turn an interesting research frontier. In what follows, when we refer to Dirac matter we mean systems described by the QFT massless Dirac equation with linear dispersion relation.

\item {\bf The vacuum}

The QFT vacuum refers to fields at zero temperature $T$ and chemical potential $\mu$.  Originally, anomalies were formulated in vacuum and the role at finite temperature $T$ and chemical $\mu$ has only been fully appreciated recently. In condensed matter realizations, $T=0$ and $\mu=0$ are fined tuned, possibly unrealistic, limits.
Most importantly, the QFT fermionic vacuum is the  Dirac sea,  an infinite tower of negative energy occupied levels. In contrast, the number of occupied electronic states below the Fermi surface  (the condesned matter vacuum), is always finite (see Fig. \ref{fig:graphene}). So in the QFT intuitive description of the chiral anomaly, left (right) particles are annihilated and created from the vacuum.  Strong enough fields can also create pair of states from the vacuum as in the Schwinger mechanism \cite{Schwinger51}. 
In the condensed matter realization,  the two chiralities are connected through the bottom of the band   (see Fig.~\ref{fig:graphene}) so the ``spectral flow" of QFT \cite{Landsteiner:2016led} reduces to a shift of particles in the same band.

\item {\bf A note on Lorentz invariance}

The Dirac equation~\eq{eq_Dirac} assumes the invariance of the theory under the group of Lorentz transformations, which include boosts and spatial rotations as well as time-reversal and parity transformations. The Lorentz invariance reflects itself in the form of the kinetic term, $\gamma^\mu\partial_\mu \equiv  \gamma^0 \partial_t + c {\vec \gamma} \cdot {\vec \nabla}$, where $c$ is the speed of light. Since both the massless fermions and photons propagate with the same velocity $c$, the full interacting theory, which includes the kinetic terms for photons and particles, is invariant under  Lorentz boosts.

In Dirac and Weyl semimetals, however, the quasiparticle excitations propagate with the Fermi velocity $v_F$, which is much slower than the speed of light. The quasiparticles are described by the same Dirac equation~\eq{eq_Dirac} which is, however, modified by the substitution $c \to v_F$ in the derivative term: $\gamma^\mu\partial_\mu \to  \gamma^0 \partial_t + v_F {\vec \gamma} \cdot {\vec \nabla}$, where we assumed for a moment the isotropy and homogeneity of the spacetime as seen by the low-energy quasiparticles. Due to the inequivalence of the velocities of matter fields and interaction carriers, $v_F \neq c$, the kinetic term of quasiparticles~\eq{eq_DiracL} loses the invariance under the Lorentz transformations. 

The Lorentz symmetry breaking in condensed matter enriches the physics of the Dirac quasiparticles  without spoiling the self-consistency of the appropriate low-energy models. We refer to Refs.~\cite{AMV18,Grushin:2019uuu} for a more detailed discussion on this topic which is touched in our review only in bypassing.

\item {\bf The Bloch theorem}

Intuitively a (transport) current moves "something" from a place A in space to another place B. To make this statement a bit more meaningful we also demand that the "something" is stable and does not decay in the timespan in which the transport process takes place. A current transports therefore (approximately) conserves charges. Energy and electric charge are exactly conserved, spin or axial charge might be only approximately conserved.  If the physical system under consideration is in its lowest energy state, e.~g. is in thermodynamic equilibrium, one would not expect such motion to be possible, since this would be a realization of a form of a perpetuum mobile. This intuition is formalized in a theorem\footnote{One should not confuse this non-current theorem with a better known Bloch theorem on the quasimomentum description of solutions of Schr\"odinger equation in a periodic potential.} stated by Felix Bloch in the 1930s which was later summarized by D. Bohm in Ref.~\cite{Bloch49}. Originally it states that it is not possible for a condensed matter system residing in a thermodynamic limit to support an electric current in equilibrium~\cite{Bloch49}. A more modern formulation states that the net current corresponding to an exactly conserved $U(1)$ charge has to vanish in thermodynamic equilibrium. A quick argument goes as follows. The coupling of the current in the Hamiltonian is
\begin{equation}
\label{eq:jA}
    H_{\mathrm{int}} = \int d^3r \vec{A}\cdot\vec{J}\,.
\end{equation}
For an exactly conserved $U(1)$ charge the field $\vec A$ is a gauge field. This means that a spatially constant and time independent value of the gauge potential $\vec A$ should have no physical consequence. If the current has however an expectation value $\langle {\vec J}\rangle \neq 0 $ then an infinitesimally small variation of the gauge potential $\delta {\vec A}$ results in the energy change
\begin{equation}
    \langle \delta H_{\mathrm{int}} \rangle = \delta {\vec A} \cdot\int d^3x \langle \vec{J} \rangle ,
\end{equation}
which can be negative and thus would allow to lower the energy of the system. This contradicts the assumption that the system is in the lowest energy state. It follows that for an exactly conserved $U(1)$ current
\begin{equation}
\int d^3 r \, \vec{J}(\vec{r}) =0    \,.
\end{equation}
We note that this condition does not rule out circular (persistent)  magnetization currents nor does it rule out superconducting currents. In the former case the system resides in a finite (topologically equivalent to a ring) volume far from the thermodynamic limit. The geometry of the system imposes a quantization constraint on the gauge field ${\vec A}$ via the Aharonov-Bohm phase which does not allow for infinitesimally small variations of the gauge potential $\delta {\vec A}$. Therefore, the Bloch no-current theorem does not work for persistent currents. In a superconductor, the gauge symmetry is spontaneously broken and a state with a non-vanishing supercurrent can only be a metastable state.

On the other hand if the current is not exactly conserved or is affected by an anomaly the vector $\vec{A}$ is strictly speaking not a gauge field but a physical observable by itself. Therefore the system with $\vec A \neq 0$ is not equivalent to the ground state in the system $\vec{A} =0$  but rather a different physical system with a new Hamiltonian. Therefore the Bloch theorem does not apply in this situation
and equilibrium currents are possible for this type of currents. 
A recent quantum field theory discussion of the Bloch theorem can be found in \cite{Yamamoto:2015fxa}, and a modern formulation for lattice Hamiltonians is in \cite{Watanabe19}. A generalization of the Bloch theorem to the energy current of Hamiltonian lattice systems has been given in \cite{Kapustin19}.
\end{itemize}

\begin{tcolorbox}
\begin{itemize}
\item {\bf The massless Dirac Hamiltonian in odd space dimensions, (Dirac semimetal) may split into two Weyl Hamiltonians (Weyl semimetal). The two Weyl eigenspinors have well defined, and opposite chiralities.}
\item {\bf There is no chiral anomaly in even spacial dimensions (graphene).}
\item {\bf The Bloch theorem (no transport currents in equilibrium in the thermodynamic limit) applies only to 
exactly conserved currents associated with the gauge symmetry.}
\end{itemize}
\end{tcolorbox}

\section{Thermal and electro-thermal transport}
\label{sec_TT}

Thermal transport refers generically to transport of energy, charge, or any other property under the effect of a gradient of temperature. In metals most transport phenomena are dominated by the free electrons that conduct  energy, and charge. In principle phonons can also contribute to the thermal conductivity. Since we are mainly interested in the thermal transport properties of fermions (electrons) described by the Dirac equation we will not consider the phonon contribution.
For standard metals describable as Fermi liquids, the thermal conductivity is proportional to the electric conductivity (Wiedemann-Franz law). Generically this can be seen as a sign of the existence of long lived quasi-particles. When these quasi particles obey a Dirac or Weyl equation they are subject to the effects of anomalies. As we will see in section \ref{sec_chiralanoms} the Wiedemann-Franz law holds for the anomaly induced conductivities in the presence of a magnetic field (magneto--thermal transport).

We start from the phenomenological equations describing electric and energy transport as a response to electric fields and gradients of chemical potential and temperature \cite{Luttinger1964} 
\begin{gather}
{\vec J}_{\textrm{tr}} = \bm L^{(1)}\left({\vec E}-T{\vec\nabla}\frac{\mu}{T}\right)+\bm L^{(2)}T{\vec\nabla}\frac{1}{T},\label{eq transport 1}\\
{\vec J}_{\textrm{tr}}^{\,\mathcal{E}} =\bm L^{(3)}\left({\vec E}-T{\vec\nabla}\frac{\mu}{T}\right)+\bm L^{(4)}T{\vec\nabla}\frac{1}{T}.
\label{eq transport 2}
\end{gather}
The currents $\vec{J}_{tr}$ and $\vec{J}^\mathcal{E}_{tr}$ are the particle current and energy current respectively, with the particle current related to the electrical current as $\vec{J}^e_{tr}=-e\vec{J}_{tr}$ ($-e$ is the electron charge). The tensors $\bm L^{(i)}$ are the transport coefficients, and $\vec{E}$ is the electric field. The Onsager relations \cite{31Onsager} relate the second and third transport coefficients as $\bm L^{(2)}=\bm L^{(3)}$, or $L^{(2)}_{ij}(B)=L^{(3)}_{ji}(-B)$ in the presence of a magnetic field (see Sec.~\ref{sec_magnet}).
The first terms inside the parenthesis in Eqs.~\eqref{eq transport 1} and \eqref{eq transport 2} represent the Einstein relation \cite{S05,E05,S06,Luttinger1964}, which states that in equilibrium an electric field is compensated by an opposing gradient of chemical potential and temperature $\vec{E}=T\vec{\nabla} (\mu/T)$. The energy current ${\vec J}_{\textrm{tr}}^{\,\mathcal{E}}$ results from the combination of heat current $\vec{J}^Q$ and energy transported by the particle current ${\vec J}_{\textrm{tr}}$; the heat current can therefore be written as
\begin{equation}
{\vec J}_{\textrm{tr}}^{\,Q} ={\vec J}_{\textrm{tr}}^{\, \mathcal{E}} - \mu{\vec J}_{\textrm{tr}}.
\label{eq heat current 1}
\end{equation}
The currents in Eqs. \eqref{eq transport 1} and \eqref{eq transport 2} are transport currents which means they represent net transport through a cross section of the given material. As such, they should vanish in equilibrium. It is important to carefully distinguish between the transport currents $\vec{J}_{\textrm{tr}}$ relevant to Eqs.~\eqref{eq transport 1} and~\eqref{eq transport 2}, and the total local currents $\vec{J}$ at any given point in the sample \cite{Smrcka_1977,Cooper:1997gp,BR15}. While in time-reversal invariant systems these coincide, breaking time-reversal symmetry can lead to the appearance of magnetization electric and energy currents. These are defined as divergenceless circulating currents which average to zero and therefore do not carry any transport \cite{Cooper:1997gp}. We will discuss the presence of magnetization currents in more detail in Sec.~\ref{sec_magnet}.

Let us now fix the chemical potential and allow for gradients of temperature, and consider a sample disconnected from current leads such that no net electric current flows. In this situation, the transport electric current vanishes and an electric field is generated. From the transport Eqs. (\ref{eq transport 1}) and (\ref{eq transport 2}), and using Eq. (\ref{eq heat current 1}) we have
\begin{gather}
{\vec E}= {\bf S}{\vec\nabla} T\label{eq voltage},\\
{\vec J}^Q_{\textrm{tr}}=-{\bf K}{\vec\nabla} T,
\end{gather}
where the thermopower $\bm S$ (or Seebeck coefficient) and thermal conductivity $\bm K$ tensors are functions of the transport coefficients:
\begin{gather}
\bm S=\frac{1}{T}\big[\bm\rho\bm L^{(2)}-\mu\big],\label{eq_thermopower}\\
\bm K=\frac{1}{T}\big[\bm L^{(4)}-\bm L^{(3)}\bm\rho\bm L^{(2)}\big].
\label{eq thermal conductivity}
\end{gather}
where $\rho$ is the resistivity tensor $\bm\rho=\bm\sigma^{-1}$, and we have renamed $\bm L^{(1)}\equiv\bm\sigma$, which is the most established notation for the electric conductivity. 
\\

Eq. \eqref{eq voltage} gives rise to the {\bf Seebeck effect}, which is an expression of the longitudinal components of \eqref{eq transport 1}:  generation of
an electric potential in the direction of a temperature gradient.
\\

\begin{figure}[!thb]
\begin{tabular}{cc}
\includegraphics[scale=0.2,clip=true]{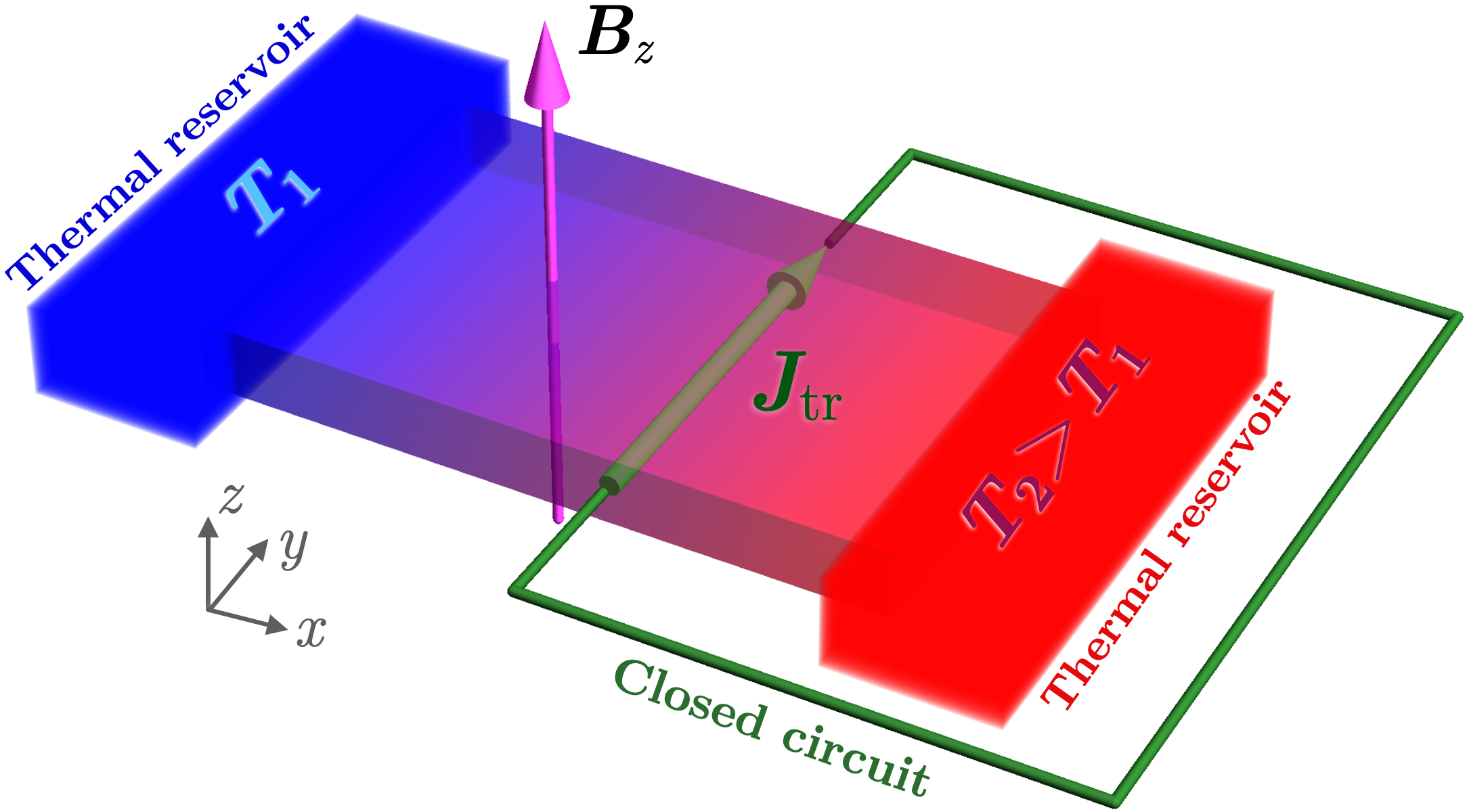}
& \hspace{0.5cm}
\includegraphics[scale=0.2,clip=true]{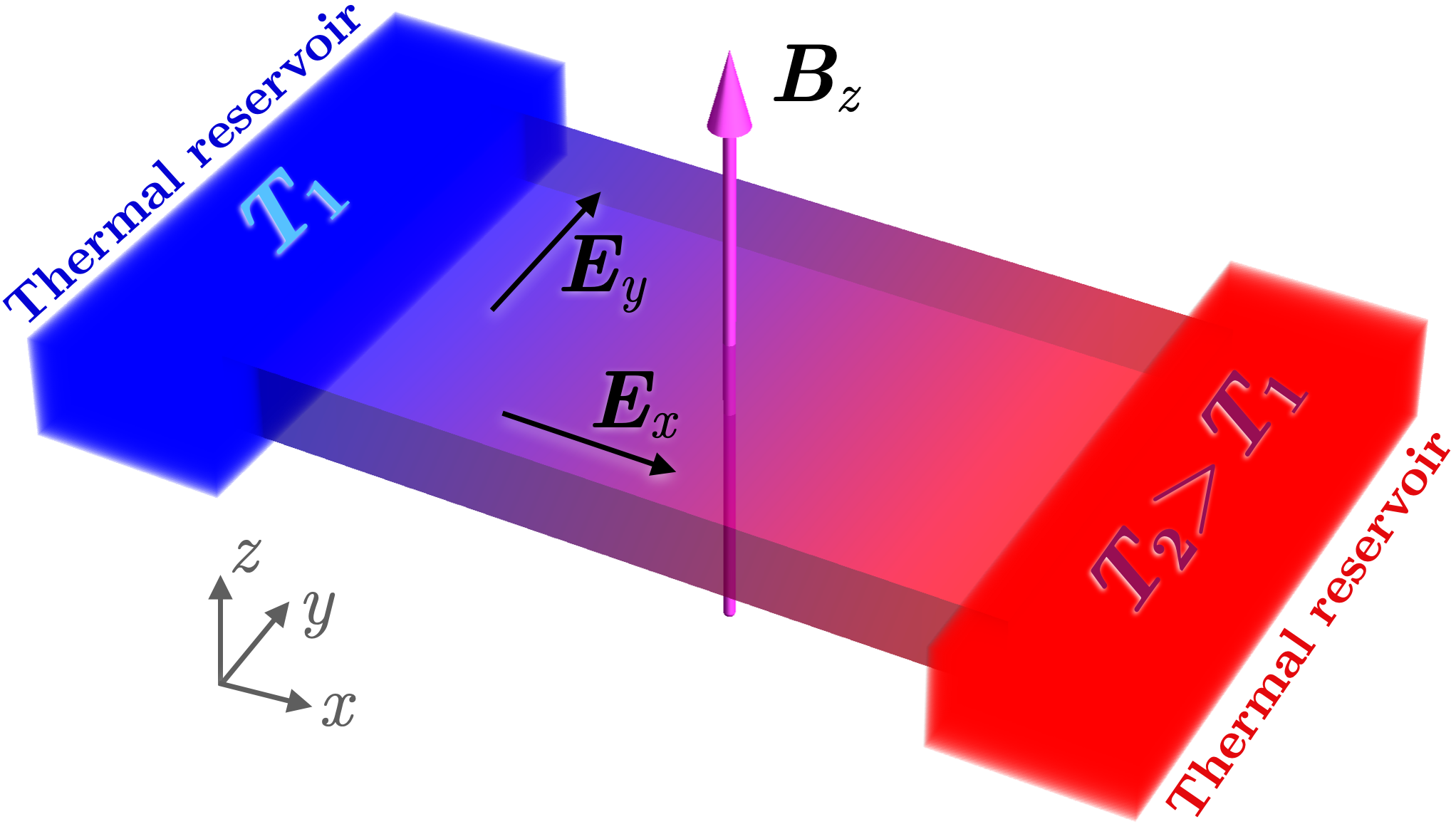}
\\[2mm]
(a) & (b)
\end{tabular}
\caption{A system consisting of a sample (a) connected to and (b) disconnected from current leads, in an external magnetic field, attached to two thermal reservoirs at different temperatures such that a thermal gradient is generated through the sample. (a) A transport current flows in the direction perpendicular to both the magnetic field and gradient of temperature, corresponding to the Nernst current, Eq. \eqref{eq:Nernst};
(b) An electric current cannot flow, so the electric fields $E_x$ and $E_y$ are generated in the plane perpendicular to the magnetic field direction, corresponding to the Nernst-Ettingshausen \eqref{eq_Ettingshausen-Nernstcoefficient} and Nernst \eqref{eq_Nernstcoefficient} coefficients respectively.}
\label{fig_thermal_transport}
\end{figure}
~\\

The {\bf  Nernst-Ettingshausen effects} occur in  the presence of a magnetic field. The Nernst effect  refers to the generation of an electric current perpendicular to a temperature gradient and to the applied magnetic field, Fig. \ref{fig_thermal_transport}(a):
\beqn
\vec{J} \propto \vec{B}  \times \vec{\nabla} T.
\label{eq:Nernst}
\eeqn
For a sample disconnected from current leads, a compensating electric field must necessarily appear such that the transport current vanishes. This situation is schematically described in Fig.~\ref{fig_thermal_transport}. If we choose the gradient of temperature to point, say, along the $\hat{x}$ direction, $\nabla_x T$, and the magnetic field $\vec{B}$ to point along $\hat{z}$, two coefficients are usually defined: the Nernst coefficient
\beqn
{\cal N}_{yx}\equiv\frac{E_y}{B_z \partial_x T}=\frac{\rho_{yi}L^{(2)}_{ix}}{B_z}.
\label{eq_Nernstcoefficient}
\eeqn
and the Ettingshausen-Nernst coefficient:
\beqn
{\cal N}_{xx}\equiv\frac{E_x}{B_z \partial_x T}=\frac{\rho_{xi}L^{(2)}_{ix}-\mu}{B_z},
\label{eq_Ettingshausen-Nernstcoefficient}
\eeqn
see Fig. \ref{fig_thermal_transport}(b).

\subsection{Thermoelectric relations}

In the standard model for metals, the Fermi liquid, the thermo--electric coefficients in Eqs.  \eqref{eq transport 1},\eqref{eq transport 2} obey certain  phenomenological relations, the main ones being the Wiedemann-Franz law and the Mott relation \cite{Ziman,JM80}:
\begin{gather}
\bm K=LT\bm\sigma(T=0),
\label{eq Wiedemann-Franz}\\
\bm S=LT\bm\rho(T=0)\frac{d\bm\sigma(T=0)}{d\mu}{\biggl\vert}_{\mu=\epsilon_F/e}.
\label{eq Mott}
\end{gather}
The Wiedemann-Franz law \eqref{eq Wiedemann-Franz} establishes  that the ratio
of the thermal to the electrical conductivity is the temperature times a universal number,
the Lorenz number $L=\pi^2k_B^2/3e^2$, where $k_B$ is the Boltzmann constant. The Mott relation \eqref{eq Mott} relates the  thermopower with the derivative of the electrical conductivity with respect to the chemical potential at the Fermi level. Note that the electric conductivity in Eqs.~\eqref{eq Wiedemann-Franz} and~\eqref{eq Mott} is evaluated at zero temperature, as both the Mott rule and the Wiedemann-Franz law are obtained as an expansion in $k_B T \ll |\mu|$. The Mott and Wiedemann-Franz relations hold for transport currents only \cite{Smrcka_1977,JG84,Cooper:1997gp,QTN11}, which means that in systems that break time-reversal symmetry the local currents do not generally fulfill them (see Sec.~\ref{sec_magnet}). The validity of these laws has been established for any system which can be described as a Fermi liquid, provided the quasiparticles do not exchange energy during collisions. The validity has also been proven to be valid  when a semi-classical description (as the Boltzmann formalism) of the electronic system is justified. Deviations from these phenomenological relations in conventional matter are normally attributed to electron--electron interactions inducing departures from Fermi liquid behavior or to the emergence of a new phase regime. In particular the relations might not apply when the systems are  in the hydrodynamic  regime.

Topological materials have anomalous conductivities (particularly the Hall conductivity) similar to that occurring in ferromagnetic materials, induced by the Berry curvature of the bands. The question  of whether or not these anomalous transport coefficients obeyed the Wiedemann-Franz and Mott relations, arose soon after the recognition of the topological properties. Berry curvature effects were included in the Boltzmann transport formalism  in a way to fulfil the rules \cite{XY06,RML12}, but the experimental situation is more uncertain and violations of the thermoelectric relations  have been reported in some materials \cite{YCetal08,Kim14,LGetal14,Lietal15}.

Typically Dirac materials in two and three dimensions are expected to follow the standard relations in the low $T$ regime  and deviate from it at larger temperatures \cite{LLetal17,GMSS17,JFetal18}. Violations of the Wiedemann-Franz law have been described in these systems associated to the presence of a hydrodynamic regime where departures from the standard Fermi liquid behavior are also found~\cite{LDS16,GMetal18}. In sec. \ref{sec_magtrans} we will see that the anomaly-related transport coefficients fulfil these relations although a different claim has been made recently in Ref. \cite{KA20}.

\subsection{Luttinger theory of thermal transport coefficients \label{sec:luttinger}}

Transport coefficients are computed in the framework of linear response theory: the currents are obtained as a response to an applied field, usually by the derivation of the corresponding Kubo formulas. The electrical current results from the response to an applied electric field, which, if the given system is in equilibrium, is compensated by a gradient of chemical potential: $\vec{E}=\vec{\nabla}\mu$. This relation, the Einstein relation, is what sets the link between linear response to a field (electric field) and the response to a statistical force (the gradient of the chemical potential) (see Eq. \eqref{eq transport 1}). It permits to compute the response to a statistical force through linear response theory.

A question therefore naturally arises in the context of thermal transport: what is the ``thermal field" needed to develop a linear response theory of thermal transport? What is the equilibrium relation for the other statistical force, the gradient of temperature? In 1964, Luttinger solved these problems by showing that the gravitational field can play the role of the ``thermal field"~\cite{Luttinger1964}. By coupling the gravitational field to the electronic degrees of freedom he developed the linear response theory for thermal transport. In this section we review and contextualize Luttinger's theory and ideas.
\\

\subsubsection{Tolman--Ehrenfest pioneering work: Thermodynamic equilibrium in gravitational fields}

The earliest link that opened the door to relate gravitational effects with thermal transport in condensed matter systems 
was set up by the works of  Tolman and Ehrenfest in 1930 \cite{Tolman1930,TolmanE30} when trying to adapt the thermodynamic relations 
to general relativity. The underlying idea that ``heat has weight" let them to conclude that, 
in a gravitational field, thermal equilibrium must be accompanied by a temperature gradient 
in order to prevent the flow of heat from regions of higher to
those of lower gravitational potential. 
The basic equation defining equilibrium in a  curved  background was set to be
\beqn
T\sqrt{g_{tt}({\vec r})}={\rm constant}.
\eeqn
In a weak gravitational field $\phi$, 
\beqn
g_{tt}({\vec r}) = 1 + 2\phi({\vec r})/c^2, 
\label{eq_g_tt}
\eeqn
where we restore $c$ as the speed of light, we get the relation 
\beqn
\frac{1}{T} {\vec \nabla} T = - \frac{1}{c^2} {\vec \nabla}  \phi\,.
\label{eq:Luttinger}
\eeqn
 This relation was first obtained by Tolman \cite{Tolman1930} for a spherical mass distribution and it
was later generalized by Tolman and Ehrenfest \cite{TolmanE30} to general -static-  gravitational fields. As a consequence, the temperature variation in a gravitational field inducing  the acceleration $g$ was 
\beq
 \frac{1}{T} \frac{d  T}{d r}=-\frac{g}{c^2}.
\eeq  
The estimate of the temperature variation at the surface of the earth given by Tolman in Ref. \cite{Tolman1930} was
\beq
\frac{1}{T}\frac{d T }{d r}=-\frac{g}{c^2}\sim - 10^{- 18} {\mathrm{cm}}^{-1}. 
\eeq
Despite the small magnitude of the effect, the underlying ideas provided the basis for the Luttinger theory of thermal transport widely used in condensed matter~\cite{Luttinger1964}. 

The relation \eqref{eq:Luttinger} defining thermodynamic equilibrium in gravitational fields with specific distributions of matter, was later extended  to the presence of a finite chemical potential~\cite{Klein49}. This issue have been revised in recent works  \cite{SV18,SV19,Lima2019}. We will see a formal derivation in Sec. \ref{subsec_formalLuttinger}.

\subsection{Kubo formulas for the thermal and thermoelectric conductivities}

The previous ideas were used by Luttinger to derive a Kubo formula for the thermal transport coefficient
\cite{Luttinger1964}. In linear response theory \cite{GV05}, the transport coefficients  can be written in terms of the retarded current-current two-points function. A common example is the expression of the electrical 
conductivity \cite{Kubo57}:
\beq
J^i=\sigma^{ij} E_j \quad, \quad E_j=i\omega A_j;
\eeq
\beq
\sigma^{ij}(\omega) = \frac{1}{i\omega} \langle J^i J^j \rangle (\omega, k=0).
\label{eq_Kubo}
\eeq

The Kubo formula is based on the possibility of adding a local source ($A_\mu$) to the current ($J^\mu$) in 
the action in the standard way:
\beq
S[J]= S_0 - \int d^4 x J^\mu A_\mu,
\eeq
so that
\beq
J^\mu= - \frac{\delta S}{\delta A_\mu}.
\label{eq_j_vs_A}
\eeq

The difficulty to apply this approach to thermal transport lies in finding the appropriate local source 
coupling to heat (or energy) density. Early attempts  were  made by considering macroscopic thermodynamic variables as local variables by dividing the system in small portions and making some assumptions about how these variables develop in time \cite{Green52,Kubo57b,Mori58}.

Luttinger \cite{Luttinger1964} aimed to give a mechanistic sustain to what he called Green-Kubo-Mori formulas. Proceeding by analogy with the electrical conductivity, he proposed the introduction of a (fictitious) inhomogeneous gravitational field  $\phi(r)$ which  causes energy or heat currents to flow as a response to gradients of $\phi(r)$. This formalism is clear in special relativity, where the energy momentum tensor of a generic matter field $\phi$ is the conserved current associated
to metric variations due to Lorentz invariance: 
\beq
T^{\mu\nu}(\Phi) = - \frac{2}{\sqrt{-g}} \frac{\delta S[\Psi,g]}{\delta g_{\mu\nu}},
\label{eq_energy_momentum_tensor}
\eeq
where $\Psi$ represents matter fields.
In particular, the energy current is given by the components $J_{\mathcal{E}}^i=T^{0i}$. For small deviations from flat space the gravitational potential $\phi$ is proportional to  the zero-zero component of the metric: $g^{00}=1+2\phi$, which couples to the energy density $T^{00}$.  Here we follow \cite{Luttinger1964} and absorbed the velocity $c$ into the gravitational potential $\phi \rightarrow c^2 \phi$. This rescaled gravitational potential is simply the field that couples to the energy density.
The perturbed Hamiltonian to be used in the linear response formalism is \cite{Luttinger1964}
\beq 
H_{\mathrm{int}} (t) =  \int d^3r \; h({\vec r},t) \phi (t,\vec r),
\label{eq_source}
\eeq 
where $h({\bf r},t)$ is the Hamiltonian density\footnote{This follows since the metric variation $\phi$ couples  to the energy density $T^{00}$. On the other hand the operator measuring the energy density is (equivalent to) the Hamiltonian density. For example for a Dirac field $T^{00} = i \bar \psi \gamma^0 \partial^0\psi$ which is equivalent to the Hamiltonian density once the equations of motion are imposed on the field operator $\psi$.}. We can write the coefficients for the local current densities as responses to the external fields $A_0$ and $\phi$
\begin{gather}
\vec{J}={\bs{\mathcal{L}}}^{(1)}\vec{E} - {\bs{\mathcal{L}}}^{(2)}\vec{\nabla}\phi,\label{eq local currents 1}\\
\vec{J}^\mathcal{E} = {\bs{\mathcal{L}}}^{(3)}\vec{E} - {\bs{\mathcal{L}}}^{(4)}\vec{\nabla}\phi,
\label{eq local currents 2}
\end{gather}
which can in turn be written in terms of correlations of the currents. Note we use ${\bs{\mathcal{L}}}^{(i)}$ in Eqs. \eqref{eq local currents 1}, \eqref{eq local currents 2} to denote the coefficients that determine the local currents obtained from Kubo formulas, and $\bm L^{(i)}$ in Eqs. \eqref{eq transport 1}, \eqref{eq transport 2} to denote the coefficients that determine transport currents. The tensors ${\bs{\mathcal{L}}}^{(i)}$ and $\bm L^{(i)}$ may differ if magnetization currents are present, which are discussed in section \ref{sec:Luttheatmag}. The thermal and thermoelectric coefficients ${\bs{\mathcal{L}}}^{(2)}$ and ${\bs{\mathcal{L}}}^{(4)}$ can be computed as \cite{Luttinger1964}
\beq
{\mathcal{L}}^{(2)}_{ij} =\lim_{\omega\to 0} \frac{1}{i\omega} \langle J^\mathcal{E}_i J_j \rangle (\omega, k=0),
\label{eq_kappa}
\eeq
\beq
{\mathcal{L}}^{(4)}_{ij} =\lim_{\omega\to 0} \frac{1}{i\omega} \langle J^\mathcal{E}_i J^\mathcal{E}_j \rangle (\omega, k=0).
\label{eq_kappa2}
\eeq

The use of the gravitational field coupling to the energy density  whose variations induce energy currents is the first main contribution of Luttinger's work. Still the issue remains as how to couple the statistical force (temperature gradient) to the energy current. The passage from Eqs. \eqref{eq local currents 1}, \eqref{eq local currents 2} to Eqs. \eqref{eq transport 1}, \eqref{eq transport 2},  (this is, the relation between ${\bs{\mathcal{L}}}^{(i)}$ and $\bm L^{(i)}$) was first presented in the original work by Luttinger~\cite{Luttinger1964}. In what follows we will show a different re--formulation of the Luttinger approach  more suitable for making the connection with other aspects of the review.


\subsection{Formal derivation of generalized Luttinger relations}
\label{subsec_formalLuttinger}

Here we present a formal
approach to generalized Luttinger relations. It is based on introducing external fields to mimic variations in thermodynamic parameters, the temperature $T$, chemical potential $\mu$ and velocities $u_a$ (see Appendix \ref{sec_velocities}).
This allows to derive Kubo formulas for electric and thermal conductivities from a unifying principle or, more generally, for transport coefficients appearing in an effective hydrodynamic description of a physical system. 

We start from the very basic assumption that we have an action given as a functional
of the metric and a gauge field
\begin{equation}
S = S[g, A]\,.
\end{equation}
The precise form of this function will depend on the underlying physical theory but we are only interested in the way a system described by $S$ reacts to changes in the external fields, the metric and the gauge field. 
We can define the energy-momentum tensor through variations of the metric (see Eq.~\eq{eq_energy_momentum_tensor}) and the electric current through variations of the gauge field:
\begin{align}\label{eq:defcurrent}
J^\mu & = - \frac{1}{\sqrt{-g}} \frac{\delta S}{\delta A_\mu(x)}\,.
\end{align}
If the action is invariant under diffeomorphisms and gauge transformations the current and energy-momentum tensor (\ref{eq_energy_momentum_tensor}) obey conservation laws. 
The conserved momentum and charge can be obtained by integrating over the spatial volume
\begin{align}
\label{eq:momentum}
P_\mu &= \int d\Sigma\, T^t_\mu \,,\\
\label{eq:charge}
Q &= \int d\Sigma\, J^t\,.
\end{align}
The time component of the four-vector $P_\mu$ is the total energy (the Hamiltonian) and the spatial components are the (spatial) momenta. 
Now we can form the general statistical operator (see Appendix \ref{sec_velocities})

\begin{equation}\label{eq:densitymatrix}
\rho = \exp \left[ - \frac{1}{T}\left(u^\nu P_\nu - \mu Q \right) \right] \,,
\end{equation}
where $u^\nu$ is a four velocity, $T$ the temperature and $\mu$ the chemical potential.
We will assume an ensemble at rest with $u^\nu=(1,0,0,0)$. 
The four-velocity is constrained to be a time-like unit vector $u^2 = 1$. This constraint is solved by writing 
\begin{equation}
    u^\mu = \left(\frac{1}{\sqrt{1-\vec\beta^2}}, \frac{\vec{\beta}}{\sqrt{1-\vec\beta^2}}\right)\,,
\end{equation}
in terms of a purely spatial velocity $ \vec \beta = \vec v/c$. Here $c$ can be interpreted as the speed of light in the context of high-energy physics or as the Fermi-velocity in a Dirac material. 

We are interested on the effect that a variation of the gauge field and metric has on the thermal ensemble. For simplicity and because it is also the most relevant case in condensed matter applications we consider the case of a small metric perturbation on top of a flat background geometry $g^{\mu\nu}=\mathrm{diag(1 + \delta g^{tt},-1,-1,-1)}$.
By the definitions \eq{eq_energy_momentum_tensor} and \eq{eq:defcurrent}, we can find the change of the Hamiltonian under small and constant variations of the temporal component $A_t$ of the gauge field and the $g^{tt}$ component of the metric. Since in terms of the kinetic ($T$) and the potential ($V$) the action and the Hamiltonian can be written as $S=T-V$ and $H=T+V$, respectively, it follows that $\delta S = - \delta H$ if the variation $\delta$ only concerns $V$, as it does here. From this it follows that
\begin{equation}
\delta H = -\frac{1}{2} \delta g^{tt} H + \delta A_t Q\,,
\end{equation}

So far we have changed the microscopic Hamiltonian by switching on small increments in gravitational and electrostatic potential.
Now we want to ask if we can mimic the effects of these increments by introducing small variations in
the thermodynamic variables $T$ and $\mu$.
In other words we now keep the microscopic Hamiltonian fixed but change that state thermodynamic state characterised by increments $T+\delta T$ and $\mu+\delta\mu$.
We demand therefore that the statistical operators of the new system with $\delta g^{tt}$ and $\delta A_t$ are equivalent to the statistical operator of the old system in the new state $T+\delta T$ and $\mu +\delta \mu$
\begin{equation}
\exp\left[-\frac{1}{T}
(H+\delta H -\mu Q)\right] =
\exp\left\{ -\frac{1}{T}\left[\left(1 - \frac{1}{2}\delta g^{tt}\right)H -(\mu-\delta A_t) Q\right] \right\} =
\exp\left[-\frac{1}{T+\delta T}( H - (\mu+\delta\mu) Q)\right].
\end{equation}
To first order in $\delta T$ and $\delta\mu$ and introducing the Newtonian potential $ \phi = - \delta g^{tt}/2$ this leads to the conditions
\begin{align}\label{eq:Luttg}
    \phi &=  -\frac{\delta T}{T} =  T \delta\left(\frac{1}{T} \right)\,,\\ \label{eq:LuttA}
    \delta A_t &= -\delta \mu + \mu\frac{\delta T}{T} = - T\delta\left(\frac{\mu}{T}\right)\,.
\end{align}
Finally, we can also study the consequences of the variation $\delta g^{xt}$. It leads to 
\begin{equation}
\label{deltagxt}
\delta H =  -\delta g^{xt} P_x\,,
\end{equation}
which leads to a shift in the fluid velocity 
\begin{equation}\label{eq:Luttingerv}
\delta u^x = -\delta g^{xt}\,.
\end{equation}
In this way variations in the external fields can be translated into
variations in the thermodynamic variables.
While these considerations apply to equilibrium and for spatially uniform variations they also apply in the presence of spatial gradients if we assume that the system is in local thermal equilibrium. The assumption of local thermal equilibrium implies
that gradients are small on length scales of the order of the mean free path and that thermodynamic relations hold locally.
More precisely a system in local thermal equlibrium can be described by local and time dependent temperature $T(t,\vec x )$, chemical potential $\mu(t,\vec x)$ and fluid velocity $u^\mu(t, \vec x)$.
Under these conditions it is legitimate to substitute $\delta \rightarrow \vec\nabla$ in (\ref{eq:Luttg}), (\ref{eq:LuttA}) and (\ref{eq:Luttingerv}).
These relations imply that the condition for remaining in equilibrium are
\begin{align}
\label{eq:relA-chem-pot}
   -\vec{\nabla}A_t - T \vec{\nabla}\left(\frac{\mu}{T}\right) =&0 \,,\\
     \vec{\nabla} \phi- T\vec{\nabla}\left(\frac{1}{T}\right) =&0
    \label{eq:lutt-rel-temp}
\end{align}
These equations are however neither gauge invariant nor relativistically covariant. Whereas it is clear that gauge invariance implies the substitution $-\vec\nabla A_t \rightarrow \vec E$,
it is not immediately clear how a covariant term for the gravitational potential can be written. The problem is that there is no covariant tensor that can be constructed out of the metric and its first derivatives. Indeed the curvature tensor $R_{\alpha\beta\mu\nu}$ is second order in derivatives on the metric. 
It is possible however to write down a term of the correct form with the help of the velocity field $u^\mu$.
An appropriate phenomenological constitutive relation that is first order in derivatives, gauge invariant and covariant is 
\begin{equation}
\label{eq:currhydro}
     J^\mu = \rho u^\mu + P_{\perp}^{\mu\nu} \left[ \sigma \left( E_\nu -  T \nabla_\nu (\frac{\mu}{T})\right) + \kappa T \nabla_\nu (\frac 1 T) + \xi u^\lambda \nabla_\lambda u_\nu \ \right]\,.
\end{equation}
Here $\rho$ is the charge density
(or energy density if we consider the energy current) and $P_\perp^{\mu\nu}= g^{\mu\nu}-u^\mu u^\nu$ is the projector transverse to $u^\mu$. 
This ensures the $u_\mu J^\mu = \rho$ without any first order term in derivatives. 
 The covariant electric field is $E_\mu = F_{\mu\nu} u^\nu$.
 The last term depends on the frame vector $u^\mu$ only.
 Expanding around a flat metric $g_{\mu\nu} = \eta_{\mu\nu} + \delta g_{\mu\nu}$ 
 and in the rest frame $u^\mu = (1,0,0,0)$ this last term becomes
\begin{equation}
\label{eq:gravitoelectricfield}
    P^{\mu\nu}_\perp u^\lambda\nabla_\lambda u_\nu = -\delta^{\mu i} \nabla_t u_i =
    -\delta^{\mu i} (\delta\dot{g}_{it} - \frac 1 2 \partial_i  
    \delta g_{tt}) = - \delta^{\mu i} E_i^g
\end{equation}
where $E_i^g$ is the gravito-electric field (see Appendix \ref{sec_GEM}). It is therefore the covariant term that encodes the action of the gravitational field in analogous way
of the electric field. The condition that in equilibirium the current should vanish implies then $\xi = -\kappa$.
The physical interpretation of $u^\lambda \nabla_\lambda u_\nu$ is a covariant version of the acceleration. 

We note that $h_{ti}$ can be viewed as a gravito-magnetic vector potential $A_{g,i}$. A vector field as source for the energy current will further be discussed in section \ref{sec:Luttheatmag}.
\\

{\bf Luttinger's transport equations}
\\

We can, at this point, complement Eqs. \eqref{eq transport 1}, \eqref{eq transport 2} by including Luttinger's gravitational potential. From the relation \eqref{eq:lutt-rel-temp} we can write
\begin{gather}
\vec{J}_{\textrm{tr}}=\bm L^{(1)}\left(\vec{ E}-T\vec{\nabla}\frac{\mu}{T}\right)+\bm L^{(2)}\left(-\vec{\nabla}\phi+T\vec{\nabla}\frac{1}{T}\right),\label{eq transport 21}\\
\vec{J}_{\textrm{tr}}^\mathcal{E}=\bm L^{(3)}\left(\vec{ E}-T\vec{\nabla}\frac{\mu}{T}\right)+\bm L^{(4)}\left(-\vec{\nabla}\phi+T\vec{\nabla}\frac{1}{T}\right).
\label{eq transport 22}
\end{gather}
From Eqs. \eqref{eq:relA-chem-pot} and \eqref{eq:lutt-rel-temp} we see that, in equilibrium, $\vec{ J}_{tr}=\vec{ J}^\epsilon_{tr}=0$. This is correct, since $\vec{ J}_{tr}$, $\vec{ J}^\epsilon_{tr}$ are transport currents and should vanish in equilibrium. It is clear now what is the relation between the coefficients computed by the Kubo-formulas (${\bs{\mathcal{L}}}^{(i)}$), Eqs. \eqref{eq local currents 1} and \eqref{eq local currents 2}, and the transport coefficients:
\begin{equation}
    {\bs{\mathcal{L}}}^{(i)}=\bm L^{(i)}\,, \qquad \mbox{(no magnetisation current)}\,,
    \label{eq kubo-transport rel. 1}
\end{equation}
There is still a small, or perhaps not that small, caveat here. The relation ${\bs{\mathcal{L}}}^{(i)}=\bm L^{(i)}$ only generally holds in time-reversal invariant systems, where magnetization currents are absent. The Kubo-formula coefficients capture the total local current at any given point in the sample. This includes contributions from magnetization currents. But because magnetization currents do not contribute to transport, ${\bs{\mathcal{L}}}^{(i)}=\bm L^{(i)}$ no longer holds. We will discuss this further in section \ref{sec_magnet}.
\\

\begin{tcolorbox}
\begin{itemize}
\item{\bf An inhomogeneous gravitational field  can be used to  generate energy (heat) currents to derive Kubo formulas of thermal transport coefficients.}
\item{\bf 
In a system residing in a local thermodynamic equilibrium, small gradients in the background gravitational and gauge potentials generate gradients in temperature, chemical potential and flow velocity:
\begin{align}
    \vec{\nabla} \phi = -\frac{\vec{\nabla} T}{T}\quad, \quad
    \vec{\nabla} A_t = - T \vec{\nabla} \left( \frac{\mu}{T}\right) \quad, \quad
    \vec{\nabla} g^{it} = -\vec{\nabla} u^i\,.\nonumber
\end{align}
}
\end{itemize}

\end{tcolorbox}


\section{Transport and magnetization currents}
\label{sec_magnet}

\subsection{General remarks}

In systems where time-reversal symmetry is broken, the definition of transport currents, measured in transport experiments becomes more subtle due to the presence of magnetization currents. Magnetization currents are circulating currents that arise when time-reversal symmetry is broken. These appear for instance in the presence of an external magnetic field, and do not contribute to the net transport through the sample. Therefore, in general, the local particle and energy currents $\vec{J}$, $\vec{ J}^\epsilon$, as given by the Kubo formulas, are written as the sum of transport plus magnetization currents
\begin{equation}
    \vec{ J}=\vec{ J}_{\mathrm{tr}}+\vec{ J}_{\mathrm{mag}},\quad \vec{ J}^\epsilon=\vec{ J}^\epsilon_{\mathrm{tr}}+\vec{ J}^\epsilon_{\mathrm{mag}}.
\label{eq_magnetization_currents}
\end{equation}
The magnetization currents can be defined as the rotational of a (energy) magnetization density
\begin{equation}
\label{eq:energy-mag-current}
    J_{i,\mathrm{mag}}=\epsilon_{ijk}\partial_j M_k,\quad  J^\epsilon_{i,\mathrm{mag}}=\epsilon_{ijk}\partial_j M^\epsilon_k,
\end{equation}
The magnetization vanishes outside the sample, and therefore $J_{i,\mathrm{mag}}$, $J^\epsilon_{i,\mathrm{mag}}$ average to zero over a cross-section of the sample. At vanishing external fields $A_\mu=\phi=0$, we can define the unperturbed magnetization densities $M_0^i(\mu,T)=M^i(A_\mu=\phi=0)$, $M_0^{\epsilon,i}(\mu,T)=M^{\epsilon,i}(A_\mu=\phi=0)$. In this situation, the magnetization is constant in the bulk, but because it vanishes outside the material, it will have a pronounced variation at the boundary giving rise to boundary magnetization currents. For non-vanishing external fields, $A_t,\phi\neq 0$, the total magnetization density can be written in terms of the unperturbed magnetization~\cite{Cooper:1997gp}:
\begin{equation}
    \vec{ M}=(1+\phi)\vec{ M}_0(\mu,T), \quad \vec{ M}^\epsilon=(1+2\phi)\vec{ M}_0^\epsilon(\mu,T)+A_t\vec{ M}_0(\mu,T),
\end{equation}
and the bulk magnetization currents are then given by \cite{Cooper:1997gp}
\begin{equation}
\label{eq magnetization 1}
    J_{i,\mathrm{mag}}=-\epsilon_{ijk}\left(M_{0,j}\partial_k\phi+\frac{\partial M_{0,j}}{\partial\mu}\partial_k\mu+\frac{\partial M_{0,j}}{\partial T}\partial_kT\right),
\end{equation}
\begin{equation}
\label{eq magnetization 2}
    J^\epsilon_{i,\mathrm{mag}}=-\epsilon_{ijk}\left(-M_{0,j}E_k+2M^\epsilon_{0,j}\partial_k\phi+\frac{\partial M_{0,j}^\epsilon}{\partial\mu}\partial_k\mu+\frac{\partial M_{0,j}^\epsilon}{\partial T}\partial_kT\right).
\end{equation}

To obtain the transport coefficients $\bm L^{(i)}$ defined  in Eqs. \eqref{eq transport 1} and \eqref{eq transport 2}, one should subtract the magnetization currents (for $\vec{\nabla}\mu=\vec{\nabla} T=0$) from the local currents \eqref{eq local currents 1}, \eqref{eq local currents 2}. The transport coefficients are then given in terms of the coefficients obtained from the Kubo-formulas, ${\bs{\mathcal{L}}}^{(i)}$, and the unperturbed magnetization densities
\beqn
\label{eq transport coeff 1}
    \bm L^{(1)} & = & {\bs{\mathcal{L}}}^{(1)}=\bm\sigma,
    \hskip 15mm 
    L^{(2)}_{ij}  = {{\mathcal{L}}}^{(2)}_{ij}+\epsilon_{ijk}M_{0,k},\\
\label{eq transport coeff 2}
    L^{(3)}_{ij} & = & {{\mathcal{L}}}^{(3)}_{ij}+\epsilon_{ijk}M_{0,k}, 
    \qquad 
    L_{ij}^{(4)} = {{\mathcal{L}}}_{ij}^{(4)}+2\epsilon_{ijk}M^\epsilon_{0,k},
\eeqn
The transport coefficients are related under time-reversal by Onsager relations \cite{31Onsager} as: $\sigma_{ij}(B)=\sigma_{ji}(-B)$, $L^{(4)}_{ij}(B)=L^{(4)}_{ji}(-B)$, $L^{(2)}_{ij}(B)=L^{(3)}_{ji}(-B)$, where $B$ denotes the magnetic field or any other parameter that breaks the time-reversal symmetry. It is important to state that it is the transport coefficients $\bm L^{(i)}$, and not the ones in the local currents, ${\bs{\mathcal{L}}}^{(i)}$, which fulfill the Mott rule and the Wiedemann-Franz law~\cite{Smrcka_1977,JG84,Cooper:1997gp,QTN11}.

Transport currents are then obtained as a response to gradients of the statistical fields ($\mu$, $T$) by combining Eqs. \eqref{eq transport coeff 1}, \eqref{eq transport coeff 2} and \eqref{eq transport 21}, \eqref{eq transport 22}, and setting $\vec{ E}=\vec{\nabla}\phi=0$. To obtain the local currents as a response to $\vec{\nabla}\mu$, $\vec{\nabla}T$, the magnetization currents \eqref{eq magnetization 1}, \eqref{eq magnetization 2} have to be added to the transport currents. This will give the total current locally at any point in the sample.
\\

\subsection{Luttinger, thermal transport and energy magnetization}
\label{sec:Luttheatmag}

In section \ref{subsec_formalLuttinger} we introduced the Luttinger relation Eq.~\eqref{eq:Luttinger} as a gradient of a gravitational potential, and showed that we can include a thermal vector potential, in analogy with an electromagnetic field. 
The inclusion of a thermal vector potential turns out to be related to several issues and necessary extensions that concern the original work by Luttinger to capture a broader set of phenomena. First, unphysical divergences at $T\to 0$ have been encountered using the Luttinger derivation of the Kubo formula for thermal transport currents~\cite{QTN11,Tatara:2015bf}. Second, Luttinger did not incorporate magnetization currents, which were added later~\cite{Cooper:1997gp,Smrcka_1977,QTN11}, resulting in \eqref{eq magnetization 1} and \eqref{eq magnetization 2}. Lastly, the difference between diamagnetic and paramagnetic currents, explained below, is not as evident in the Luttinger formalism as they are in electromagnetism~\cite{Tatara:2015bf}. 

The first two issues were addressed~\cite{Cooper:1997gp,Smrcka_1977,QTN11} by noticing that the subtraction of the equilibrium magnetization currents from the result of linear response theory leads to well defined transport currents, see Eqs. \eqref{eq transport coeff 1}, \eqref{eq transport coeff 2}, even as temperature approached zero~\cite{QTN11}. Tatara observed~\cite{Tatara:2015bf} that a proper treatment of the diamagnetic heat current term will lead to this conclusion as well. In electromagnetism, the paramagnetic terms arise from terms in the Hamiltonian proportional to the vector potential $\vec{A}$, while diamagnetic terms stem from terms proportional to the square of $\vec{A}$. While the Dirac Hamiltonian only gives rise to paramagnetic terms because it is linear in momentum, the quadratic Schr\"{o}dinger Hamiltonian will generically contain both paramagnetic and diamagnetic terms. In electromagnetism both terms are important as the paramagnetic current contains an equilibrium contribution determined by the electrons below the Fermi level which exactly cancels the diamagnetic current~\cite{Tatara:2015bf}. By formulating the Luttinger relation in terms of a vector potential, $\vec{A}_g$, Tatara was able to isolate the diamagnetic and paramagnetic contribution similarly to electromagnetism. This formulation reinterprets the results in~\cite{Cooper:1997gp,Smrcka_1977,QTN11}, sheds light into the issues and extensions discussed above related to the Luttinger approach, and emphasizes the parallelism between electromagnetism and gravitoelectromagnetism.

The emergence of a vector thermal potential $\vec{A}_g$ can be understood as arising from conservation laws~\cite{Tatara:2015bf}. In the Luttinger formalism the coupling to the gravitational field is imposed through the Hamiltonian density $h(\mathbf{r})$ given by $H_{\mathrm{int}}$ as written in \eqref{eq_source}
\begin{equation}
\label{eq:TataraLutt}
    H_\mathrm{int} = \int d^3{r}\;  \phi(\vec{r}) h(\vec{r}).
\end{equation}
As discussed in Sec.~\ref{subsec_formalLuttinger} the responses caused by a thermal gradient can be calculated in powers of $\nabla \phi$ and identifying $\vec{\nabla} \phi = -\frac{\vec{\nabla} T}{T} $~\cite{Luttinger1964}.
Using the conservation law ${\partial}_t h(\vec{r}) +\vec{\nabla} \vec{J}_\epsilon=0$, where $\vec{J}_\epsilon$ is an energy current, and restricting to steady-state properties, it is possible to rewrite this equation as
\begin{equation}
\label{eq:TataraAT}
    H_{A_g} = -\int d^3{r}\; \; \vec{J}_\epsilon(\vec{r},t)\cdot \vec{A}_g(\vec{r},t).
\end{equation}
By identifying $\partial_t \vec{A}_g= -\frac{\vec{\nabla} T}{T}$, Tatara showed that the correct diamagnetic contribution is naturally captured, since $H_{A_g}$ depends explicitly on a vector potential. 
It is important to note that both formulations, either $H_{\mathrm{int}}$ or $H_{A_g}$ are equivalent. Both can incorporate magnetization currents and can be regular at $T\to 0$. The formalism in terms of $H_{A_g}$ allows to distinguish the diamagnetic contribution explicitly, and thus establishes a strong link between gravitoelectric fields and electromagnetic fields. Taking both $H_{\mathrm{int}}$ and $H_{A_g}$ together, a gauge degree of freedom emerges. So long as we are interested in steady-state properties, we are free to choose $\partial_t \vec{A}_g$ and $\vec{\nabla} \phi$ independently, so long as we identify $\vec{\nabla}\phi + \partial_t \vec{A}_g = -\frac{\vec{\nabla} T}{T}$~\cite{Tatara:2015bf}.

The emergence of a vector thermal potential $\vec{A}_g$ has been described by resorting to the Newton-Cartan formalism~\cite{Shitade2014,Gromov2015,BR15,Geracie2015}. The conservation equations for momentum and energy densities follow from global space and time translations. To promote these to local symmetries one can define a local frame in terms of the vierbeins $e^{\mu}_a$. The resulting theory is similar to general relativity complemented by torsion, known as the Einstein-Cartan formalism  described in sec. \ref{sec_torsion} and in Appendix \ref{sec_curved}.
If Lorentz symmetry is absent, as natural in condensed matter systems, it was suggested that the Newton-Cartan formalism is appropriate to describe the geometrical properties of the theory. Refs.~\cite{Shitade2014,Gromov2015,BR15,Geracie2015} showed that by taking variations of the action with respect to  $e^{\mu}_a$ and the metric, one can obtain the correct thermal transport and magnetization currents. In this formalism both $\vec{A}_g$ and $\phi$ appear naturally, as a consequence of local coordinate transformations.

It is important to point out that the results of Refs.~\cite{Shitade2014,Gromov2015,BR15,Geracie2015} can be interpreted in terms of derivatives of the velocity field, $u_\mu$, in general curved, but torsion-free, backgrounds. This is suggested by the results of a recent work~\cite{FL20}. At finite temperature and chemical potential, gravito-electromagnetic fields can appear, even in torsion-free backgrounds, as derivatives of the velocity field, see Eqs. \eqref{eq:currhydro} and \eqref{eq:gravitoelectricfield} and the associated discussion. This can not happen in vacuum, where the only covariant tensor involving first order derivatives of the vielbein is the torsion tensor. Therefore, at finite chemical potential and temperature, responses to gravito-electromagnetic fields can naturally appear in the absence of torsion.
\\

To emphasize this point explicitly let us assume that we have a magnetized material. The magnetization is a term in the free energy of the form
\begin{equation}\label{eq:Fmag}
   F_B = \int d^3x \,\vec{B}\cdot\vec{M}\,. 
\end{equation}
Instead of the free energy we switch now to the action formalism. This is always preferable if we want to make the symmetries manifest. The action should be a covariant object and is integrated over space and time. The magnetization part of the action is then the covariant form of (\ref{eq:Fmag})
\begin{equation}\label{eq:Smag}
    S_B =
    \frac 1 2 \int d^4x\, \epsilon_{\mu\nu\rho\lambda}  F^{\mu\nu} u^\rho M^\lambda
\end{equation}
 Now we can compute the current by differentiating with respect to the gauge field
\begin{equation}
    J_{\mu,\mathrm{mag}} = \epsilon_{\mu\nu\rho\lambda}\partial^\nu\left(
     u^\rho M^\lambda
    \right)\,.
\end{equation}
In the rest frame $u^\mu=(1,0,0,0)$ this gives the well-known result
\begin{equation}
    \vec{J}_\mathrm{mag} = \vec\nabla \times \vec M \,.
\end{equation}
Now we want to generalize this to the energy current. The gauge field for energy (and momentum) currents is the metric. We face however again the problem that there is no covariant tensor built out of the metric and its first derivatives. Terms with a Riemann or Ricci curvature tensor would lead to higher order terms in derivatives. 
The way out of this problem is to consider the velocity field as defined in Appendix \ref{sec_velocities}. For example, let us consider a system under rotation with a covariant angular velocity $\Omega_\mu =\frac{1}{2} \epsilon_{\mu\nu\rho\lambda} u^\nu \nabla^\rho u^\lambda$. Then we can write down the action
\begin{equation}
\label{eq:actionrot}
    S_\mathrm{rot} = -\int d^4x\, \Omega_\mu M^\mu_\epsilon = -\frac{1}{2}\int d^4x\, \epsilon_{\mu\nu\rho\lambda}  M^\lambda u^\mu \nabla^\nu u^\rho\,,
\end{equation}
where now we take a general covariant derivative 
\begin{equation}
    D^\nu u^\rho = g^{\nu\sigma} ( \partial_\sigma u^\rho + \Gamma^\rho_{\sigma\kappa} u^\kappa )\,.
\end{equation}
In the restframe $u^\mu=(1,0,0,0)$ the action is therefore
\begin{equation}
    S_\mathrm{rot} = \frac{1}{2}\int d^4x\, \epsilon_{0ijk}\delta^{il} \Gamma^{j}_{l0} M^k\,,
\end{equation}
and to first order in the metric perturbation
\begin{equation}
   \Gamma^{j}_{l0} = -\frac{1}{2}\delta^{jm}
   \left( \partial_l \delta g_{m0} +
   \partial_0 \delta g_{ml} - \partial_m \delta g_{l0} \right)\,. 
\end{equation}
We insert this into the expression for the rotational action \eqref{eq:actionrot} noting that the term with symmetric spatial indices on the metric perturbation vanish due to contraction with the anti-symmetric epsilon tensor. We can also introduce the gravito-magnetic field $A^g_i = \delta g_{it}$. 
Since now we have specified a frame it also makes sense to write down the gravito-magnetic contribution to the free
energy defined as $\int d t F_\mathrm{g} = -  S_\mathrm{g}$
\begin{equation}
    F_{g} = \frac{1}{2}\int d^3x\, \epsilon_{ijk} \partial_i A^g_j M^k = \int d^3x\, \vec{B}_g\cdot \vec{M}_\epsilon.
\end{equation}
This leads to the gravito-magnetization current
\begin{equation}
    \vec{J}_\epsilon = \vec\nabla\times \vec{M}_\epsilon\,.
\end{equation}
We note that the gravito-magnetic potential $\vec A_g$ is the thermal vector potential coupling to the energy current, see \eqref{eq:TataraAT}~\cite{Tatara:2015bf}.
Finally we can also give a physical interpretation of the gravito-magnetization $\vec{M}_\epsilon$, which is nothing but the energy magnetization defined in Eq. \eqref{eq:energy-mag-current}.
Since it can be traced back to the coupling to angular velocity $\vec \Omega$, $\vec{M}_\epsilon$ is just the angular momentum $\vec{M}_\epsilon =  \vec L$. For the non-relativistic case, the magnetization currents will be determined by a mix of spin and orbital contributions~\cite{Gromov2015}.
\\

\begin{tcolorbox}
\begin{itemize}
\bf{
\item{\bf In the presence of magnetic fields (or other time-reversal breaking effects) magnetization currents appear, which do not contribute to net electrical and thermal transport.}
\item{\bf In this situation, the Wiedemann--Franz and Mott relations are not fulfilled by the coefficients obtained through the Kubo formulas. These relations still hold for the transport coefficients defined in Eqs.~\eqref{eq transport coeff 1} and \eqref{eq transport coeff 2}}.
\item The most general theory to derive the thermal transport and magnetization currents contains both a gravitational potential $\phi$ and a thermal vector potential $\vec{A}_g$, generalizing Luttinger's derivation.
\item The thermal vector potential $\vec{A}_g$ allows to understand the presence of magnetization currents in terms of a gravito-magnetic field.}
\end{itemize}
\end{tcolorbox}


\section{Transport and chiral anomalies}
\label{sec_chiralanoms}
\subsection{Anomalies in QFT}

\begin{figure}[!thb]
\begin{center}
\includegraphics[scale=0.50,clip=true]{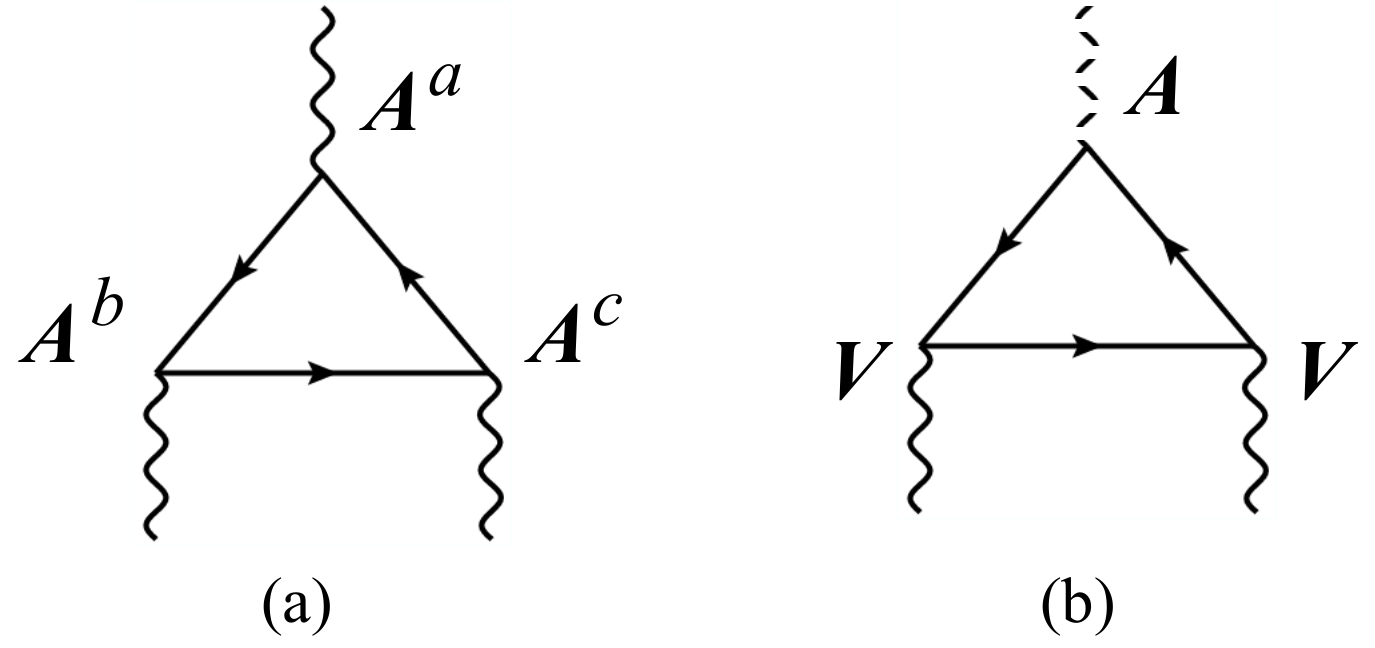}
\end{center}
\caption{Anomalies can be easily detected in triangle diagrams arising in perturbation theory. The corresponding Feynman diagrams are divergent and need to be regularized. It is the regularization that breaks the classical chiral symmetry. Depending on the details of the regulator one obtains different forms of anomalies. On the right panel we show the triangle diagram for three general symmetries. One might employ a regularization scheme that treats all three vertices in a symmetric way. On the other hand in figure (b) we show the triangle leading to the axial anomaly. The $V$-vertices are singled out in that no anomaly should be associated to the $V$ vertices since they correspond to the coupling to the electromagnetic gauge field. In that case the anomaly has to be concentrated in the $A$ vertex. An alternative is to single one one particular vertex and demand that it couples covariantly to the gauge fields sitting at the other two vertices. This scheme defines the covariant anomaly.}
\label{fig_scheme_chiral}
\end{figure}

In what follows we will describe the most general form of triangle anomalies  of  Dirac fields with non-Abelian  symmetries and the new anomaly-induced transport phenomena associated to them. Other  interesting aspects of the quantum anomalies not discussed here, including the case of (1+1) dimensional system and the topological implications of anomalies, are discussed pedagogically in Ref.~\cite{Fradkin20}

In perturbation theory anomalies appear at one loop in the three-point function correlation functions of classically conserved currents and the energy momentum tensor (triangle diagrams, see figure \ref{fig_scheme_chiral}) \cite{Adler:1969,Bell:1969,Kimura:1969wj}. Alternatively they appear in the regularized definition of the path integral measure \cite{Fujikawa:1979ay}. In four dimensions the most general form of triangle anomalies is
\begin{align}\label{eq:covanomalyJ}
D_\mu J^\mu_a &= \frac{d_{abc}}{32\pi^2} \epsilon^{\mu\nu\rho\lambda} F_{\mu\nu}^b F_{\mu\nu}^c + \frac{b_a}{768\pi^2} \epsilon^{\mu\nu\rho\lambda} R^\alpha\,_{\beta\rho\lambda} R^\beta\,_{\alpha\rho\lambda} \,,\\
D_\mu T^{\mu\alpha} &= F^{\alpha\beta}_a J^a_ \beta + \frac{b_a}{384\pi^2} \epsilon^{\mu\nu\rho\lambda} D_\beta \left(
F_{\mu\nu}^a R^{\alpha\beta}\,_{\rho\lambda}\right).
\label{eq:covanomalyT}
\end{align}
The indices denoted by latin letters enumerate different currents $J^\mu_a$ which couple to gauge fields $A_\mu^a$. 
These currents can be the currents for left- and right-handed  fermions, or the  axial or vector like combination thereof.
In general they are the Noether currents corresponding to
a symmetry.
The $F^a_{\mu\nu}$  are the field strengths of the gauge fields, $F^a_{\mu\nu}=\partial_\mu A^a_\nu-\partial_\nu A^a_\mu+ f^{abc} A^b_\mu A^c_\nu$, where $f^{abc}$ the structure constants.
The symmetries are generated by Hermitian matrices $T^a$ with commutation relations $[T^a, T^b] = i f^{abc} T^c$. 
For Abelian symmetries the structure constants vanish. The Riemann tensor $R^\beta\,_{\alpha\rho\lambda}$  accounts for curvature in space-time and is defined in eq. \eqref{eq:Riemann}.

The currents  and the energy-momentum tensor are composite operators in terms of the fundamental fermion fields. 
In quantum field theory a composite operator is one that is built out of local products of the fundamental quantum fields. For free fermions and in flat space they take the simple forms\footnote{We emphasize that these are the expression in flat space. In general curved space the energy momentum tensor depends in a non-trivial way on the metric (rather the vielbein) and this has to be taken into account in calculating the anomaly. For details see the discussion in \cite{Kimura:1969wj,AlvarezGaume:1983ig}.}
\begin{align}
J^\mu_a(x) &= \bar \Psi_f(x) (T_a)^f_g \gamma^\mu \Psi^g(x)\,,\\
T^{\mu\nu}(x) &= \frac{i}{2} [\bar\Psi_f(x) \gamma^{(\mu} \partial^{\nu)} \Psi^f(x) -\partial^{(\nu} \bar\Psi_f(x)  \gamma^{\mu)}  \Psi^f(x) ]\,.
\end{align} 
For the Abelian vector and axial currents Eqs. (\ref{eq:JVcurrent}), (\ref{eq:J5current}) of a Dirac fermion,  $T_a$ is taken from $\{\mathbf{1},\gamma_5\}$. The round brackets on the indices denote symmetrization according to $ A^{(\mu} B^{\nu)} = \frac 1 2 (A^\mu B^\nu+A^\nu B^\mu)$. Famously, quantum field theory suffers from ultraviolet divergences which must be dealt with by a regularization and renormalization scheme.
In particular a composite operator is only defined up to a specific regularization scheme. The anomaly has to be
understood as the fact that no regularization scheme exists which allows to realize all classical conservation laws in the quantum theory. However, different regularization schemes might give rise to different quantum definitions of the current operator. These ambiguities can be
fixed by demanding certain physically plausible conditions on the currents and operators.
One possible definition of regularization is to demand that the currents themselves 
transform as well defined tensors ("covariantly") under the 
symmetry transformations even if the symmetry is anomalous. It turns out that this is a unique way of regularizing the current operators and it gives rise to what is called the  covariant anomaly \cite{Bardeen:1984pm}. This regularization scheme is particularly well suited for answering the question if a theory does have anomalies. The anomalies in eqs. (\ref{eq:covanomalyJ}) and (\ref{eq:covanomalyT}) are precisely these covariant anomalies. It turns out that their existence is completely determined by two simple group invariants. These are the  anomaly coefficients 
\begin{align}\label{eq:acoeffgauge}
d_{abc} &=   \frac 1 2 \sum_r \mathrm{tr}(T_a.\{T_b,T_c\})_r -  \frac 1 2\sum_l \mathrm{tr}(T_a.\{T_b,T_c\})_l \,,\\ \label{eq:acoeffgrav}
b_a &= \sum_r \mathrm{tr}(T_a)_r - \sum_l \mathrm{tr}(T_a)_l \,,
\end{align}  
where the sums are over the representations under which the right-handed and left-handed chiral fermions transform and curly brackets denote the anti-commutator.
For a $U(1)$ symmetry, the generator $T_a$ in a specific representation is just as the charge $q_a$. 
The Adler-Bardeen non-renormalization theorem \cite{Adler:1969er,Adler:2004qt} guarantees that there are no further quantum corrections to these numbers. The existence of anomalies can be unambiguously determined by the calculations of the anomaly coefficients $d_{abc}$ and $b_a$.
\\

As an example, let us consider the single massless  Dirac fermion. At the classical level a Dirac fermion is the
direct sum of a right-handed and a left-handed fermion. We denote the
charge of the left- and right-handed fermions by $(q_r,q_l)$. In a vector-axial basis we have two symmetries with charges $U(1)_V$: $(1,1)$ and $U(1)_5$: $(1,-1)$ . Accordingly the non-vanishing anomaly coefficients are $d_{5VV}=d_{V5V}=d_{VV5} = 2$, $d_{555}=2$ and $b_5 = 2$. We note here that in this case all these anomaly coefficients take the same value. 

Now we run however into something that appears troublesome. The anomaly coefficient $d_{abc}$ is completely symmetric, and this
means that for a Dirac fermion the  covariant vector current is anomalous as well
\begin{equation}\label{eq:AVcov}
\partial_\mu J^\mu_V = \frac{1}{8\pi^2} \epsilon^{\mu\nu\rho\lambda} F_{\mu\nu}^V F_{\rho\lambda}^5\,.
\end{equation}
This equation depends on the field strength formed by the axial gauge fields $A^\mu_5$ which couples to the axial current. 
One might argue that in nature there are no axial gauge fields so there is no such anomaly. This argument is however
problematic. First axial gauge fields can effectively appear as happens in the low energy description of Weyl semi-metals 
(see Sec. \ref{sec_axialfields}) 
and secondly in a quantum theory
charge conservation should be an operator equation that holds in generic correlation functions.  

Since the problematic \eqref{eq:AVcov} is derived from (\ref{eq:covanomalyJ}),(\ref{eq:covanomalyT}), one might ask how the anomalous conservation equation for the currents (\ref{eq:covanomalyJ}),(\ref{eq:covanomalyT}) can be interpreted. There are at least four related but slightly different possibilities
\begin{enumerate}
\item First, we can interpret the anomaly equations as the non-conservation of the current corresponding to the classical symmetry $A$ when appropriate classical background gauge fields are switched on. An example is the non-conservation of axial charge in a homogeneous background of parallel electric and magnetic fields. In this case the anomaly equation becomes 
\begin{equation}
\dot \rho_5 = \frac{1}{2\pi^2} \vec E \cdot \vec B\,.
\end{equation}
Axial charge is created "out of the vacuum" at a constant rate determined by the electric and magnetic fields.
Here the electromagnetic fields are thought of as external to the quantum dynamics of the Dirac fermions. In what concerns the
quantum dynamics they are just classical c-numbers and not operators. This is the situation that arises in measurements of
magneto conductivities of Weyl or Dirac semi-metals. 

\item We can also interpret the anomaly equations as a statement of how the operator ``$\partial_\mu J^\mu(x)$", the divergence of the current or the energy momentum tensor, behaves when inserted in correlation function with other currents. Here we interpret the left-hand side of an anomaly equation as the expectation value of the divergence of the anomalous current and the right-hand side as a (local) functional of the classical sources that couple to the currents and the energy-momentum tensor. The gauge fields and the metric on the right-hand side of the anomaly equation are not physical fields but simply act as sources for insertions of the current into correlation functions according to Eqs. (\ref{eq_energy_momentum_tensor}) and (\ref{eq:defcurrent}).  

As an example we consider the axial anomaly and write it as
\begin{equation}
    \langle \partial_\mu J_5^\mu(x) \rangle = \frac{1}{16\pi^2} \epsilon^{\mu\nu\rho\lambda} F^V_{\mu\nu}(x) F^V_{\rho\lambda}(x)\,.
\end{equation}
Functionally differentiating twice with respect the  vector gauge field, i.e. applying $\frac{\delta^2}{\delta V_\rho(y) \delta V_\lambda(z)}$,   gives the anomalous Ward identify 
\begin{equation}
    \langle \partial_\mu J_5^\mu(x) J^\nu (y) J^\lambda(z)\rangle = \frac{1}{2\pi^2} \epsilon^{\mu\nu\rho\lambda} \partial_\mu^x\delta(x-z) \partial_\rho^x\delta(x-z)\,.
\end{equation}
This makes it clear that the anomaly appears even in the absence of physical external electric and magnetic fields as a contact term in a correlation function. 

The same logic holds for the gravitational contribution to the axial anomaly. Even in flat space the gravitational contribution
to the axial anomaly does not vanish.  It is manifest in correlation functions of the divergence of the axial current obtained
by differentiating eq. (\ref{eq:covanomalyJ}) twice with respect to the metric \footnote{This comprises a triangle diagram with two
energy-momentum tensor insertions and a two point function with an operator that arises due to the non-linearity of the
gravitational coupling. The latter is in complete analogy to the diamagnetic term that arises in Kubo formulas for the electric
conductivity in models of charged bosonic matter. \cite{Kimura:1969wj, AlvarezGaume:1983ig}}. This type of interpretation is also often called 't Hooft anomaly. 
The anomaly is not absent even if there are no gauge fields present or if one is in flat space. The external gauge fields only make the anomaly visible in the expectation value of the divergence of a current.
Anomalies are properties of the quantum theory. The non-renormalization theorem also implies that the results
are independent of the coupling to other matter fields. In particular anomalies are present in the free theory 
and can be detected by computing correlations functions of composite operators (the currents) in the free theory. 

\item The anomaly can also be interpreted as an identity between two a priori different operators in a quantum field theory. This interpretation is
appropriate whenever the
the quantum dynamics of the gauge fields themselves is important. Whereas in the first two cases above
the vacuum expectation value of the divergence of the current still vanishes if the right hand side is set to zero, this
is not possible anymore  if the gauge fields are dynamical. The right hand side 
in Eqs.~\eqref{eq:covanomalyJ},\eqref{eq:covanomalyT} contains now fluctuating quantum fields and therefore under no circumstance
the corresponding current is really conserved. It is always subject to the quantum fluctuations of the gauge fields\footnote{This lies at the heart of the resolution of the so-called $U(1)_A$ (A for axial here) puzzle in QCD \cite{tHooft:1986ooh}}. Because of the quantum fluctuating gauge field there are quantum corrections to the divergence of the current due to higher-order loop diagrams, even in Abelian gauge theories \cite{Adler:2004qt}.

\item In another situation the anomaly has to be interpreted as an inconsistency in the quantum theory.  This happens when a gauge current is anomalous. What is meant here is that the current coupling to dynamical gauge fields suffers from an anomaly. This is a problem indeed. Even without quantizing the gauge fields the inconsistency can be seen from the equations of motion of a $U(1)$ gauge field 
\begin{equation}
\partial_\mu F^{\mu\nu} = J^\nu\,.
\end{equation}
Taking the divergence of both sides leads to an inconsistency unless the operator $\partial_\mu J^\mu$ is identically zero.
This inconsistency carries over to the quantum theory and to non-Abelian gauge fields. In the quantum theory a consequence of this
type of anomaly is that it would allow transverse physical photon states to scatter into unphysical longitudinal polarizations of zero norm in the Hilbert space. Thus unitarity is lost\footnote{In principle, at least for $U(1)$ fields, a possible cure is to make the gauge field massive by
a Stueckelberg mechanism \cite{Preskill:1990fr} since in this case the transverse polarization becomes a physical state}.  
Therefore anomalies in currents which couple to dynamical gauge fields have to vanish identically. Sometimes this can be achieved by using the ambiguities in the regularization prescription as we will exemplify below in the case of the vector like current coupling to the dynamical electro-magnetic gauge field.
\end{enumerate}

These different interpretations of anomalies are often referred with specific nomenclature. For example the so-called ``'t Hooft anomalies" correspond to our second point of view whereas "ABJ anomalies" (after Adler-Bell-Jackiw) to our third point and "gauge anomalies" to our fourth point. \\

Having established that anomalies can not be eliminated by going to flat space and switching off all electric and magnetic fields we still have to deal the worrisome Eq. (\ref{eq:AVcov}). It is an anomaly in a gauge current and therefore should actually vanish.
So far we have insisted that the currents transform
covariantly under the symmetries even if the symmetry is anomalous. This is indeed too much to ask for. It is possible to define a vector current that is exactly conserved, even taking the axial background
gauge fields into account, by changing the regularization scheme. This is the electromagnetic consistent conserved vector current. The name "consistent" has nothing to do with
the fact that this current can be consistently coupled to dynamical electromagnetic gauge fields but rather has a technical meaning in that the corresponding anomaly solves the so-called Wess-Zumino consistency conditions \cite{Wess:1971yu}. Without going further into
details we note that (\ref{eq:AVcov}) is a so called mixed anomaly that involves two symmetries: the axial and the vector-like ones.
In such a situation of mixed anomalies it is possible to 
shift the consistent anomalies between the different symmetries.
In particular one can completely eliminate the anomaly (\ref{eq:AVcov})
by choosing a regularization scheme that explicitly preserves the vector-like symmetry in question (e.g. Pauli-Villars) 
or by adding appropriate counterterms (Bardeen counterterms) to the quantum effective action \cite{Bardeen:1969md}. It is however not possible
to eliminate the anomaly totally. Therefore it is sufficient to calculate the symmetric anomaly coefficients $d_{abc}$ and $b_a$ with a covariant regularization scheme to establish the presence of anomalies. In the case of the anomaly in the vector current eq. (\ref{eq:AVcov}) the relation
of the conserved consistent current $\mathcal{J}^\mu$ to the covariant current $J^\mu$ is
\begin{align}\label{eq:consistentJ}
\mathcal{J}^\mu &= J^\mu - \frac{1}{4\pi^2} \epsilon^{\mu\nu\rho\lambda} A^5_\mu F_{\rho\lambda}\,,\\
\partial_\mu \mathcal{J}^\mu &= 0\,.
\end{align}
The additional Chern-Simons current arises from the regularization as a local finite counterterm to the current (Bardeen-Zumino Polynomial) \cite{Bardeen:1984pm}. The Chern-Simons current precisely cancels the anomaly in the covariant current. Note that this also means
that now the axial gauge field is not really a gauge field anymore. It is a physical field and observable as a current in
electric and magnetic fields. This conserved consistent current can now be coupled to a Maxwell field and has a good
physical interpretation as electric current. On the other hand the anomaly in the axial current can not be cancelled in the
same way since it contains the term
\begin{equation}
\partial_\mu J^\mu_5 = \dot{\rho}_5 + \vec\nabla \cdot \vec J_5 = \frac{1}{16\pi^2} \epsilon^{\mu\nu\rho\lambda} F^V_{\mu\nu}  F^V_{\rho\lambda} = \frac{1}{2\pi^2} \vec E \cdot \vec B\,.
\label{eq:axialanomaly}
\end{equation}
The corresponding Chern-Simons current  that would cancel this anomaly depends explicitly on the (true) electromagnetic gauge field $V_\mu$ and thus has to be considered unphysical. The axial anomaly is there because of the non-zero $d_{VVA}$ coefficient. We note that also the $d_{AAA}$ coefficient is non-vanishing. This essentially means that the axial symmetry is hopelessly lost in the quantum theory. Therefore the best possible result is that the vector-like symmetry is preserved in the quantum theory. The corresponding  consistent current is conserved and can be interpreted as electric current. 

In the case of the gravitational anomaly the covariant form indeed makes the energy-momentum tensor a non-conserved current. But again one can add counterterms to the quantum effective action which cancel the
anomaly in the energy-momentum tensor\cite{Bardeen:1969md,AlvarezGaume:1983ig}. In theory one could also keep the anomaly in the energy-momentum tensor and cancel the
anomaly in the axial current, but it is not possible to keep the conservation laws for both of them simultaneously. 
\\
\begin{tcolorbox}
\begin{itemize}
\item {\bf 
{\bf Anomalies arise in quantum field theory because there is no possible regularization which preserves all the symmetries present in the classical theory.}
\item 
{\bf Chiral anomalies can be detected by computing the one-loop exact anomaly coefficients $d_{abc}$ and $b_a$ Eqs. (\ref{eq:acoeffgauge}) and (\ref{eq:acoeffgrav}). }
\item }
{\bf Anomalies are intrinsic properties of the operators. They are manifest in correlation functions even in the absence of external gauge fields and in flat space-time.}
\item 
{\bf Anomalies in gauge currents lead to mathematical inconsistencies.}
\item{\bf Anomalies suffer from ambiguities which need to be fixed by specific regularization conditions. These ambiguities give rise to different forms of anomalies (covariant or consistent anomalies) but also allow to preserve a subset of anomalous symmetries.}
\end{itemize}
\end{tcolorbox}

\subsection{Anomalies in matter}
\label{sec_transport}

Now we discuss anomaly induced transport, in particular the so-called chiral magnetic and chiral vortical effects \cite{Vilenkin:1979ui, Vilenkin:1980fu, Alekseev:1998ds, Giovannini:1997eg, Newman:2005hd,Fukushima:2008xe, Son:2009tf, Erdmenger:2008rm, Banerjee:2008th}.
The chiral magnetic effect (CME) denotes the generation of a current in parallel to an applied magnetic field $\vec{B}$ and the chiral vortical
effect (CVE) is the generation of a current along a vortex $\vec\Omega$ (for a review see \cite{Landsteiner:2016led,Kharzeev:2013ffa},).
\begin{align}\label{eq:cmecveJ}
\vec{J}_{a} &= \sigma_{ab} \vec{B}_b + \sigma^\Omega_a \vec{\Omega}\,,\\ \label{eq:cmecveT}
\vec{J}_\epsilon &= \sigma_{a}^\epsilon \vec{B}_a + \sigma^{\epsilon,\Omega} \vec\Omega \,.
\end{align}
The gauge fields are not necessarily physical fields
but can simply be sources for the currents as outlined before. 
The vorticity is $\vec\Omega = \frac{1}{2} \vec\nabla\times \vec v$.  Explicit calculations in free fermion theories \cite{Landsteiner:2011cp} and in holographic field theories  \cite{Landsteiner:2011iq} give the results\footnote{We will specialize these equations to the cases that affect the condensed matter systems in the next section.}
\begin{align}\label{eq:cmeJ}
\sigma_{ab} &= \frac{d_{abc}\mu_C}{4\pi^2}\,,\\
\label{eq:cveJ}
\sigma_a^\Omega &= \frac{d_{abc}\mu_b\mu_c}{4\pi^2} + \frac{b_{a}}{12} T^2 \,,\\\label{eq:cmeT}
\sigma_a^\epsilon &= \frac{d_{abc}\mu_b\mu_c}{8\pi^2} + \frac{b_{a}}{24} T^2\,,\\
\label{eq:cveT}
\sigma^{\epsilon,\Omega} &= \frac{d_{abc}\mu_a\mu_b\mu_c}{12\pi^2} + \frac{b_{a}\mu_a}{12} T^2
\end{align}
The remarkable fact is that these transport coefficients are completely determined by the anomaly coefficients (\ref{eq:acoeffgauge}) and (\ref{eq:acoeffgrav}). It means that these currents are non-zero when the theory has the
anomalies (\ref{eq:AVcov}).  Since anomalies are subject to non-renormalization theorems one expects non-renormalization theorems to hold for these transport coefficients as well. Perturbative proofs of non-renormalization have been presented in \cite{Golkar:2012kb,Hou:2012xg}. 

It is important to realize that these non-renormalization theorems hold only when the gauge fields on the right hand side of the anomaly equation are non-dynamical classical fields ('t Hooft anomalies).  
On the other hand if the gauge fields are taken as quantum 
fields (ABJ anomalies) as well they give rise to non-trivial loop corrections. This was first pointed out in \cite{Hou:2012xg} for the $T^2$ terms and
later shown to hold for all chiral conductivities (\ref{eq:cmeJ})-(\ref{eq:cveT}) in \cite{Jensen:2013vta}. 
In hindsight this is not surprising since also the divergence of
an anomalous current is renormalized in this situation \cite{Adler:2004qt}. The non-renormalization properties of the chiral conductivities are therefore the same as the ones of the anomalous divergence of the current itself. 

The dependence on the chemical potentials in Eqs. (\ref{eq:cmeJ}) - (\ref{eq:cveT}) is also fixed by demanding the existence of an entropy current with positive divergence (generalized second law) \cite{Son:2009tf,Neiman:2010zi} in the presence of chiral anomalies.
The dependence on the temperature on the other hand can not be fixed by 
this criterion. Rather the temperature enters as an integration constant \cite{Neiman:2010zi}. More advanced consideration are necessary to establish
a general relation between the gravitational contribution to the anomalies and the temperature dependence in (\ref{eq:cveJ})-(\ref{eq:cveT}).
This has first been established in \cite{Jensen:2012kj}. 
The approach of \cite{Jensen:2012kj} was to study Euclidean quantum field theory on cone-like geometries.
 A related setup this time however in a Lorentzian signature space-time containing a black hole 
was developed in \cite{Stone:2018zel}.  
This involves integration of the anomaly equations (\ref{eq:covanomalyJ}), (\ref{eq:covanomalyT}) in a black-hole space time with the boundary condition such that the currents and the energy current vanish on the horizon. We will briefly review this argument in the Appendix \ref{app_bh}.
Finally the relation of the temperature dependence of the anomalous transport coefficients to a global version of gravitational anomaly has been established in \cite{Golkar:2015oxw} and further studied in \cite{Chowdhury:2016cmh,Glorioso:2017lcn}.
\\

It might come as a surprise that the gravitational contribution to the chiral anomaly can enter the chiral conductivities. One should keep in mind that the theory is anomalous even in flat space, in the absence of gauge fields and without interactions. It still is somewhat surprising since the gravitational anomaly is a term of fourth order in derivatives and naively can not influence terms in transport that are only first order. It is certainly true that in vacuum  one can not form candidate covariant terms of lower degree in derivatives. In vacuum the only available tensor is the metric and  there exists no covariant tensor that can be formed with only one derivative on the metric. The chiral transport expression apply however not in vacuum but in a gapless theory of chiral fermions at finite temperature and chemical potential. This means that in the low energy
theory one has more effective fields, not only the metric or the gauge fields but also the effective low energy fields temperature $T(x)$,
chemical potential $\mu(x)$ and the velocity field $u^\mu(x)$. This allows now to form new covariant objects, such as the vorticity $\Omega^\mu$ with low degrees of derivatives. We have emphasized in section \ref{sec_TT} that temperature and velocity variations have the same effect in local thermal equilibrium as variations in the gravitational background. This gives an intuitive reason why gravitational effects can be captured by temperature and vorticity. In particular if a gravito-magnetic field (see Appendix \ref{sec_GEM}) is present as a small perturbation of otherwise flat space-time $\delta g_{ti} = A^g_i$, even a fluid at rest $u^\mu=(1,0,0,0)$ can have vorticity $2\vec\Omega = \vec{B_g} = \vec\nabla \times \vec{A}^g$. The vorticity is induced via frame dragging by the gravito-magnetic potential $\vec{A}_g$. In this sense the chiral vortical effect can also be seen as a chiral gravito-magnetic effect \cite{Landsteiner:2011tg}.

As already mentioned the anomaly induced conductivities (\ref{eq:cmeJ}) - (\ref{eq:cveT}) are also constrained by a generalized second law of thermodynamics. It is possible to define an entropy current such that the anomalous currents do not contribute to entropy production. 
This anomalous contribution to the entropy current is given by \cite{Son:2009tf,Neiman:2010zi, Stephanov:2015roa}
\begin{equation}\label{eq:CMEVentropy}
    \vec{J}_{S,\mathrm{anom}} =  \frac{T}{12} b_a \vec{B}_a + \frac{\mu_a b_a}{6} \vec{\Omega}
\end{equation}
Later we will show that using this  expression one can derive the Wiedemann-Franz law and the Mott formula for anomalous transport in Weyl semimetals by defining an appropriate heat current based on this entropy current
as $J^\mu_Q = T J^\mu_{S}$.
For clarity let us emphasize that we have not considered dissipative transport here. Including also dissipative transport coefficients the total entropy current fulfills $\partial_\mu J_S^\mu \geq 0$. 

We note that the expression for the entropy current (\ref{eq:CMEVentropy}) depends also on the frame choice. The one quoted here
corresponds to the so-called "no-drag" frame \cite{Stephanov:2015roa}. This is the frame in which the fluid is at rest for $u^\mu=U^\mu$ and $U^\mu$ denotes the frame field of a impurities on on which the fluid looses momentum. In a condensed matter context there are always impurities which imply the existence of a preferred frame. For applications of anomalous transport to condensed matter physics the no-drag frame is therefore the physically preferred one. 

A violation of the relation between triangle anomalies and the chiral vortical effect happens for relativistic spin $3/2$ fermions (gravitinos) \cite{Loganayagam:2012pz}. In that case it was shown in \cite{Golkar:2015oxw} that a global version of the gravitational anomaly implies 
the chiral vortical effect (see also \cite{Chowdhury:2016cmh,Glorioso:2017lcn}). Finally we point out that recently the relation between the gravitational anomaly in Maxwell theory and a photonic analogue of the chiral vortical effect has been shown to hold if both are calculated with the same regulator \cite{Prokhorov:2020npf}.

\subsection{Anomalies and magnetotransport in Weyl semi-metals}
\label{subsec:WSMano}
\begin{figure}[!thb]
\begin{center}
\includegraphics[scale=0.25,clip=true]{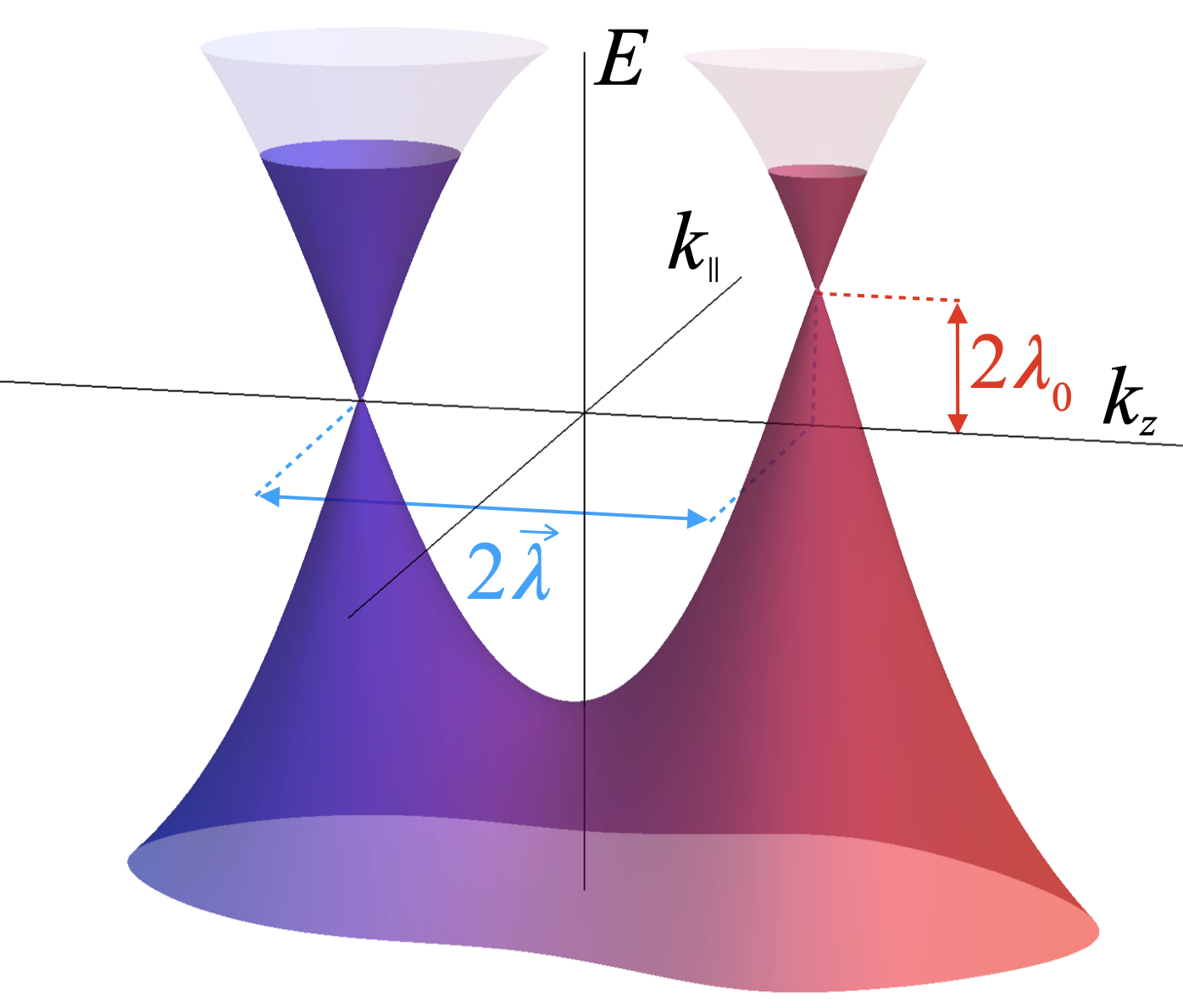}
\end{center}
\caption{A schematic view of a Weyl semimetal's band structure. }
\label{fig:WSM}
\end{figure} 

\label{sec_magtrans}
In Figure \ref{fig:WSM} we show a schematic view of the band structure of a Weyl semi-metal. In this simple description, the band structure has two band crossing points near the Fermi level.  By the Nielsen-Ninomiya theorem the two nodal points have opposite chirality. To lowest order in a derivative expansion near the Fermi points, the electronics can be effectively described by two Weyl fermions of opposite chirality. 
A useful equivalent view is to combine both Weyl fermions into one Dirac fermion. Then the Dirac like wave equation for this four-component spinor is~\cite{Zyuzin2012a,Grushin2012,Zyuzin2012,Goswami2012}
\begin{equation}
    (i\gamma^\mu \partial_\mu + \gamma^\mu \lambda_\mu \gamma_5 ) \psi =0\,.
\end{equation}
The four vector $\lambda_\mu = (\lambda_0, \vec \lambda)$ describes the separation of the Weyl nodes of opposite chirality in momentum space. A useful view is to consider
$\lambda_\mu$ to be the background value of an axial gauge field $A_\mu^5 = \lambda_\mu$.
As we have already emphasized the description in terms of chiral fermions in a crystal is not exact. Due to the compactness of the Brillouin zone the bands hosting left- and right-handed chiral fermions are connected such that a hard scattering process might scatter a fermionic quasiparticle from one Weyl cone to another one of opposite chirality. This inter-valley scattering will have as a consequence that in equilibrium the Fermi energies of left- and right-handed Weyl cones are equal $\epsilon_F^R = \epsilon_F^L=\epsilon_F$. In order to apply the quantum field theoretical transport formulas we need to measured energies with respect to the normal ordered vacuum. This is defined with
respect to the band crossing points at which the density of states within a Weyl cone vanishes. Let us denote these points with $(E_{R,L},\vec{K}_{R,L})$.
We can define now the four vector $\lambda_\mu$ by $(\lambda_0, \vec \lambda) = \frac{1}{2}(E_R-E_L, \vec K_R - \vec K_L)$.
In particular the chemical potentials have to be measured relative to the locations of the Weyl points such that $\mu_{R,L} = \epsilon_F - E_{R,L}$. The total chemical potential is $\mu = \frac 1 2 (\mu_R+\mu_L)$. 
For the axial chemical potential we have $\mu_5 = \frac 1 2 (\mu_R - \mu_L) =  \lambda_0$ in equilibrium. 

We now review how the anomalous transport formulas lead to a theory of magneto-transport in Weyl semi-metals. First we want to get the so called Bloch theorem out of our way. Bloch argued that a system can not sustain an electric current in thermal equilibrium. As we briefly reviewed in the introduction, the Bloch theorem applies only to currents which are exactly conserved gauge currents. Thus we need not be concerned about currents that are not gauge currents, e.g. the axial current $J_5^\mu$ can have an equilibrium expectation value without being in conflict with Bloch's argument. This applies to the so-called Chiral Separation Effect (CSE)
\begin{equation}
\label{eq:cse1}
\vec J_5 = \frac{\mu}{2\pi^2} \vec B\,.
\end{equation}
It is a special case of the general anomaly transport formulas (\ref{eq:cmecveJ}) with $d_{5VV}=2$.
More worrisome is the expression
\begin{equation}\label{eq:cse}
\vec J = \frac{\mu_5}{2\pi^2} \vec B\,,
\end{equation}
the Chiral Magnetic Effect proper. In this form it is determined by the same totally symmetric anomaly coefficient $d_{5VV}$.
Again it simply is a special case of eq. (\ref{eq:cmecveJ}). If we were to interpret the current in eq. (\ref{eq:cse}) as electric current it would contradict the Bloch theorem. Since the Bloch theorem is a consequence of gauge invariance it would also indicate a subtle violation of gauge symmetry. We need to remember however, that there is an ambiguity in the definition of the current and we need to make sure that we actually use the properly defined quantum operator. The electric current should be conserved under all circumstances. Therefore it is correct to identify the conserved consistent current (\ref{eq:consistentJ}) as electric current with 
the identification $A^5_\mu = \lambda_\mu$!
This conserved consistent current has an additional contribution stemming from the Chern-Simons current. The Bloch theorem
can be expected to hold only for the conserved consistent current $\mathcal{J}^\mu$. If we add this contribution we find
\begin{equation}
\vec{\mathcal{J}} = \frac{\mu_5 - \lambda_0} {2\pi^2} \vec B\,.
\end{equation}
As we explained before in equilibrium we have the relation $\lambda_0 = \mu_5$ and thus the Bloch theorem is obeyed. The chiral magnetic current vanishes identically in strict equilibrium. This result has first been obtained in a holographic calculation \cite{Rebhan:2009vc,Gynther:2010ed}. In \cite{Gynther:2010ed} it was also shown that it is a consequence of any gauge invariant regularization in quantum field theory. Later it was shown to hold in a tight-binding approach in \cite{VF13} which from the  point of view of quantum field theory can be viewed as a lattice regularization. The presence of the Chern-Simons term in a tight binding approach has been  explicitly calculated in \cite{Gorbar:2017wpi}. The identification of consistent and covariant currents in lattice models becomes more involved in the presence of axial gauge fields~\cite{Behrends2019}. 
These issues arise if one seeks a quantum field theoretical description. However, this is not strictly necessary as momentum-space in a crystal is compact. So, instead of doing quantum field theory with a normal ordered vacuum, one might just do many body quantum mechanics an measure
energies with respect to the bottom of the band. This point of view has been emphasized in Refs.~\cite{Basar:2013iaa,Behrends2019}. In any case
it is re-assuring that both points of view give rise to the same results. 

Since we can not expect to get any measurable effect by simply applying a magnetic field to a Weyl semi-metal we need to 
do something more to drive the system slightly out of equilibrium. To activate the CME we must break the relation
$\mu_5=\lambda_0$ in some way. Here it is important that the chemical potential is a state variable whereas the axial gauge
field is an external parameter. The quantum dynamics of the fermions alone can not change it the external field $A^5_0=\lambda_0$ but only the state variable $\mu_5$. If we do work on the fermions we can change their state, e.g. applying an electric field leads to an increase of temperature due to Joule heating. In an analogous way the anomalous transport phenomena have an impact on the state of the fermionic system and can drive them out of equilibrium leading to measurable electric and heat currents.

We may imagine injecting electrons at one end of a Weyl semi-metal and extracting fermions on the other side. This will 
effectively lead to a gradient in the chemical potential $\vec \nabla\mu$. Alternatively we can apply an electric field $\vec E$. Now we use the anomalous conservation of the axial current (\ref{eq:axialanomaly}) to obtain the increase in axial charge
\begin{equation}
\dot{\rho}_5 = \frac{1}{2\pi^2}\vec E \cdot \vec B - \vec \nabla\cdot\vec{J}_5 =
\frac{1}{2\pi^2} (\vec E - \nabla \mu)\cdot \vec B\,.
\end{equation}
We also have taken into account the contribution of the CSE current eq. (\ref{eq:cse1}) proportional to the chemical potential.
Now we relate the axial charge density to the axial chemical potential via an axial susceptibility $\chi_5 = \partial \rho_5/\partial\mu_5$
\begin{equation}
\dot \rho_5 = \chi_5 \cdot\mu_5\,.
\end{equation}
Fourier transforming in time leads to an increase in the axial chemical  potential away from its equilibrium value
\begin{equation}
\delta \mu_5 = i\frac{(\vec E -\vec\nabla \mu) \cdot \vec B}{2\pi^2  \chi_5 \omega} \,.
\end{equation}
Since we are interested in a steady state situation we have to take the limit $\omega \rightarrow 0$ in which the
axial charge diverges. In a crystal axial charge is however destroyed due to large momentum scatterings from one Weyl cone
into the other. This takes place at a time scale $\tau_5$ which is called the inter-valley scattering time~\cite{Son:2012bg}. This is also
the time scale at which the growth of the axial chemical potential will be cut off. 
Formally this can be represented by the substitution.
\begin{equation}
 \omega \rightarrow \omega + i/\tau_5\,,   
\end{equation}
with $\tau_5$ the inter-valley scattering time.
In this way one predicts a finite
value of the axial chemical potential which eventually can be plugged into the formula for the CME to obtain
\begin{equation}
\vec J = \tau_5 \frac{(\vec E - \vec \nabla \mu) \cdot \vec B}{4 \pi^4 \chi_5} \vec B\,.
\end{equation}
This reasoning predicts a large enhancement of the electric conductivity in a magnetic field if the inter valley scatting time is a long time scale \cite{Nielsen:1983rb,Li:2014bha}. A microscopic calculation in a kinetic theory approach was presented in \cite{Son:2012bg}. Here we have emphasized that it is based on universal results, anomalous conservation laws and the corresponding anomaly induced transport phenomena. We also made it clear that even when there are no external electric fields the effect can be triggered by driving the system in a state with a gradient in the chemical potential. This is just a reflection of the well known fact that in linear response the effect of an electric field is the same as the effect of 
a gradient in the chemical potential. 

\subsection{Anomaly induced magneto-electric and -thermal transport for Weyl 
semi-metals~\label{eq:anomalytransport2}}
\begin{figure}[!thb]
\begin{tabular}{cc}
\includegraphics[scale=0.25,clip=true]{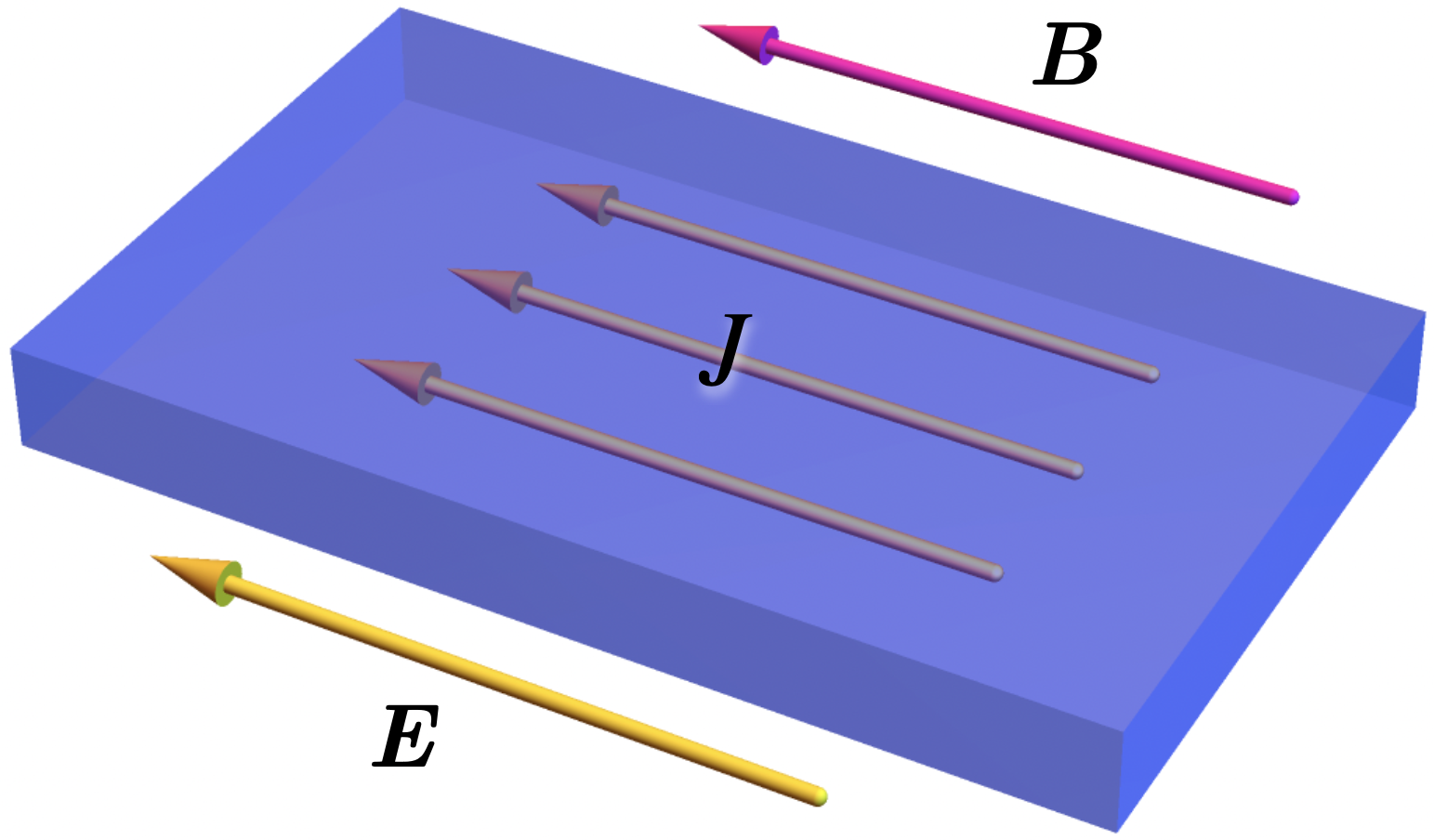}
& \hspace{0.5cm}
\includegraphics[scale=0.25,clip=true]{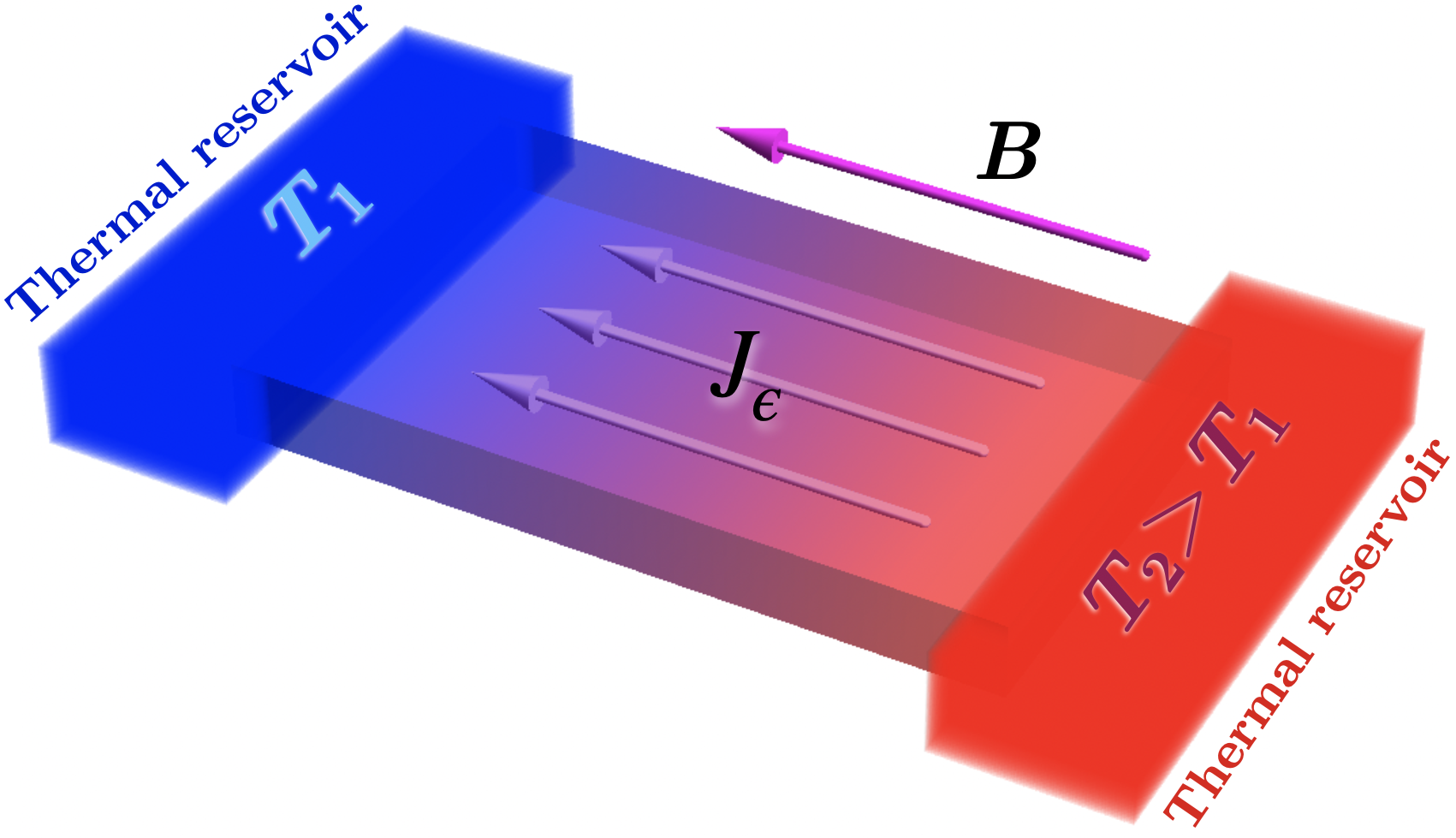} \\[2mm]
(a) & (b)
\end{tabular}
\caption{Schematics of the enhanced transport in an magnetic field due to anomalies. In panel (a) in addition to the magnetic field a parallel electric field is applied. This leads to an enhanced current due to the chiral magnetic effect. In panel (b) a temperature gradient parallel to the magnetic field activates the temperature component of the chiral magnetic effect in the energy current (see Eq. \eqref{eq:cmeT}) proportional to the gravitational anomaly coefficient.}
\label{fig_transport_trains_planes_and_automobiles}
\end{figure}
Let us now discuss anomaly induced magneto-electric and -thermal transport for Weyl semi-metals (see Figure \ref{fig_transport_trains_planes_and_automobiles}). This can and has been studied in a microscopic approach based on (chiral) kinetic theory and the corresponding Boltzmann equation \cite{AndreevSpivak, Lundgren:2014gy}. The relation of the thermal conductivities to the anomalies is however not evident in such an approach. In contrast, the effective transport theory based on the anomalous transport currents Eqs. (\ref{eq:cmeJ})-(\ref{eq:cveT}), can track the anomaly contribution.
This has been shown first in a hydrodynamic model in \cite{LDS16}. It is however unclear if a hydrodynamic regime with independent flow for different chiralities can be achieved in a Weyl semi-metal. Fortunately, an effective theory of coupled charge and energy transport in a magnetic field based on anomaly induced transport can be developed  without assuming a hydrodynamic regime, as we will review now.

First, we elaborate on the role played by the different relaxation processes. In general we can distinguish between intravalley, intervalley and electron-electron or electron-phonon scattering. A key assumption is that there exists a hierarchy of relaxation times $\tau < \tau_5 < \tau_{ee}\approx \tau_{e\phi}$.  The intravalley scattering time  $\tau$ is the shortest. 
It is the timescale in which electronic particles are scattered but
the change is their momentum is small such that they stay within the same Weyl cone in momentum space. This scattering
distributes the electronic quasiparticles randomly in the phase space corresponding to a single Weyl cone.  Second, we assume an inter-valley scattering time scale $\tau_5$.  It is supposed to be significantly longer than $\tau$. Finally we also assume that electron-electron and electron-phonon interactions are even rarer than inter-valley scattering events. Note that these assumptions exclude hydrodynamic behavior of the electrons\footnote{For hydrodynamic electron flow the electron-electron interactions must be dominant and thus $\tau_{ee}$ must be the shortest time scale.}. Formally we will work with an idealized situation in which $\tau=0$ and $\tau_{ee},\tau_{e\phi}=\infty$. That means that momentum is not a relevant low energy degree of freedom whereas the energies and charges of each individual Weyl cone are approximately conserved, on a time scale given by the intervalley scattering time $\tau_5$.

To set up an effective transport model we concentrate on the charges conserved by the fast elastic intravalley scattering. These are the energy and charge densities. A transport theory should include temperatures and chemical potentials for each Weyl cone. It corresponds to the quantum numbers that are conserved up to a timescale of order $\tau_5$. In the absence of external driving forces the temperatures and chemical potentials  will equilibrate at times longer than $\tau_5$. 

The currents induced by the anomalies are dissipationless and therefore we expect them to be insensitive to momentum relaxing elastic scattering. Indeed model calculations in holography \cite{Copetti:2017ywz} and in a hydrodynamic setup \cite{Stephanov:2015roa} have shown anomalous transport formulas to be robust against momentum relaxation. In a quasi-particle picture the robustness of anomalous transport can be argued on the grounds that the anomalous transport stems from the lowest Landau level. Then we can view the relevant degrees of freedom as $1+1$ dimensional chiral fermions which naturally suppresses backscattering.

We base our treatment on anomalous transport of right- and left-handed fermions in a  magnetic field.  The corresponding symmetry is the vector-symmetry $U(1)_V = U(1)_L \otimes U(1)_R$. The magnetic field acts on both chiralities in the same way.

The anomaly coefficients determining the chiral currents in a magnetic field are therefore $d_{LLV}$, $d_{RRV}$, $b_L$ and $b_R$ defined in eqs. \eqref{eq:covanomalyJ},\eqref{eq:covanomalyT}. Due to the Nielsen-Ninomiya theorem the number of left-handed Weyl cones has to be equal to the number of right-handed Weyl cones ($n_R=n_L$). Under the left-right symmetry $U(1)_L\otimes U(1)_R$ the left-handed fermions have charges $(1,0)$ whereas the right-handed fermions have charge $(0,1)$. The anomaly coefficients are then
\begin{equation}
\label{eq:WSManomalycoeffs}
d_{RRV}=-d_{LLV}= b_R= -b_L=1\,.
\end{equation}
for a single Weyl cone depending on its chirality. The Nielsen-Ninomiya theorem guarantees that the vector symmetry is anomaly free, i.e. $d_{VVV}=b_V=0$. Due to this simple structure it is sufficient to consider only one Weyl cone. The total current can be obtained by summing over the currents from each Weyl cone.

The anomaly induced energy and charge currents in a magnetic field in a single Weyl cone are
\begin{align}\label{eq:cmecurenergy}
\vec J^\epsilon_{L,R} &=   \pm\left(\frac{\mu^2}{8\pi^2} +  \frac{T^2}{24} \right)\vec B\,,\\ \label{eq:cmecurdensity}
\vec J_{R,L} & = \pm\frac{\mu}{4 \pi^2} \vec B\,.
\end{align} 
The sign is determined by the anomaly coefficients (\ref{eq:WSManomalycoeffs}).
These expression are obtained from (\ref{eq:cmecveJ}) and (\ref{eq:cmecveT}) by going to the restframe $u^\mu=(1,0,0,0)$ and identifying the energy current as $T^{0i}=J_\epsilon^i$.
Now we want to study the effect of applying an external electric field (or a gradient in the chemical potential) and a thermal
gradient to the system. Because the anomaly acts on each Weyl cone we find from anomalous current conservation
\begin{align}
\delta\dot \epsilon &= -\vec\nabla\cdot \vec J_\epsilon + \vec E\cdot \vec J\,,\\
\delta\dot\rho &= -\vec \nabla\cdot\vec J \pm \frac{1}{4\pi^2} \vec E \cdot \vec B\,.
\end{align}
The first line is energy conservation taking into account the Joule-heating term and the second line is the anomalous conservation
law of the number density of Weyl fermions in a single cone. Using Eqs.~(\ref{eq:cmecurenergy}) and (\ref{eq:cmecurdensity}) 
allows to compute the growth rates of energy and number densities in a single Weyl cone as
\begin{align}
-i (\omega+\frac{i}{\tau_5} ) \delta\epsilon &= \mp \left(\frac{\mu}{4\pi^2} (\vec\nabla \mu - \vec E)\cdot\vec B +  \frac{T}{12} \vec\nabla T\cdot\vec B \right)\,,\\
-i (\omega+\frac{i}{\tau_5}) \delta\rho &= \mp \frac{1}{4\pi^2} (\vec \nabla \mu - \vec E) \cdot \vec B\,.
\end{align}

We also have already performed a Fourier transform in time. As before we cut off the infinite growth at zero frequency
by the intervalley relaxation time. This allows us to simply set $\omega=0$ and compute $\delta \epsilon$ and $\delta \rho$.
Now we express the energy and number densities by the
chemical potentials using the susceptibility matrix $\chi$
\begin{equation}\label{eq:deltaTmu}
\left(\begin{array}{c}
\delta \epsilon \\
\delta \rho
\end{array} \right)
=
\chi \cdot \left(\begin{array}{c}
\delta T \\
\delta \mu
\end{array} \right),
\end{equation}\
where 
\begin{equation}
\chi = \left(
\begin{array}{cc}
\frac{\partial \epsilon}{\partial T} & \frac{\partial \epsilon}{\partial \mu}\\
\frac{\partial \rho}{\partial T} & \frac{\partial \rho}{\partial \mu}
\end{array}
\right).
\end{equation}
We can plug these increments of the temperatures and chemical potentials into the anomaly induced currents 
\begin{align}
\delta \vec J^\epsilon &= \pm \left(\frac{\mu \delta \mu}{4\pi^2}  +  \frac{T}{12} \delta T \right) \vec B\,,\\
\delta \vec J &= \pm \frac{1}{4\pi^2} \delta \mu \vec B\,.
\end{align}
Finally we can put everything together and compute the
induced energy and charge current for each single Weyl cone. 
In matrix notation it is conveniently expressed as 
\begin{align}\label{eq:condmat}
\left(
\begin{array}{c}
\delta J^\epsilon_i\\
\delta J_i
\end{array}
\right)
&=
\Sigma_{ij} \cdot
\left(\begin{array}{c}
-\partial_j T  \\
E_j-\partial_j \mu 
\end{array}
\right)\,,\\
\Sigma_{ij} &= \tau_5
\left(
\begin{array}{cc}
\frac{T}{12} & \frac{\mu}{4\pi^2} \\
0 & \frac{1}{4\pi^2}
\end{array}
\right)
\cdot\chi^{-1}\cdot
\left(
\begin{array}{cc}
\frac{T}{12} & \frac{\mu}{4\pi^2} \\
0 & \frac{1}{4\pi^2}
\end{array}
\right)
B_i B_j\,.
\end{align}
The final result does not depend on the chirality. Both chiralities 
contribute in the same way. 

\subsection{Weak magnetic field}
Let us apply this result now to a gas of free chiral fermions. We also assume here that the magnetic field is small compared to chemical potential $\mu$ and and temperature $B$, 
i.e. $B/T^2\ll 1$ and $B/\mu^2\ll 1$.
The free energy density is \cite{le2000thermal}
\begin{equation}
F = -\frac{1}{24\pi^2} \left(\mu^4 + 2 \pi^2 T^2 \mu^2 + 7/15 \pi^4 T^4 \right),
\end{equation}
from which the energy $\epsilon = F-\frac{\partial F}{\partial T} + \mu \rho$ and the number density $\rho = -\frac{\partial F}{\partial \mu}$ and the \
susceptibility matrix $\chi$ can be calculated. The conductivity matrix as defined in (\ref{eq:condmat}) is

\begin{equation}
\label{eq:sigmamatrix}
\Sigma_{ij} = \left(
\begin{array}{cc}
\kappa_{ij} & \beta_{ij} \\
\alpha_{ij} & \sigma_{ij}
\end{array}\right),
\end{equation}
where the different entries are the anomaly induced contributions to the different conductivities
\begin{align}
    \label{eq:thermalcondB}
    \mathrm{thermal~conductivity}: \kappa_{ij} &=\tau_5 B_i B_j\frac{15\mu^2 T - 5 \pi^2 T^3 }{24\Delta}   \\ 
    \label{eq:electro-thermalcondB}
    \mathrm{electro-thermal~conductivity}: \beta_{ij} &= \tau_5  B_i B_j \frac{11\pi^2 T^2 \mu + 15\mu^3}{8\pi^2 \Delta} \\ 
    \label{eq:thermo-electriccondB}
    \mathrm{thermo-electric~conductivity}:\alpha_{ij}&= -\tau_5 B_i B_j \frac{5 \mu T}{4 \Delta} \\
    \label{eq:electriccondB}
    \mathrm{electric~conductivity}:\sigma_{ij} &= \tau_5 B_i B_j \frac{21 \pi^2 T^2 + 15 \mu^2}{8\pi^2\Delta}\,,
\end{align}
with $\Delta = 7 \pi^4 T^4 + 6 \pi^2 T^2 \mu^2 + 15 \mu^4$.

\subsection{Thermoelectric transport relations of the anomaly-induced transport coefficients}
We will now study how the anomaly induced conductivities collected in the matrix $\Sigma_{ij}$ realize some standard electronic transport properties such as the Onsager reciprocity relation, the Wiedemann-Franz law and the Mott relation. Let us start with the Onsager reciprocity relations.
\label{sec_ano_relations}
\begin{itemize}
\item {\bf Onsager relations}. 
 They are a consequence of time-reversal symmetry and should hold on general grounds. We note that the magnetic field breaks time-reversal symmetry which means that Onsager's relations hold if one replaces $\vec B \rightarrow -\vec B$. The anomaly induced conductivity matrix $\Sigma_{ij}$ is however quadratic in the magnetic field and therefore insensitive to the sign of the magnetic field. 
The conductivity matrix $\Sigma_{ij}$ is the response due to gradients in temperature, chemical potential and electric field.
Onsager relations are manifest however if the conductivities are written as responses due to gradients in the thermodynamic driving forces (Langrange multipliers) $\beta = 1/T$ and $\gamma = -\mu/T$. 
The Onsager reciprocity relations state then that the
conductivity matrix should be symmetric and positive.
By transforming to the $\beta$, $\gamma$ basis 
\begin{align}
\left(
\begin{array}{c}
J^\epsilon_i\\
J_i
\end{array}
\right) &= L_{ij}\cdot \left(
\begin{array}{c}
\partial_j(\frac{1}{T})\\
\frac{E_j}{T} - \partial_j(\frac \mu T )
\end{array}
\right),
\end{align}
and we find
\begin{align}
\label{eq:lowBLij}
L_{ij} &= \Sigma_{ij}\cdot\left( \begin{array}{cc}
T^2 & 0\\
\mu T &  T 
\end{array}\right) = 
\frac{\tau_5 B_i B_j}{\Delta}
\left(\begin{array}{cc}
\frac{1}{24\pi^2}(5 \pi^4 T^5 + 18 \pi^2 T^3 \mu^2 + 45 T \mu^4) & \frac{T\mu}{8 \pi^2} ( 11 \pi^2 T^2 + 15 \mu^2)\\
 \frac{T\mu}{8 \pi^2} ( 11 \pi^2 T^2 + 15 \mu^2) & \frac{3 T}{8\pi^2}(7 \pi^2 T^2 +5\mu^2)
\end{array}
\right),
\end{align}
which is indeed symmetric and positive definite.
\\

\item {\bf Wiedemann-Franz law and the Lorentz-ratio from anomalies.} 
One of the basic paradigms of quasiparticle transport is the Wiedemann-Franz law. It states that at low temperature the ratio of the thermal conductivity for the heat current and the electric conductivity is given by a constant, the Lorenz number $L=\pi^2 k_B^2/(3 e^2)$ times the temperature $T$. Here we sill show that also the anomaly induced magnetoconductivities of a chiral fermion obey the Wiedemann-Franz law.
Let us define the heat current by
\begin{equation}\label{eq:heatcur1}
    \vec{J}_Q = \vec{J}^\epsilon - \mu \vec{J}
\end{equation}
where we use the currents from eq. (\ref{eq:condmat}). 
The thermal conductivity in the heat current is defined by
\begin{equation}
    J_{Q,i} = -\hat\kappa_{ij} \partial_j T
\end{equation}
Alternatively we could use the  definition of the anomaly induced entropy current (\ref{eq:CMEVentropy}).
Note that the net entropy current in equilibrium vanishes since the contributions form left-handed and right-handed Weyl fermions cancel. But upon applying the temperature gradient and electric field they are shifted out of their equilibrium values as in (\ref{eq:deltaTmu}). In particular the increment in  temperature is 
\begin{equation}
    \delta T = \mp \tau_5 \frac{5 \pi^2 T^3 + 15 \mu^2 T}{2\Delta} \vec{B}.\vec\nabla T \,.
\end{equation}
Using this expression for the entropy current 
\begin{equation}\label{eq:heatcur}
    J_{S,i} = \pm \frac{\delta T}{12} B_i = - \tau_5 \frac{5 \pi^2 T^3 + 15 \mu^2 T}{24\Delta} B_i B_j \partial_j T \,.
\end{equation}
From this expression for the anomaly induced entropy current we can give another definition of heat current as
\begin{equation}\label{eq:heatcur2}
    \vec{J}_Q = T \vec{J_S}\,.
\end{equation}
While the two definitions of heat current look very different the do indeed give the same result eq. (\ref{eq:heatcur})!

Now we can build the ratio of the diagonal parts of the thermal and electric conductivities (we also restore the Boltzmann constant $k_B$ and the unit of electric charge $e$)
\begin{equation}
\frac{\hat\kappa_{ii} }{  \sigma_{ii}} = \frac{5 k_B^2 \pi^2  (\pi^2 k_B^2 T^3 + 3 \mu^2 T)}{e^2 9(7\pi^2 k_B^2 T^2 +5\mu^2)} = 
\frac{\pi^2 k_B^2}{3 e^2} T + O(T^2/\mu^2)\,.
\end{equation}
To leading order in $\mu/T$ this expression gives the usual Lorenz ratio. The anomaly induced magnetic conductivities fulfil the
Wiedemann-Franz law in the regime of small temperature and large chemical potential. 
\\

\item {\bf Seebeck coefficient and Mott formula.} The Seebeck coefficient describes how much voltage is built up when applying
a temperature gradient at vanishing electric current. From
\begin{equation}
0 = \vec{J} = \sigma \vec{E} - \alpha \vec\nabla T\,.
\end{equation}
we see that the Seebeck coefficient $S$ is the ratio of the thermo-electric conductivity and the electric conductivity. Again we consider for simplicity only the diagonal parts describing the conductivities along with the magnetic field
\begin{equation}
S  =  \frac{\sigma_{ii}} {\alpha_{ii}}\,.
\end{equation}
At large chemical potential the Mott formula states that
\begin{equation}
S = -\frac{\pi^2 k_B^2 T}{3 e} \frac{dc(\mu)/d\mu}{c(\mu)} ~~,~~ \sigma(\mu,T) = c(\mu) + O(T^2/\mu^2)\,.
\end{equation}
Again we  insert the electric charge unit and the Boltzmann constant as before.
To leading order in $\mu/T$ from Eqs. (\ref{eq:thermo-electriccondB}) and (\ref{eq:electriccondB}) we find
\begin{align}
c(\mu) &= \frac{ e^2 \tau}{8\pi^2 \mu^2}\,,\\
S &=  -\frac{ 2 k_B^2 \pi^2 T}{3 e \mu}
\end{align}
which indeed fulfills the Mott formula.
\end{itemize}

\subsection{Large magnetic field}
So far we have assumed that the magnetic field is small. 
In the quantum regime, in which the magnetic field sets the dominant energy scale
$B\gg T^2$, $B\gg\mu^2$
we may assume that all fermions are in the lowest Landau level. 
The free energy is now\footnote{This is easily calculated by noting that in the lowest Landau Level we are dealing with fermions in one spatial dimension. Their free energy is 
F=$\int \frac{dk}{2\pi} k [n_F(k;T,\mu) + n_F(k;T,-\mu)] = - (\pi^2 T^2 + 3 \mu^2)/(12\pi)$
with $n_F(k;T,\mu) = (\exp( (k-\mu)/T ) +1 )^{-1}$ the Fermi-Dirac distribution.
Here we use that in two dimensions by conformal symmetry the energy density and pressure are related by $\epsilon=p$ and that $F=-p$.
This has to be multiplied by the degeneracy of factor of the Landau level $|B|/2\pi$.}
\begin{equation}
F = -\frac{|B|}{24\pi^2} (\pi^2 T^2 + 3\mu^2)\,. 
\end{equation}

Going through he same considerations as before we find the conductivity matrix $\Sigma$
with the entries
\begin{equation}
\Sigma_{ij} = \frac{\tau_5}{4 \pi^2 |B|}
\left( \begin{array}{cc}
 \frac{\pi^2 T}{3} & \mu \\
 0 & 1 
\end{array} \right) B_i B_j
\end{equation}

\begin{align}
    \label{eq:thermalcondlargeB}
    \mathrm{thermal~conductivity}: \kappa_{ij} &=\tau_5 \frac{B_i B_j}{|B|} \frac{T}{12}   \\ 
    \label{eq:electro-thermallargeB}
    \mathrm{electro-thermal~conductivity}: \beta_{ij} &= \tau_5 \frac{B_i B_j}{|B|} \frac{\mu}{4\pi^2} \\ 
    \label{eq:thermo-electriclargeB}
    \mbox{thermo-electric conductivity}:\alpha_{ij}&= 0 \\
    \label{eq:electriclargeB}
    \mathrm{electric~conductivity}:\sigma_{ij} &= \tau_5 \frac{B_i B_j}{|B|}\frac{1}{4\pi^2}\,,
\end{align}

The result for the electric conductivity has first been obtained by \cite{Nielsen:1983rb}.
In particular we notice the vanishing of the thermo-electric conductivity and consequently the Seebeck coefficient. The Wiedemann-Franz law holds in this case independently of temperature. 
The symmetric conductivity matrix in the basis relevant for the Onsager relations is
\begin{equation}
\label{eq:highBLij}
L_{ij} = \frac{\tau_5}{4 \pi^2 |B|}
\left( 
\begin{array}{cc}
T \mu^2 + \frac{\pi^2 T^3}{3}  & T \mu \\[1mm]
 T \mu & T 
\end{array} 
\right) 
B_i B_j.
\end{equation}

\subsection{Concluding remarks on magneto-transport}

We only have briefly considered the cases of weak and large magnetic fields. At intermediate values of the magnetic field one expects to observe quantum oscillations as one varies the magnetic field strength~ \cite{Monteiro:2015mea,Kaushik:2017rtj,KA20}. At weak magnetic fields many materials show a suppression of the magneto-conductivity with a non-analytic cusp behavior at zero field strength. Weak anti-localization has been suggested as an explanation but this effect was found to be too weak \cite{Ong20}. As discussed in Sec.~\ref{sec_exp}, experiments measuring magneto-conductivities are affected by so-called current jetting which can contribute or mask completely a possible anomaly related signal. In contrast to the electric case, magneto-thermal conductivities are not affected by current jetting and can probe the thermal part and thus gravitational anomaly induced transport. 

We close this chapter with some remarks on the relation of the anomaly coefficients to the Wiedemann-Franz law and the Mott formula. Our derivation based on anomalous transport depended on the anomaly coefficients of the chiral anomaly $1/(32\pi^2)$ and the gravitational contribution to the chiral anomaly $1/(768\pi^2)$. These numbers come deeply from within the inner workings of relativistic quantum field theory. On the other hand the Wiedemann-Franz law is a rather mundane piece of 19th century condensed matter phenomenology. Depending on one's point of view one can find the validity of these laws for anomalous transport rather interesting or amusing but it seems certainly remarkable. The Wiedemann-Franz law and the Mott formula have been obtained here in the limit $T/\mu \rightarrow 0$. It is known that violations appear for high temperatures in all metals. In \cite{Das:2019non} it was shown that these temperature dependent violations are stronger for the anomalous conductivities than for the usual Drude type conductivities.

\begin{tcolorbox}
\begin{itemize}
\item{\bf Chiral magnetic and chiral vortical effects are direct consequences of chiral anomalies in matter (finite temperature and chemical potential).}
\item{\bf Anomaly induced transport can be used as effective theory for magnetic transport in Weyl semi-metals}
\item{\bf Both chiral and gravitational anomalies play a role in thermal transport, generally related to the gravitational contribution to the anomaly. }
\item {\bf Anomalous transport coefficients fulfill the Wiedemann--Franz and Mott relations both at weak and strong magnetic field strengths.}
\end{itemize}

\end{tcolorbox}


\section{Axial gauge fields and the chiral anomaly.}
\label{sec_axialfields}

As we mentioned in section \ref{sec_chiralanoms}, there are no axial gauge fields in particle physics (see, in particular, the comments   after eq. \eqref{eq:AVcov}). Nevertheless  anomalies related to these gauge fields have been described and the anomalous coefficients are contained in equations \eqref{eq:acoeffgauge}. 
In certain materials, typically those with pseudo-relativistic dispersion relations, the variation of microscopic material parameters, like the hopping, in space or time leads to the emergence of effective, axial pseudo-gauge fields~\cite{VKG10,LBetal10,CFLV15,Rachel2016,Strain20,Strain21}. One of the most studied examples are the  elastic axial gauge fields created by straining graphene, a 2D material with low-energy Dirac quasi-particles. The discovery of Weyl semimetal materials\cite{AMV18}, with 3D pseudo-relativistic dispersion relations, opened the exploration of anomaly induced transport in the presence of, or due to, axial pseudo-electromagnetic fields. 

In this section we briefly review the emergence of axial pseudo-gauge fields in condensed matter and the new anomaly-induced transport phenomena that these fields can induce in Dirac and Weyl semimetals. 
We refer the interested reader to the focused reviews on axial gauge fields in 2D-materials \cite{VKG10} and 3D Weyl semimetals~\cite{Strain20,Strain21}.

\subsection{ Axial gauge fields in Weyl semimetals.}

As described in section \ref{subsec:WSMano},  the separation between the nodes in energy-momentum space $\lambda_\mu$ in a Weyl semimetal,  couples to the electronic degrees of freedom as an axial (constant) gauge field. The low energy effective action for the quasi-particles is
\begin{equation}
\label{eq:Weylact}
S = \int d^4x \bar{\psi}\gamma^{\mu}[i\partial_\mu - \gamma_{5}\lambda_{\mu}]\psi.
\end{equation}

Since the electronic  band structure depends crucially on the lattice structure,  it was soon realized, by analogy with graphene~\cite{VKG10,LBetal10}, that  elastic deformations of the lattice would change the separation between the Weyl nodes. Inhomogeneous lattice deformations induce a  space-dependent $\lambda_\mu(\vec{x})$ \cite{CFLV15,CKLV16} while time-dependent deformations as phonons would provide a  $\lambda_\mu(t)$. As we will see, the boundaries of the sample provide also a natural space dependence for the nodes separation \cite{CCetal14}. As mentioned before, although Weyl semimetals are typically harder to deform than graphene, their 3D nature makes these axial gauge fields much more interesting under a fundamental point of view due to their potential contributions to the various anomalies~\cite{Karl18} that are absent in 2D. 

Elasticity theory, and specifically the coupling of lattice deformations to electronic excitations,  allows to understand how pseudo-electromagnetic fields induced by strain emerge in Weyl semimetals.  In elasticity theory \cite{LL71b} the main geometric objects are the deformation vector ${\vec u}$
and the gradient deformation or deformation tensor whose linearized expression is $u_{ij}=1/2(\partial_i u_j+\partial_j u_i)$. 
In an effective action approach, the coupling of lattice deformations to  electronic excitations  can be constructed  by including terms involving the  strain tensor and the electronic degrees of freedom, compatible with the symmetries of the system\footnote{For a detailed description of the effective action construction in graphene see \cite{MJSV13}.}. For a generic  Weyl semimetal described by \eqref{eq:Weylact} a first order term is 
\begin{equation}
S_{\mathrm{e}-\mathrm{ph}}^1=\beta\int d^4x \bar{\psi}_{x}[\gamma^{\mu}\gamma_{5}A_\mu^5(x)]\psi_{x},
\label{eq:ephonon}
\end{equation}
with 
\begin{equation}
A^5_0(x)= \lambda_0 {\mathrm {Tr}}(u(x))\quad, \quad A^5_i(x)=u_{ij}(x)\lambda_j. 
\label{eq_elasticGF}
\end{equation}
In the former equations, Tr($u$)=$u^i_i$, is the trace of the deformation tensor that, being a scalar quantity, can couple to every term in the original action. In particular, it gives rise to a deformation potential which  plays a major role in the experimental consequences of the elastic-electronic interaction 
\cite{AMORIM16,SCetal16,AV18}. $\beta$ is a material dependent coupling constant that encodes the strength of the electron-phonon coupling. This is the simplest way to understand and  generate 
axial pseudo-gauge fields\footnote{The fields defined in Eq. \eqref{eq_elasticGF}  are called  pseudo-gauge  fields because, although the coupling \eqref{eq:ephonon} is invariant under a generalized (chiral) gauge transformation, 
\beqn
\Psi \to e^{i\gamma_5 \lambda}\Psi \quad , \quad A^5_\mu\to A^5_\mu-i\partial_\mu\lambda\,,
\label{eq:pseudog}
\eeqn
the kinetic term of the pseudo-gauge fields is given by elasticity theory which, unlike the Maxwell Lagrangian, is not invariant under the displacement of the vector field in \eqref{eq:pseudog}. In what follows we will omit the prefix pseudo and call the fields elastic gauge fields.}. The expressions \eq{eq_elasticGF} were obtained in a tight-binding calculation in the original references \cite{CFLV15,CKLV16}.
When the crystal lattice has additional spatial symmetries, e.g. supporting an invariant 3-rank tensor $f_{ijk}$ as happens in graphene \cite{VKG10}, additional elastic axial gauge fields can be constructed as $A^5_i=f_{ijk}u_{jk}$.
Axial electromagnetic fields are defined from the elastic gauge fields  in the standard way: 
\begin{equation}
E_i^5 = \partial_i A_0^5-\partial_0 A_i^5\;, \;
B_i^5=\epsilon_{ijk}\partial_j A^5_k. 
\end{equation}

Experimentally there are some promising directions to observe  elastic gauge fields arising from strain. Ref.~\cite{StrainLL19} reported the observation of pseudo-Landau levels potentially associated to corrugations of the sample in a Weyl semimetal thin film. 
Alternatively, pseudo-Landau levels where also observed in acoustic realizations of band structures hosting Weyl points~\cite{PSetal19}. 
In this experiment a Weyl band structure is achieved by a 3D-printed periodic structure of holes, which imposes a Weyl-like dispersion relation for sound waves. In this platform it is then possible to imitate the effect of strain by spatially varying the size of the holes, by analogy with theoretical predictions for cold-atomic systems\cite{Roy_2018}. This spatial-variation  translates into a dispersion relation for sound-waves governed by pseudo-Landau levels, observed in Ref.~\cite{PSetal19}.

Beyond strain, there are several promising alternatives to realize axial fields in experiment.
In a time-reversal breaking Weyl semimetal, a non-uniform magnetization also can enter to lowest order as 
a space-dependent  $\vec{\lambda}$\cite{Strain13}. 
The reason is that, for a time-reversal breaking Weyl semimetal, the spatial part of the coupling $\lambda_i\gamma_5\gamma_i$ can be rewritten in the chiral basis as $\vec{\lambda}\cdot \vec{\sigma}$, coupling with the same sign to both chiralities. The latter expression is of the form of a Zeeman coupling term $\vec{M}\cdot \vec{\sigma}$ if we identify the magnetization $\vec{M}$ with the Weyl node-separation in momentum space $\vec{\lambda}$. Hence, if the magnetization varies inhomogeneously in space, as occurs in the Weyl semimetal candidate PrAlGe~\cite{Destraz2020},  $\vec{\lambda}$ may vary in space.

Interfaces between Weyl semimetals with different Weyl node separation naturally lead to a spatial change in $\lambda_\mu(x)$, and hence also generate pseudo-electromagnetic fields. A simple example is the interface between a Weyl semimetal, where $\lambda_\mu$ is finite, and vacuum, where $\lambda_\mu=0$\cite{CCetal14}. It is at the interface where $\lambda_\mu$ changes, which implies a localized pseudomagnetic field at the boundary, due to the definition $\vec{B}_5=\vec{\nabla}\times \vec{\lambda}$. At the boundary $\vec{B}_5$ generates pseudo Landau-levels, which are nothing but a reinterpretation of the celebrated Fermi arc surface states of Weyl semimetals~\cite{GVetal16}.

The axial pseudo-electromagnetic fields
have some particular symmetry properties.  
The most important difference with standard electromagnetic fields is that, since they couple with opposite signs to opposite chiralities, 
the total flux of $\vec{B}_5$ over the sample is necessarily zero, as the band structure has no net chirality. 
As a general consideration, since lattice deformations do not break time-reversal symmetry,  strain-induced axial fields in  time reversal invariant Weyl semimetals will give rise to time reversal invariant pseudomagnetic fields. In the case of systems with  inhomogeneous magnetization, time-reversal is always broken.

\subsection{New anomaly induced transport phenomena in the presence of axial vector fields}

Pseudo-electromagnetic fields can lead to anomalous transport phenomena. For example, a pseudomagnetic field can lead to a pseudo-chiral magnetic effect:
\begin{equation}
\label{eq:pCME}
    \vec{j}^\mathrm{pCME} = \dfrac{\mu}{2\pi^2} \vec{B}_5,
\end{equation}
which unlike the chiral magnetic effect, introduced in Eq.~\eqref{eq:cse}, can exist in equilibrium. The reason is that $j^\mathrm{pCME}$ is not a transport current but a magnetization current $\vec{j}_\mathrm{mag}$. This can be seen by recalling that $\vec{j}_\mathrm{mag} \propto \vec{\nabla} \times \vec{M}$,  $\vec{B}_5=\vec{\nabla}\times \vec{\lambda}$ and by identifying the magnetization $\vec{M}$ with the Weyl node separation in momentum space $\vec{\lambda}$. Since the origin of $j^\mathrm{pCME}_i$ is the magnetization, we learn that this current can only exist in time-reversal broken Weyl semimetals.

\begin{figure}[!thb]
\begin{center}
\includegraphics[scale=0.30,clip=true]{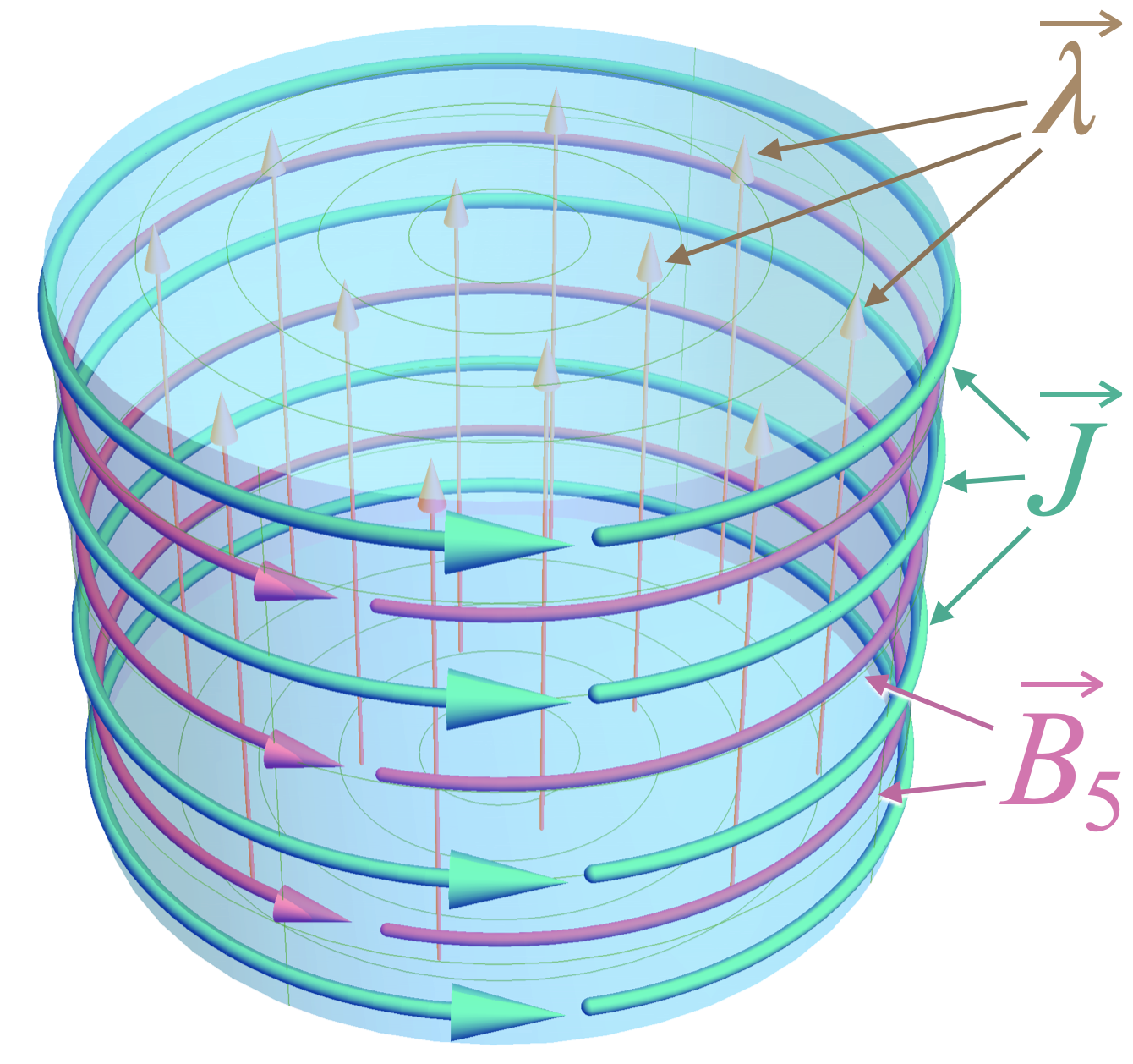}
\end{center}
\caption{An AME energy current $\vec J$ (the light green arrows) around a cylinder of a Weyl semimetal induced by the axial gauge field ${\vec B}_5$ (purple arrows) generated 
due to the presence of 
a boundary as described in the text.}
\label{fig_AME}
\end{figure}

The axial magnetic effect   is an anomaly induced transport phenomenon 
which consists on
the generation of an energy current in the direction of
an applied axial magnetic field ${\vec B}_5$:
\beq
T^{0i}=J^i_\epsilon=\sigma_{AME} B^i_5.
\label{ame}
\eeq
For a single (massless) Dirac fermion, i.e. one pair of
Weyl-cones, the axial magnetic conductivity  is
\beq
\sigma_{AME}=\frac{\mu^2+\mu^2_5}{4\pi^2}+\frac{T^2}{12}.
\label{sigmaAME}
\eeq
As discussed in Sec. \ref{sec_chiralanoms}, the temperature dependent term in the axial magnetic effect (AME) conductivity coefficient  comes 
from the gravitational contribution to the axial anomaly.
Numerical evidence of the  AME was reported in  \cite{BCLetal13} but,
since there are no axial magnetic fields in particle physics, this effect did not receive as much attention as the 
chiral magnetic effect. 

The simplest example of a boundary pseudomagnetic field was proposed in the early work  \cite{CCetal14}. Consider a 3D cylinder of Weyl semimetal of height $L$ and basal radius $a$ as the one shown in Fig. \ref{fig_AME} and choose the axis of the cylinder in the direction of the node's separation, say OZ.  Then, due to the boundary, the nodes separation will be 
\begin{equation}
\vec{\lambda}=\hat{z}b_{z}\Theta(a-|r|).
\end{equation}
We saw that the vector $\vec{\lambda}$ couples to the fermions as an axial gauge field. The corresponding axial magnetic field will point in the azimuthal direction as shown by the purple arrows in Fig. \ref{fig_AME} and is proportional to the distance between the nodes:
\begin{equation}
B_\theta\sim \lambda_z\delta(|r|-a). 
\end{equation}
An  energy current
\beq
J^i_\epsilon=\sigma_{AME} B^i_5
\eeq
runs around the cylinder (green arrows in Fig.  \ref{fig_AME}) with the coefficient given by eq. \eqref{sigmaAME}. This is an equilibrium current with zero divergence, another example of  a magnetization energy  current  \cite{SA21}.

In the original reference \cite{CCetal14}, the temperature dependent of $\sigma_{AME}$ was used to propose a device which would be a direct evidence for the gravitational anomaly.  In the setup of Fig. \ref{fig_AME}, the energy current would induce an angular momentum 
$L_k=\int_{\cal V} \varepsilon_{ijk} x_i T_{0j}$
along the axis of the cylinder (of volume ${\cal V}$):
\beq
\label{lz}
L_{z} =\int_{\cal V} \varepsilon_{z r \theta} \;r\;J_{\epsilon}^j= 2\pi \sigma_{AME} a^2\;L \lambda_{z}.
\eeq
Plugging in the expression  \eqref{sigmaAME} for $\sigma_{AME}$ it follows that, at zero density and in the absence of a chiral imbalance the states at the edge of the cylinder posses an angular momentum of magnitude
\beq
\label{Lz}
L_z=\frac{N_f}{6} T^2 \lambda_z {\cal V},
\eeq
where $N_f$ is the total number of pairs of Weyl cones in the material.
If the system, initially at a given temperature $T_{i}$, is heated to $T_{f}=T_{i}+\Delta T$ the angular momentum due to the AME will increase. 
Since the total angular momentum is conserved, the cylinder has to rotate in the opposite direction to compensate. 

That the axial gauge fields would give rise to equilibrium magnetization currents was explicitly  recognized in  \cite{GVetal16} were the Fermi arcs of Weyl semimetals were  identified as the zeroth Landau level arising from the boundary pseudomagnetic field described here.
\\

Other generalizations of anomalous transport phenomena due to axial gauge fields have been analyzed in the literature, see e.g. Refs~\cite{LPF17,Guanetal17,ACV17,GMetal17a,GMetal17b,GMetal17c,CV19-2,WC19,Ojanen19,BCLV20,Asgari20,Dima21}.
For an in-depth discussion of these issues we refer the reader to the focused reviews~\cite{Strain20,Strain21}. It is important to keep in mind that these fields must average to zero over a finite sample, which at times complicates their interpretation in lattice models~\cite{Behrends2019}. 

\subsection{Curvature and elastic deformations}

In the geometric formulation of elasticity, the strain tensor is identified with the metric of the slightly deformed flat space-time as:
\begin{equation}
g_{ij}  =\delta_{ij}+\frac{1}{2}u_{ij} . 
\end{equation}
The identification of the underlying lattice supporting the electronic fluid with the external spacetime leads to the suggestion that we could induce gravitational anomaly effects with lattice deformations. In particular it would be tempting to try to induce curvature by deforming the lattice. The formalism of quantum field theory in curved space has a long history in the graphene literature \cite{GGV92,VKG10} and, later, in Dirac and Weyl semimetals \cite{CZ16,Ojanen19}.  An important aspect of any realistc model 
for materials is that the induced metric in any dimension should have the structure 
\begin{equation}
g_{\mu\nu}=
\begin{pmatrix} 
1 & 0  \\
0 & g_{ij}(x) 
\end{pmatrix},
\quad
\end{equation}
i.e., only the spacial part  $g_{ij}(x)$ can be modified. This is a severe restriction on material realizations. Another important point to notice is that, while it is possible to induce curvature in a 2D sheet embedded in 3D by deforming the lattice in the third dimension,  elastic deformations of a 3D lattice will always induce metrics with zero curvature~\cite{Sun17}.

Finally, notice that the elastic gauge fields defined in \eqref{eq_elasticGF} are linear in the metric and would not appear in a covariant relativistic formulation. Linear terms can arise at finite temperature or chemical potential using the frame vector described in Appendix \ref{sec_velocities}. Being linear in the deformation, they are the most relevant terms for the physical implications. 
\\

\begin{tcolorbox}
\begin{itemize}
\item{\bf Axial vector fields arise in Dirac matter associated to crystal lattice deformations, boundaries, or inhomogeneous magnetization. Unlike the gauge potential, they are real fields and can give rise to equilibrium (magnetization) currents.}
\item{\bf Elastic deformations cannot induce curvature in a 3D lattice.}
\end{itemize}
\end{tcolorbox}    


\section{The scale anomaly and thermoelectromagnetic currents}
\label{sec_Scale}
\label{sec_conformal}

In this section we will analyze a set of new physical phenomena that arise in Dirac systems due to the scale anomaly, in both quantum field theory and in its material realizations. The scale or conformal anomaly~\cite{Coleman85} is the breakdown of the scale invariance of a theory after its quantization. As we will see, in a perturbative sense the scale anomaly arises due to the emergence of a scale-dependence (``running'') acquired by the coupling constants of the theory in the process of renormalization\footnote{The scale anomaly can also appear via a non-perturbative mechanism of spontaneous mass gap generation in classically scale-invariant models. While we do not consider this mechanism in the context of the condensed matter physics, we notice that in particle physics this mechanism plays an important role in the context of non-Abelian gauge theories relevant to the theory of strong interactions~\cite{Shifman:1979qcd}.}.

A theory is said to be scale invariant when the classical equations of motion are invariant under a simultaneous rescaling of all coordinates and fields (denoted below as a generic field $\Phi$), according to their canonical dimensions $\Delta_\Phi$, with a global, coordinate-independent factor $\lambda$:
\beqn
x^\mu \to \lambda^{-1} x^\mu,  \qquad \Phi(x) \to \lambda^{\Delta_\Phi} \Phi(x)\,.
\label{eq_scale_transformation_general} 
\eeqn
One important example of a scale-invariant system is the low-energy behavior of gapless fermionic excitations in Dirac semimetals.
As we mentioned in  section \ref{sec_dic}, a  continuous symmetry of a classical system leads to the existence of a conserved Noether current. In particular,  scale invariance implies the conservation of the dilatation current,
\beqn
j_D^\mu = T^{\mu\nu} x_\nu\,.
\label{eq_j_D}
\eeqn
The conservation of the current~\eq{eq_j_D} at the level of the classical equations of motion,
\begin{equation}
   \partial_\mu  j_D^\mu= x_\nu \partial_\mu  T^{\mu\nu} + T^\mu_\nu\delta_\mu^\nu = 0\,,
    \label{eq:trace1}
\end{equation}
together with the conservation of energy and momentum, $\partial_\mu T^{\mu\nu} = 0$,  implies the tracelessness of the energy-momentum tensor $T^{\mu\nu}$:
\beqn
(T^\mu_\mu)_{\cl} \equiv 0\,,
\label{eq_trace_zero}
\eeqn
which is the signature of scale invariance in a physical system.

One should remark that the concepts ``scale'' and ``conformal'' (used with the words ``transformation'', ``invariance'', ``anomaly''), are often employed interchangeably in the literature. Mathematically, these are different notions~\cite{Nakayama:2013is}. In geometry, the scale transformation corresponds to an operation that modifies all spatial dimensions of an object by multiplying them with the same global scale factor~\eq{eq_scale_transformation_general}. This global dilatation should be contrasted to the conformal transformation, constrained by a single requirement that after the conformal transformation, the angles between any intersecting curves or planes remain the same.
Mathematically, the conformal transformation corresponds to a diffeomorphism, $x^\mu \to {x^\mu}{}' = f^\mu(x)$, which leaves the metric unchanged up to an arbitrary multiplicative nowhere vanishing scalar factor $\Omega$:
\beqn
g_{\mu\nu}(x) = \Omega^2(x') g_{\alpha\beta}'(x') \frac{\partial {f^\alpha}}{\partial x^\mu} \frac{\partial {f^\beta}}{\partial x^\mu}\,. 
\label{eq_conformal}
\eeqn
A theory is said to be  conformal  in the classical sense when its action remains invariant under the conformal transformation~\eq{eq_conformal} supplemented by an appropriate transformation of the fields according to their canonical dimensions. The conformal group has a bigger parameter space as compared to the single global dilatation parameter of the scale transformation. 

Physically, the difference between the scale and conformal concepts is often ignored because all physically meaningful scale-invariant quantum field theories in three spatial dimensions exhibit conformal invariance, despite of the fact that the requirement of local conformal invariance is much stronger than the condition of global scale invariance~\cite{Nakayama:2013is}. The renormalization group effects that will be discussed in this section break both the scale and conformal groups. To keep the mathematical rigor of Eq.~\eq{eq_scale_transformation_general}, we will keep using the word ``scale'' in our review. 

Other similar concepts that often appear in the literature are ``Weyl invariance'' (or ``Weyl anomaly'' if this invariance is anomalously broken) and ``trace anomaly''. The first notion corresponds to the invariance of the action of the theory with respect to the rescaling of the metric by a local multiplicative factor, $\Omega>0$, along with the local scale transformation of the field: 
\beqn
g_{\mu\nu}(x) \to \Omega^2(x) g_{\mu\nu}(x)\,, \qquad \phi \to \Omega^{-\Delta_\phi}(x) \phi(x)\,.
\label{eq_Weyl_transformation}
\eeqn
As a local rescaling, the Weyl transformation~\eq{eq_Weyl_transformation} differs from the scale transformation~\eq{eq_scale_transformation_general}, which is defined by a global rescaling of coordinates and fields, as well as from the conformal transformation~\eq{eq_conformal}, which is given by a diffeomorphism supplemented with a field transformation (which is not a rescaling, in general). As we mentioned above, physically meaningful theories do not distinguish between the scale and conformal notions~\cite{Nakayama:2013is} in a sense that both  symmetries are either simultaneously broken or simultaneously respected. Weyl and conformal symmetries are also compatible in physical theories with certain exceptions in the form of exotic higher-derivative theories~\cite{Karananas:2015ioa} for which the conformal invariance does not automatically imply the invariance under Weyl transformations. All three (scale, conformal, and Weyl) symmetries are, however, broken by quantum fluctuations if the classically vanishing trace of the energy-momentum tensor~\eq{eq_trace_zero} acquires a nonzero expectation value, $\avr{T^\mu_\mu} \neq 0$. Hence the name ``trace anomaly''. All mentioned terms, ``conformal anomaly'', ``scale anomaly'', ``trace anomaly'', or ``Weyl anomaly'', are often used on equal footing in the literature with different names serving to stress slight but sometimes essential variations on the nature of the corresponding quantum effects.

\subsection{Scale anomaly in quantum electrodynamics}
\label{subsec_QED}

\begin{figure}[!thb]
\begin{center}
\includegraphics[scale=0.45,clip=true]{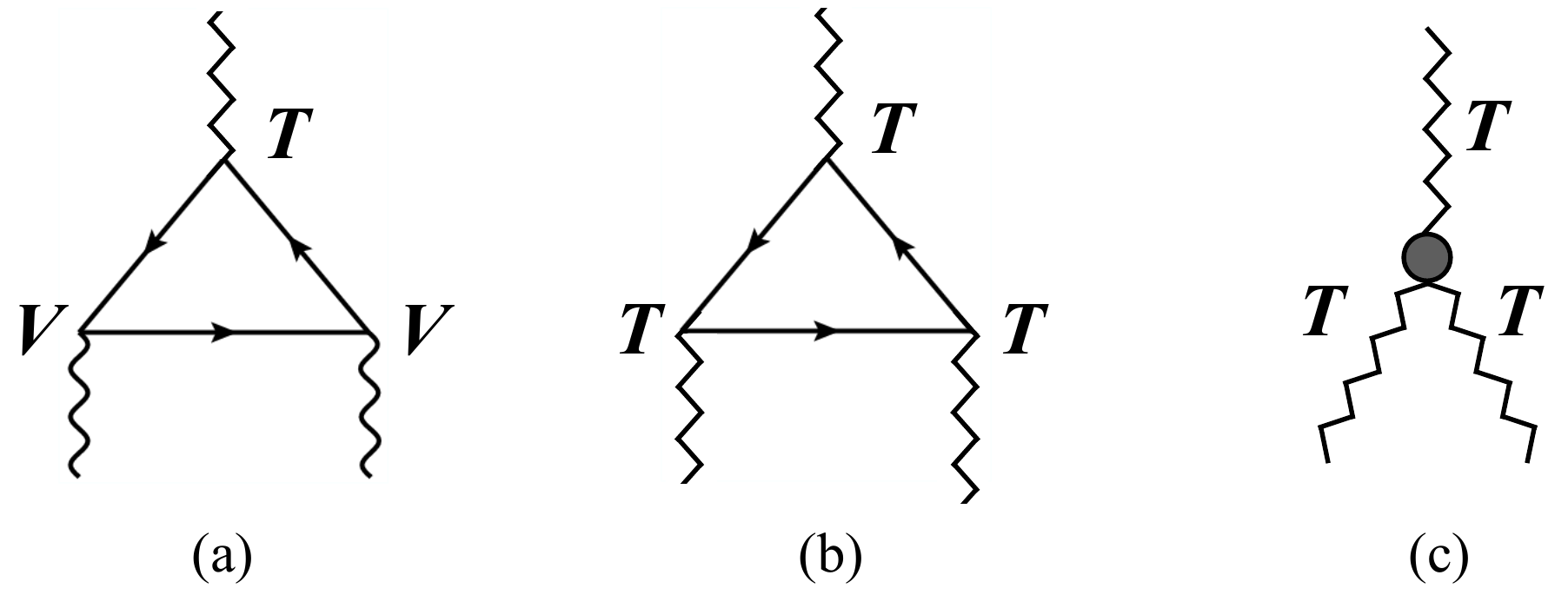}
\end{center}
\caption{The three-point triangular (a) TVV and (b) TTT diagrams as well as the contact term (c) that generate the scale anomaly. The triangular diagrams involve the vector, $V^\mu = \gamma^\mu$, and graviton, $T^{\mu\nu} = i(\gamma^\mu\partial^\nu + \gamma^\nu\partial^\mu)/2$, external lines.}
\label{fig_scheme}
\end{figure}

As  mentioned above, the  scale invariance occurs in models with no dimensionful parameters: massless particles and dimensionless coupling constants. In the standard classification of renormalization, these couplings lead to strictly renormalizable models~\cite{Zinn2021}. The scale anomaly is then associated to the running coupling constants of the model. In what follows we will describe this process in massless quantum electrodynamics, a paradigmatic model in quantum field theory which is also our  main character for the condensed matter applications. 

The Lagrangian of massless QED in a flat Minkowski space-time is
\beqn
\cL = - \frac{1}{4} \left( {\vec B}^2 - \frac{1}{c^2} {\vec E}^2 \right) +
\bar{\psi}\Bigl[\gamma^0 (i \partial_t - e A_t) + {\vec \gamma}\cdot \Bigl(i c{\vec \nabla} - e {\vec A}\Bigr)\Bigr]\psi,
\label{eq:L:QED}
\eeqn
where $\psi$ is the Dirac four spinor, $\bar \psi = \psi^\dagger \gamma^0$ and $\gamma^\mu$ are the Dirac matrices. We have split the time and space components to show where the speed of light appears. In the condensed matter applications $c$ will be replaced by $v_F$. The electronic current is given by the variation of the action of the theory~\eq{eq:L:QED} $S = \int d^4 x \, \cL$  with respect to the electromagnetic field~\eq{eq:defcurrent}:
\beqn
J^\mu = - \frac{\delta S}{\delta A_\mu} = e \bar \psi \Bigl(\gamma^0, c{\vec \gamma} \Bigr)\psi.
\eeqn
For the purpose of this section, we restore the electric coupling $e$ in front of the electromagnetic gauge field~$A^\mu$. The theory described by~\eq{eq:L:QED} is scale invariant at the classical level, which implies that the classical action~\eq{eq:L:QED} and, respectively, the classical equations of motion are invariant under a simultaneous rescaling of all coordinates and fields, according to their canonical dimensions~\eq{eq_scale_transformation_general}
\beqn
x^\mu \to \lambda^{-1} x^\mu,  \qquad A_\mu(x) \to \lambda A_\mu(x), \qquad \psi(x) \to \lambda^{\frac{3}{2}} \psi(x).
\label{eq_scale_transformation}
\eeqn

It is convenient to write the Lagrangian~\eq{eq:L:QED} in the concise form 
\beqn
\cL = - \frac{1}{4} F^{\mu\nu} F_{\mu\nu}  + {\bar \psi} i {\slashed D} \psi\,,
\eeqn
where ${\slashed D} = \gamma^\mu D_\mu$, with the covariant derivative $D_\mu = \partial_\mu - i e A_\mu$, and $F_{\mu\nu} = \partial_\mu A_\nu -  \partial_\nu A_\mu$ is the field strength tensor of the gauge field $A_\mu$. Then the (symmetric) stress-energy tensor can be obtained by the variation of the action $S$ with respect to the background metric $g_{\mu\nu}$~\eq{eq_energy_momentum_tensor},
\beqn
T^{\mu\nu} (x) & = & - \frac{2}{\sqrt{-g}}  \frac{\delta S}{\delta g_{\mu\nu}(x)} {\biggl|}_{g^{\mu\nu} = \eta^{\mu\nu}}
= - F^{\mu\alpha} F^\nu_{\  \alpha} + \frac{1}{4} \eta^{\mu\nu} F_{\alpha\beta} F^{\alpha\beta} 
+ \frac{i}{2} {\bar \psi} \left(\gamma^\mu D^\nu + \gamma^\nu D^\mu \right) \psi - \eta^{\mu\nu} {\bar \psi} i {\slashed D} \psi\,,
\label{eq_Tmunu_QED}
\eeqn 
with $g = \det (g_{\mu\nu})$.

The tensor~\eq{eq_Tmunu_QED}, is an identically traceless quantity in three spatial dimensions~\eq{eq_trace_zero} (the fermionic part vanishes due to the classical equations of motion, ${\slashed D} \psi = 0$). 

In a generic interacting quantum field theory, quantum fluctuations break the classical scale invariance~\eq{eq_scale_transformation} due to the renormalization of the couplings of the model by quantum corrections. In quantum electrodynamics~\eq{eq:L:QED} the electromagnetic coupling $e$ ``runs'', implying the emergence of a dependence of the electric coupling $e = e(E)$ on the energy scale $E$ of the interaction which involves a photon exchange~\cite{Shifman:1988zk}. This affects differently the physical phenomena that develop at different energy scales (or at different distances). This running of the coupling constant breaks the scale invariance in the quantum  theory, and is a signature of the quantum scale anomaly.

The conformal anomaly reveals itself in a non-vanishing expectation value of the trace of the energy-momentum tensor~\eq{eq_Tmunu_QED}. In the background of a classical electromagnetic field $A_\mu \equiv A^{\mathrm{cl}}_\mu$ one gets~\cite{Shifman:1988zk}\footnote{The electromagnetic part of the trace anomaly~\eq{eq_Tmunu_average} is written for the flat Minkowski metric, $g_{\mu\nu} \equiv \eta_{\mu\nu}$. In a gravitational background, $g_{\mu\nu} \neq \eta_{\mu\nu}$, the trace of the energy-momentum tensor acquires curvature-dependent terms shown explicitly in Eq.~\eq{eq_Tmunu_gravity} 
of Appendix~\ref{sec_curvedform}.}
\beqn
\avr{T^\alpha_{\ \alpha}(x)} = \frac{\beta_e(e)}{2 e}  F^{\mu\nu}(x) F_{\mu\nu}(x),
\label{eq_Tmunu_average}
\eeqn
where $\beta_e(e)$ is the beta-function associated with the running coupling $e$,
\beqn
\beta_e = \frac{{\mathrm d} e(E)}{{\mathrm d} \ln E}\,,
\label{eq_beta}
\eeqn
normalized at the energy scale~$E$. At the level of perturbation theory, the one-loop contribution to the scale anomaly~\eqref{eq_Tmunu_average} is generated by the TVV triangular Feynman diagram shown in Fig.~\ref{fig_scheme}(a). Despite its similarity with the anomalous vertices corresponding to the axial anomaly and the mixed axial-gravitational anomaly shown in Fig.~\ref{fig_scheme_chiral},
the scale anomaly is not one loop exact. This property is a natural consequence of the fact that any coupling of a generic quantum field theory acquires loop corrections at all orders of perturbation theory.

On the contrary, the pure curvature-dependent parts of the scale anomaly, discussed in Appendix~\ref{sec_curvedform}, are one-loop exact. They are generated by the three-point triangular, Fig.~\ref{fig_scheme}(b), and contact, Fig.~\ref{fig_scheme}(c), diagrams. The contact term is needed to maintain the general covariance of the anomalous contribution at the three-point level.

\begin{figure}[!thb]
\begin{center}
\includegraphics[scale=0.25,clip=true]{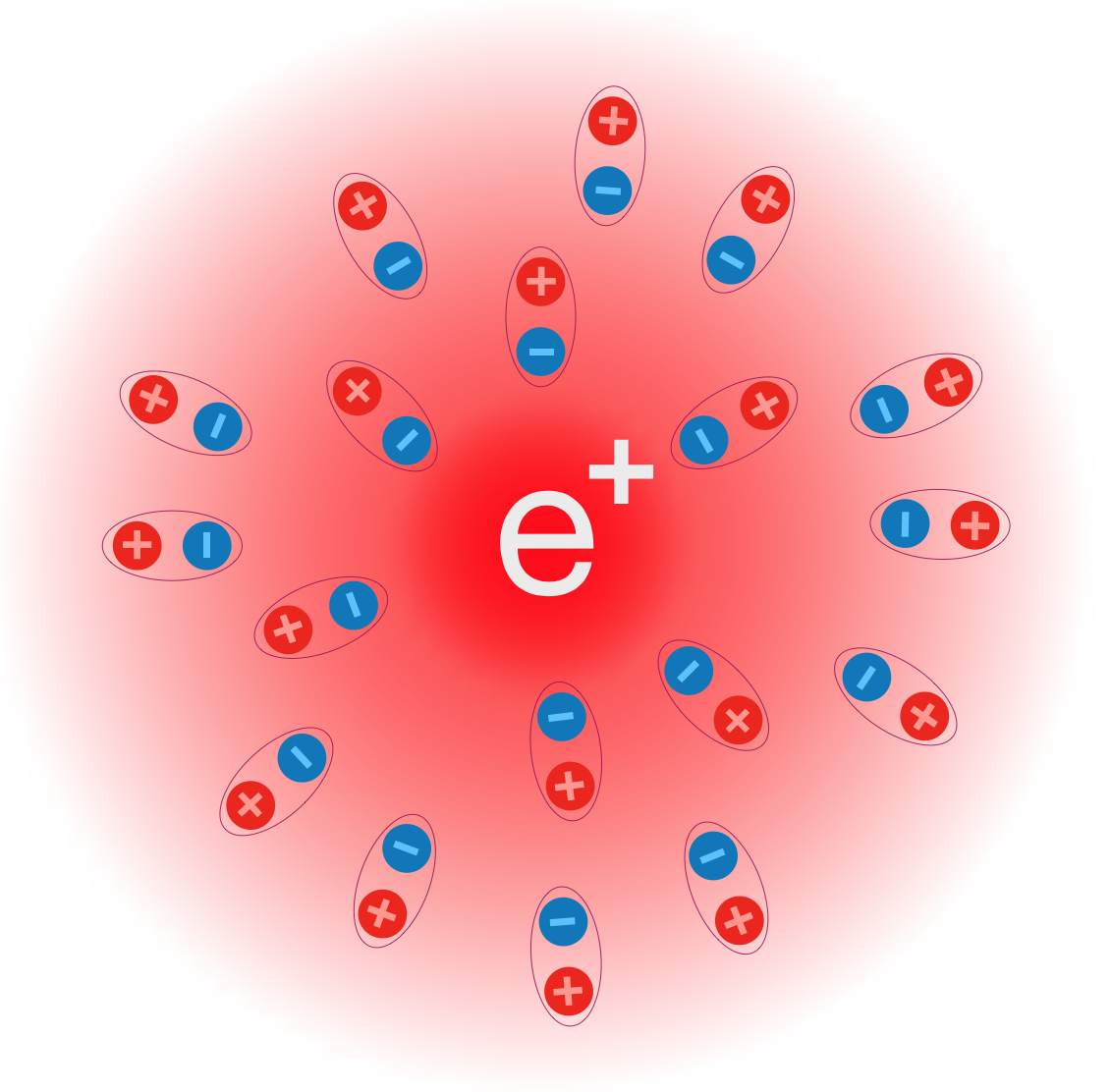}
\end{center}
\caption{Screening of a test charge in quantum electrodynamics: the electric field of the test particle gets screened in the vacuum via polarization of the short-living particle-antiparticle pairs that are associated with tiny electric dipoles.}
\label{fig_screening_bulk}
\end{figure}

Physically, the appearance of the beta function~\eq{eq_beta} in quantum electrodynamics, is associated to the polarization of the vacuum by the creation and annihilation of short-living pairs of virtual particles and antiparticles  due to quantum fluctuations. In matter it can be understood as associated to the screening of a test charge by the charges in the medium Fig.~\ref{fig_screening_bulk}. This physical screening effect, reflected in non-vanishing beta function~\eq{eq_beta}, is often associated with the ``renormalization'': a mathematical procedure that is used to calculate the screening effect.

We would like to stress that the transport effects generated by the scale anomaly  in an electromagnetic field background are exclusively associated to the running of the electric coupling $e$  with the energy scale. 
It is known that in the condensed matter systems, graphene being the paradigmatic example,  the Fermi velocity is  renormalized~\cite{Gonzalez:1993uz,Elias-et-al-2011} and the scale anomaly is  broken by the running of the Fermi velocity. But in the (2+1) version of QED describing graphene, the electric charge $e$ is not renormalized and the scale anomaly does not lead to any transport effects in response to the electromagnetic background fields. This is so even  though  the effective fine structure constant $\alpha \propto e^2/v_F$ runs with the energy in a very similar way as in QED~\cite{V11,V11b}.
Since the running of the Fermi velocity is inessential for our purposes, in what follows, we consider  a Lorentz-invariant formulation of the theory with $v_F = c$, which allows us to write formulas in a concise form. At the end of our discussion, we restore $v_F$ in the expressions.

The running of the coupling constant $e$ with the energy $E$ is governed by the renormalization group equation~\cite{Peskin:1995ev}:
\beqn
E \frac{\partial e(E)}{\partial E} =- 2 b_1 e^3, 
\label{eq_renorm_QED}
\eeqn
where the c-number parameter $b_1$  called also the bulk central charge of the model, is related to the beta function eq. \eq{eq_beta}  as $b_1 = - \beta(e)/(2 e^3)$. 
In quantum electrodynamics with $N_f$ species of Dirac fermions, the one-loop beta function Eq. \eq{eq_beta} is~\cite{Peskin:1995ev}:
\beqn
\beta_{e}^{{\text{QED}}}(e) = N_f  \frac{e^3}{12 \pi^2}.
\label{eq_beta_QED}
\eeqn
The renormalization group equation~\eq{eq_renorm_QED} allows to relate the value of the coupling constant at an energy $E$ with its value a a reference energy $E_0$ as
\beqn
e^2(E) = \frac{e_0^2}{4 b_1 \ln \left( E/E_0 \right)}.
\label{eq_e2_E:1}
\eeqn

\subsection{Scale anomaly and transport effects in the bulk}

The scale anomaly~\eqref{eq_Tmunu_average} leads to anomalous transport effects which are easiest to analyze in  a conformally flat space-time metric. A conformal transformation
\beqn
g_{\mu\nu}(x) = e^{2 \tau(x)} \eta_{\mu\nu}\,,
\label{eq_g_munu}
\eeqn
where $\tau(x)$ is a scalar conformal factor and $\eta_{\mu\nu}$ is the Minkowski metric tensor, induces, through the relation \eqref{eq_Tmunu_average}, the  electric current:
\beqn
J^\mu = - \frac{2 \beta_e(e)}{e} \partial_\nu \left( F^{\mu\nu} \tau \right)\,,
\label{eq_J_covariant_0}
\eeqn
where $e = \pm |e|$ is the elementary electron charge. 

The current~\eq{eq_J_covariant} is explicitly conserved: $\partial_\mu J^\mu = 0$. For a classical   electromagnetic field background, ($\partial_\nu \left( F^{\mu\nu} \tau \right) \equiv F^{\mu\nu} \partial_\mu \tau$) with $\partial_\mu \tau$ acting as a source of the gravitational perturbation, we get
\beqn
J^\mu(x) = - \frac{2 \beta_e(e)}{e}  F^{\mu\nu}(x) \partial_\nu \tau (x)\,.
\label{eq_J_covariant}
\eeqn

It is important to stress that, contrary to the transport phenomena generated by the axial and mixed axial-gravitational anomalies, the current~\eq{eq_J_covariant} caused by the scale anomaly is produced by the quantum vacuum and emerges at zero chemical potential and zero temperature. 

The  expression~\eq{eq_J_covariant} leads to two distinct transport effects which are realized in the presence of classical electromagnetic fields: The scale magnetic effect (SME)~\cite{Chernodub16} refers to the generation of an electric current  in a  background magnetic field $\vec B$ and a weakly, conformally curved time-independent background, $|\tau| \ll 1$ :
\beqn
{\vec J}(x) = - \frac{2 \beta_e(e)}{e} {\vec B}(x) \times {\vec \nabla} \tau(x)\,.
\label{eq_SME}
\eeqn
The electric current~\eq{eq_SME} is perpendicular to the  magnetic field ${\vec B}$ and to the gradient of the gravitational perturbation, ${\vec \nabla} \tau$. This transport phenomenon is relevant for condensed matter applications as we discuss below.

In the presence of an electric background field ${\vec E}$ the conformal anomaly leads to the scale electric effect (SEE):  An anomalous correction to the Ohmic contribution to the electric charge transport
given by
\beqn
{\vec J}_{\mathrm{SEE}} = \sigma(x) {\vec E}(x)\,,
\label{eq_SEE}
\eeqn
where the metric-dependent anomalous electric conductivity is
\beqn
\sigma(t, {\vec x}) & = & - \frac{2 \beta_e(e)}{e} \frac{\partial \tau(t,{\vec x})}{\partial t}\,.
\label{eq_sigma_anomalous}
\eeqn
Since the conductivity~\eq{eq_sigma_anomalous} takes a nonvanishing value only in a time-dependent curved background, the condensed-matter applications of the SEE transport phenomenon are less transparent than these of its magnetic counterpart \eq{eq_SME}. We discuss the SEE transport phenomenon in Appendix~\ref{sec_curvedform} mentioning briefly its particle-physics applications relevant for the physics of the early universe.

\subsection{Anomalous thermoelectric current due to the conformal anomaly}

The Nernst effect described in Section~\ref{sec_TT}, consists on the generation of an electric current perpendicular to a temperature gradient and to an applied magnetic field, Eq. ~\eq{eq:Nernst}. The conductivity, Eq.~\eq{eq_Nernstcoefficient}  is determined by the components of the ${\bs L}^{(2)}$ transport tensor defined in Eq.~\eq{eq transport 1}. While the ordinary Nernst phenomenon appears due to thermal activation of the electron and hole charge carriers, we will see how the conformal anomaly can also generate the same effect with a coefficient proportional to the beta function of QED discussed in  section \ref{subsec_QED}. 

For a metric with non vanishing $h^{00}$ component, the scale anomaly gives rise to the current \eqref{eq:anomalous_Nernst_app}
\begin{equation}
    \vec{J}=\frac{\beta(e)}{3e}\vec{B}\times\vec{\nabla}h^{00}.
\label{eq_C_SME_vF0}
\end{equation}
Using the Luttinger relation $\vec{\nabla}T/T=-\vec{\nabla}\phi$ discussed in Sec. \ref{sec:luttinger}, with $h^{00}=-2\phi$, and restoring the powers of  the Fermi velocity $v_F$ (that replaces the velocity of light in the Dirac materials) and the value of the beta function, we get
\begin{equation}
    \vec{J}=\frac{e^2v_F}{18\pi^2\hbar T}\vec{B}\times\vec{\nabla}T
    \label{eq_C_SME_vF}
\end{equation}

\begin{figure}[!thb]
\begin{center}
\includegraphics[scale=0.80,clip=true]{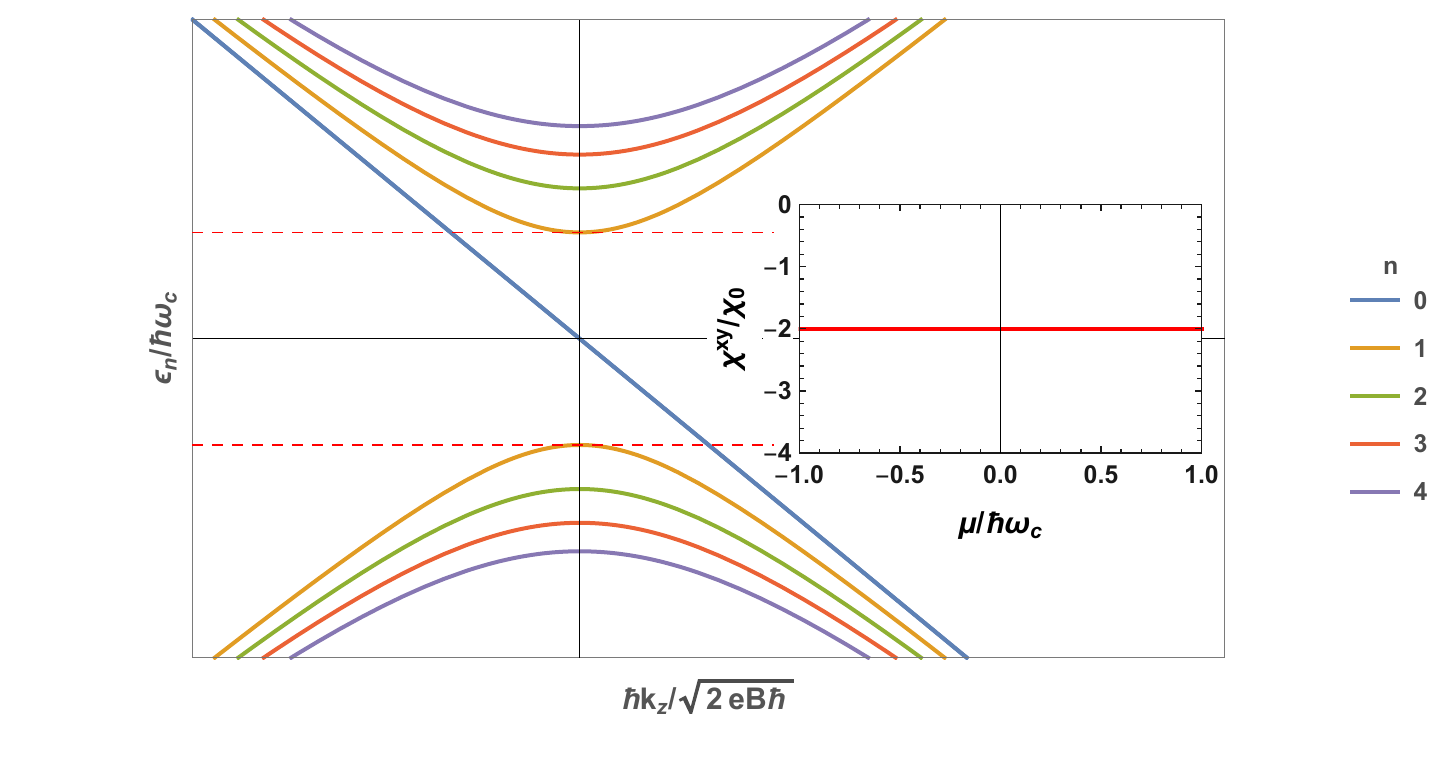} 
\end{center}
\caption{ Landau level spectrum of a single chirality Weyl system. The inset represents the flat thermoelectric coefficient as a function of the chemical potential in the interval between the first Landau levels $n=\pm 1$ normalized to its value at $\mu=0$.  }
\label{Fig:LL}
\end{figure}

A standard Kubo calculation of  the thermoelectric coefficient ${\bs L}^{(2)}$ ~\eq{eq transport 1} in the conformal limit ($T=\mu=0$)~\cite{ACV19} also gave a non zero result numerically compatible with \eqref{eq_C_SME_vF}.  Since the  conformal anomaly is not exact, the numerical coefficient of the resulting transport phenomena has to be fixed by experiments.

Although the strict  conformal situation (zero temperature and chemical potential) can not be realized in experimental conditions, there are measurable  experimental signatures of the  conformal anomaly induced transport.  An additional computation of the thermoelectric coefficient as a function of the chemical potential done in  Ref.~\cite{ACV19} predicted a plateau at large magnetic fields which will remain at its $\mu=0$ value when $\mu$ lies in the interval between the first two Landau levels (see Fig. \ref{Fig:LL}). The reason is that in the quantum limit  only the zeroth Landau level is populated. The system is projected into the (1+1) conformal Landau level whose density of states is constant. Recent experiments in ZrTe${}_5$, a Dirac semimetal~\cite{ZWS20}, show this behavior from which we  might extract the experimental value of the coefficient in eq. \eqref{eq_C_SME_vF}. 
\\

\begin{tcolorbox}
\begin{itemize}
\item{\bf  There are no non--renormalization theorems for the electromagnetic term of the scale anomaly.
The trace of energy momentum tensor is renormalized at all orders of perturbation theory in the electric coupling $\bs e$.}
\item{\bf The scale anomaly induces new transport phenomena in the bulk of Dirac semimetals at vanishing chemical potential.}
\item{\bf The scale electric effect leads to the correction to the Ohmic current in a time-dependent geometric background.}
\item{\bf The scale magnetic effect  generates an electric current perpendicular to an applied magnetic field and to a temperature gradient. This effect 
induces an anomalous contribution to the Nernst effect in Dirac semimetals.}
\end{itemize}
\end{tcolorbox}

\subsection{Scale electromagnetic effects at the boundary}

Another example of anomalous electric current generation appears near a reflective boundary of a homogeneous system. While this transport effect is not directly related to the thermal transport, the presence of the boundaries allows us to get an insight into the physics of the renormalization, which plays a role in the thermomagnetic effects produced by the scale anomaly. 

We consider a boundary that reflects the quasiparticles but is completely transparent to photons. Our motivation is that this boundary conditions give a simple model of a material in which the electrons are confined to the inside but the photons can escape to the space outside the material. 
We also emphasize that our considerations apply at the charge neutrality point in which the density of classical charge carriers vanishes. This implies that no classical screening can take place. 

\begin{figure}[!thb]
\begin{center}
\begin{tabular}{cc}
\includegraphics[scale=0.6,clip=true]{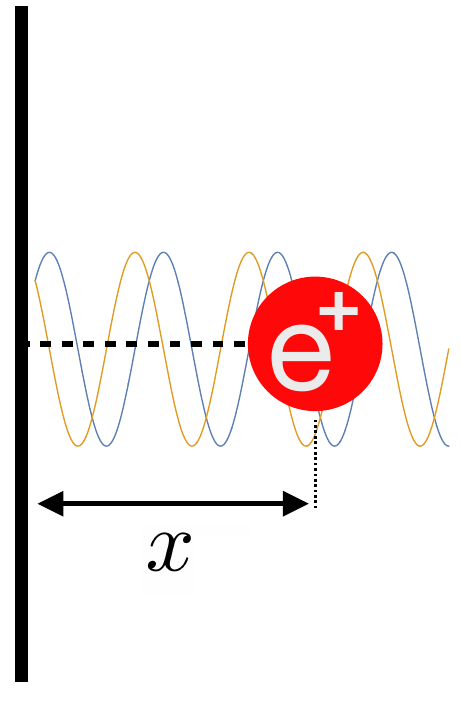} 
&
\hskip 15mm 
\includegraphics[scale=0.22,clip=true]{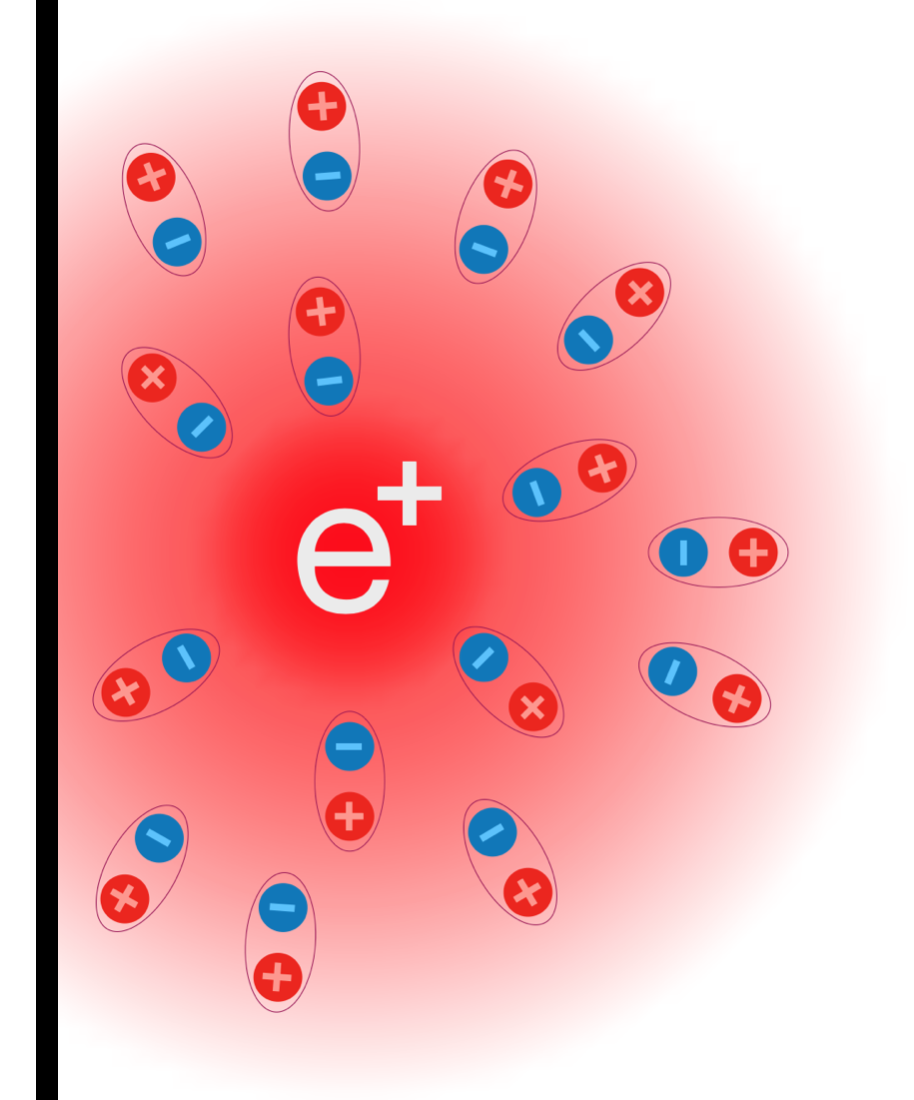} 
\\
(a) & (b) 
\end{tabular}
\end{center}
\caption{(a) Particle near the reflective non-conducting boundary. (b) Effect of the boundary on charge screening.}
\label{fig_screening_boundary}
\end{figure}

An intuitive explanation of the influence of the boundary on the renormalization of electric charge appears from the consideration of the screening in real  space. We know already that the charge screening is due to the polarization of the medium or vacuum by the electric field of the test particle, Fig.~\ref{fig_screening_bulk}. The polarization effect, however, decreases when  the test particle approaches the boundary at a distance smaller than the polarization distance. Due to a purely geometrical effect, schematically shown in Fig.~\ref{fig_screening_boundary},  part of the polarizing pairs disappear and the screening of the test charge becomes less effective. When the particle approaches the boundary too close, the screening  is incomplete due to geometrical reasons. The renormalization of the electric charge then becomes sensitive to the distance to the boundary of the system $x_\perp$.
We can translate the distance to the boundary to a momentum scale $p_\perp = \frac{\hbar v_f}{x_\perp}$. This sets the scale of the renormalized charge $e=e(p_\perp)$ implying the aforementioned dependence on the distance $x_\perp$.

Having qualitatively established that the presence of an edge  affects the renormalization of the electric charge near the edge, we can wonder if  the electromagnetic fields produce any electric transport effect close the boundary. The answer to this question was given by McAvity and Osborn about thirty years ago.
In Ref.~\cite{McAvity:1990} these authors showed that a classical electromagnetic field  acting on a bounded quantum system of charged particles, generates an electric current $J^\mu$ in the vicinity of the flat boundary. 
In a Lorentz invariant theory with $v_F = c$, the expression of the current has the form~\cite{Chu:2018ntx,Chu:2018ksb}:
\beqn
J^\mu(x) = -  \frac{2 c \beta_e}{e \hbar} \frac{F^{\mu \nu}(x) n_\nu}{x}\,,
\label{eq_J_edge}
\eeqn
where $n^\mu = (0,{\vec n})$ is the inside pointing normal vector to the edge of the system, and $x > 0$ is the spatial distance from the point where the current~\eq{eq_J_edge} is measured to the boundary along the spatial vector ${\vec n}$. The prefactor in Eq.~\eq{eq_J_edge} is given precisely by the beta function~\eq{eq_beta} which implies that the boundary effect~\eq{eq_J_edge} is generated by the scale anomaly. While we have given an intuitive interpretation of this effect with partially transparent boundary conditions it turns out that the current is universal in the sense that it actually does not depend on the particular choice of boundary conditions \cite{McAvity:1990,Chu:2018ntx,Chu:2018ksb}.

Notice the striking similarity between the coefficients in the edge electric current~\eq{eq_J_edge} and the bulk electric current induced by the scale electromagnetic effect~\eq{eq_J_covariant}. Both are proportional to the beta function and both appear at zero temperature and chemical potential.

In order to explore the possible consequences of the previous result  on a physical system, in particular on a Dirac metal, it is convenient to rewrite the edge electric current~\eq{eq_J_edge} in the reference frame of the  crystal:
\beqn
{\vec J} = - \frac{2 v_F \beta_e}{e \hbar} \frac{1}{x} {\vec n} \times {\vec B}\,,
\label{eq_SMEE}
\eeqn
 where we have also replaced the speed of light $c$ with the Fermi velocity $v_F$.
The current  is normal to the axis of the magnetic field and tangential to the boundary of the system. This scale magnetic effect at the edge (SMEE) is the direct analogue of the SME phenomenon~\eq{eq_SME} which takes place in the bulk. 

The existence of the ``scale magnetic edge effect''~\eq{eq_SMEE} was recently demonstrated in first-principles numerical simulations of scalar quantum electrodynamics with one species of massless scalar field~\cite{Chernodub:2018ihb}.

The anomalous edge current~\eq{eq_SMEE} can be interpreted in an analogous way to the edge currents that appear in the description of the classical Hall effect. In the Hall effect, an electron moves along the boundary in the form of skipping orbits which represent  circular segments joined together at the reflecting events at the boundary of the system. Contrary to the classical Hall effect, the anomalous current~\eq{eq_SMEE} emerges in the vacuum where the chemical potential is zero. The vacuum edge current appears as a result of skipping orbits of the virtual particles and antiparticles created by the quantum fluctuations near the edge of the system in the background magnetic field~\cite{Chu:2018ntx,Chu:2018ksb}. In the condensed matter context, the notion of vacuum corresponds to exact neutrality point where the sharp, zero-temperature Fermi level goes precisely through the Dirac point. Contrary to the helical edge currents that appear in quantum spin-Hall systems~\cite{QSH_graphene} and the chiral edge states in Chern insulators~\cite{Haldane1988}, the current~\eq{eq_SMEE} generated by the scale anomaly does not form an exponentially thin boundary layer.

 We also emphasize that the edge current~\eq{eq_SMEE} is a magnetization current which affects the magnetic moment of the system but does not lead to a net transport of electric charge.

The temporal component of the edge four-current~\eq{eq_J_edge} leads to the accumulation of electric charge at the edge of the system in the presence of a static electric field ${\vec E}$:
\beqn
\rho = - \frac{2 v_F \beta_e}{e \hbar c^2} \frac{{\vec n} \cdot {\vec E}}{x}.
\label{eq_SEEE}
\eeqn
This is the ``scale electric edge effect'' (SEEE). It can also be viewed as an analogue of the scale electric effect (SEE) in the bulk~\eq{eq_SEE}. While the bulk SEE phenomenon does not have so far any visible potential for the experimental verification in the condensed matter context, its edge counterpart~\eq{eq_SEEE} has a direct implication for the electrostatic charge screening in semimetals at the charge neutrality point
as we will explain now.

An ideal conductor totally screens the electrostatic field in its bulk  due to the presence of mobile charge carriers. The carriers redistribute themselves under the influence of an external electric field in such a way as to precisely compensate the external field ~\cite{ashcroft1976solid}.

In ordinary conductors, the static electric field falls down with an exponential law close to the boundary, 
\beqn
\frac{E(x)}{E(0)} \propto e^{- x/\lambda}\,,
\label{eq_screening_standard}
\eeqn
where the screening length $\lambda$ determines the width of the boundary layer of the redistributed mobile carriers. In usual metals, this width coincides with the Thomas-Fermi wavevector, which is a very small quantity in the typical range of a few nanometers. However, in the scale-invariant limit of vanishing density and temperature, all the parameters of the system reduce to dimensionless quantities and the Fermi liquid phenomenology can no more be used. In particular, the screening length $\lambda$ becomes infinite and the standard picture of the electrostatic charge screening is invalid.

However, even in the conformal limit of the exact zero-temperature neutrality point of a Dirac semimetal the electrostatic field gets  screened~\cite{CV19}. Remarkably, the screening occurs via the scale anomaly. The background electric field polarizes the bulk of the system and generates, via the SEEE~\eq{eq_SEEE}, a wide boundary layer of the redistributed mobile carriers. The accumulated boundary charge screens the background electric field in the form of a power law, 
\beqn
\frac{E(x)}{E(0)} \propto x^{-\nu}\,,
\label{eq_screening_anomaly}
\eeqn
rather then the exponential~\eq{eq_screening_standard}.
The conformal (or scale) screening exponent,
\beqn
\nu = \frac{2 \beta_e}{e c \hbar \varepsilon_0} = \frac{e^2}{6 \pi^2\hbar v_F\varepsilon \varepsilon_0},
\label{eq_nu}
\eeqn
is determined by the beta function related to the renormalization of the electric charge~\eq{eq_beta}. The last relation for the conformal exponent in Eq.~\eq{eq_nu} is given for a single cone of the Dirac semimetal with the bulk permittivity $\varepsilon$. In typical semimetals, the conformal exponent is predicted to be rather large, $\nu \sim 0.1$, which makes its experimental detection feasible via the measurement of the electrostatic charge accumulation or the electrostatic potential along the sample~\cite{CV19}. 

\begin{tcolorbox}
\begin{itemize}
\item{\bf The scale anomaly produces new phenomena at the boundaries of  Dirac semimetals in the conformal limit (zero temperature
and vanishing chemical potential). }
\item{\bf The scale magnetic edge effect  generates a magnetization Hall-like electric current along the boundary.}
\item{\bf The scale electric edge effect  leads to  accumulation of electric charge at the boundary and to experimentally detectable power law electrostatic charge screening. The  exponent is determined by the beta function  of the electric coupling~$\bs e$.}

\end{itemize}
\end{tcolorbox}


\section{Torsion and chiral transport}
\label{sec_torsion}

Torsion is a geometric property of space-time. In the theory of general relativity, space-time is assumed to be torsionless, and its geometry is entirely defined by curvature. This is so because there is seemingly no experimental evidence of torsion in the space-time of our universe. However, there are extensions of general relativity that include torsion, like Einstein-Cartan theory \cite{Hehl76}. Perhaps surprisingly, such theories have acquired special relevance in a completely different area of physics: the geometric description of point defects in crystals \cite{Nelson02} and the coupling of the electrons to the lattice degrees of freedom. 
\begin{figure}[!thb]
\begin{center}
\begin{tabular}{cc}
\includegraphics[scale=0.20,clip=true]{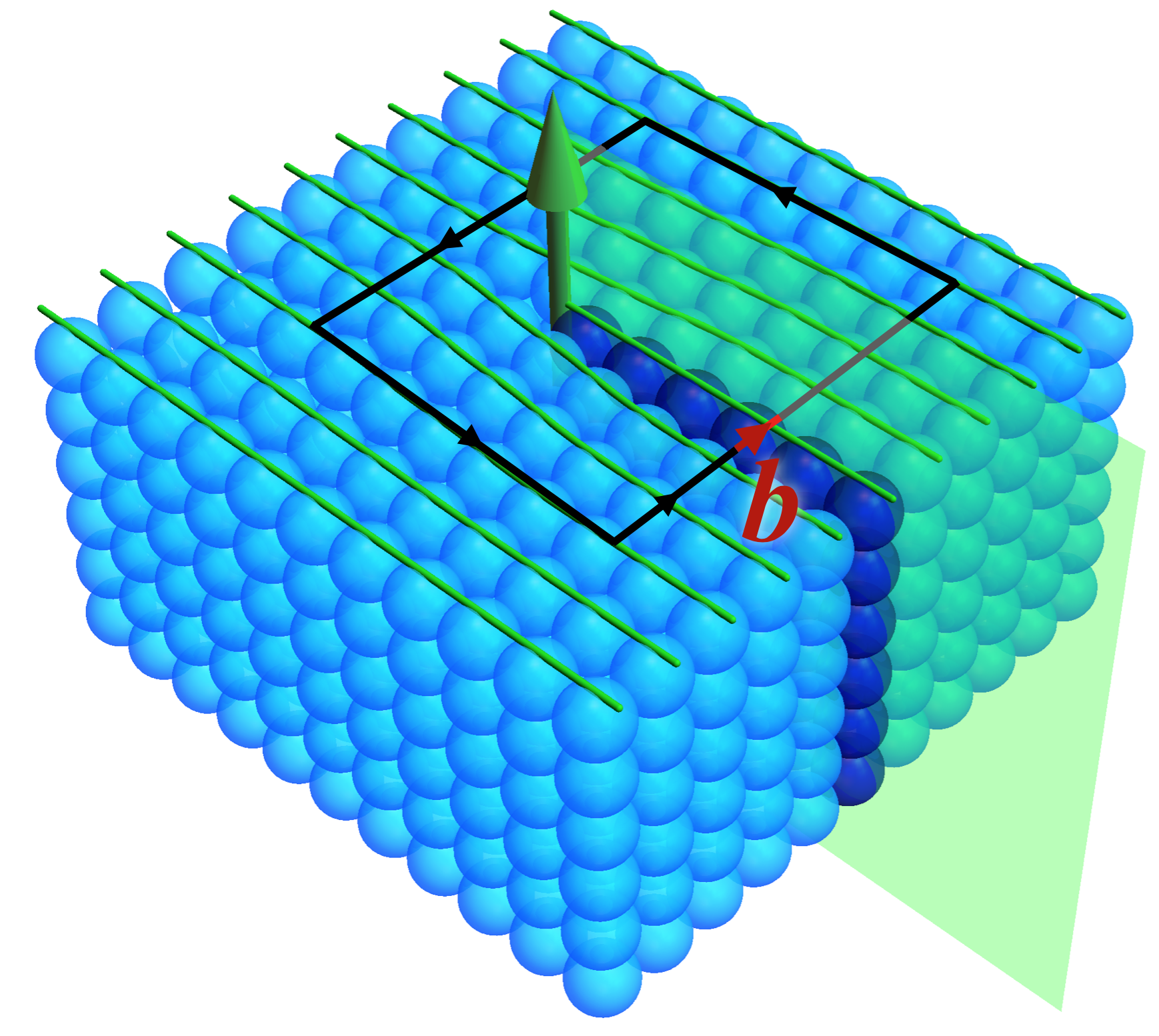} & \hskip 10mm
\includegraphics[scale=0.25,clip=true]{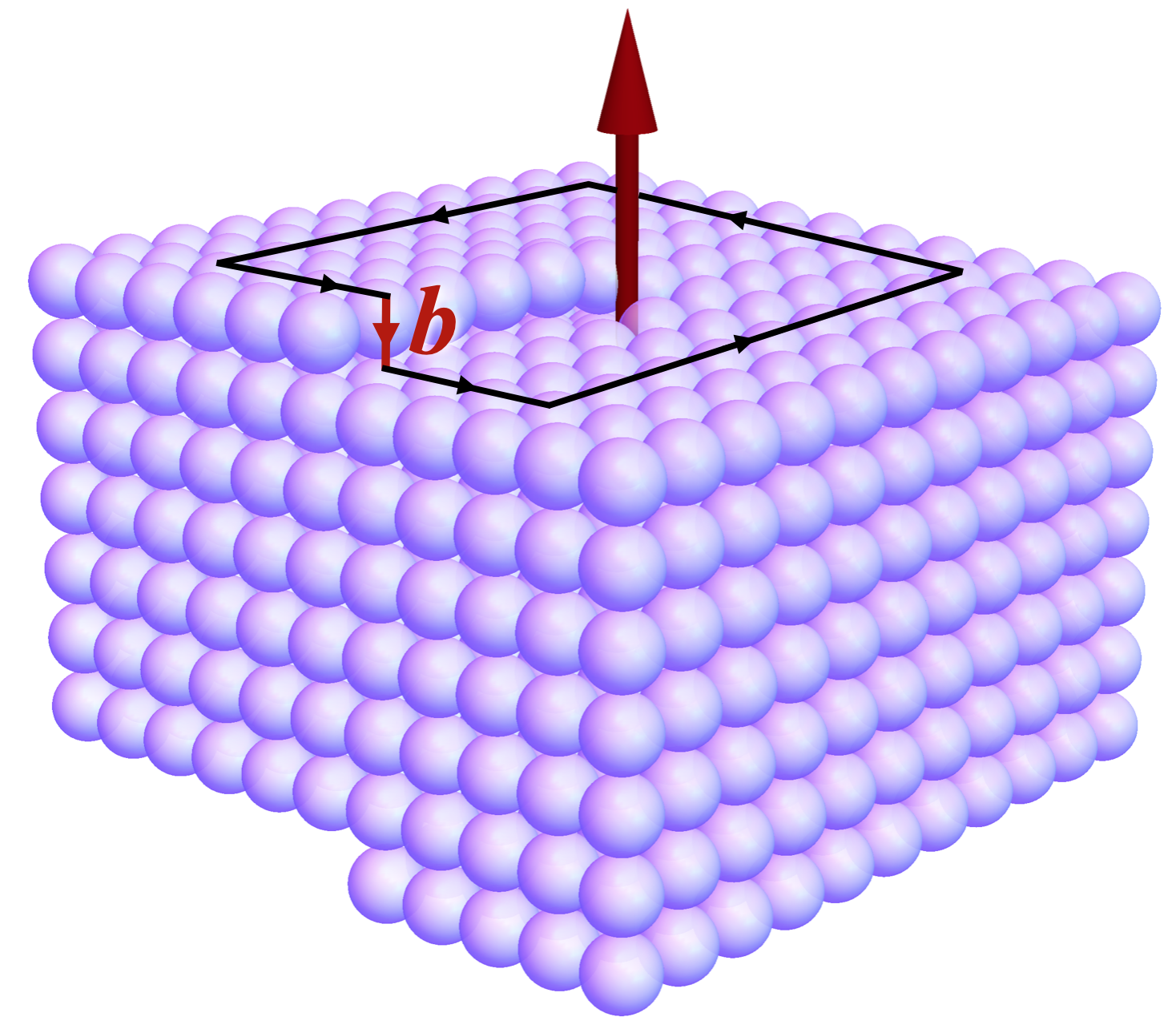}\\[2mm]
(a) & (b) 
\end{tabular}
\end{center}
\caption{(a) Edge dislocation in a square lattice. Closed loops around the dislocation result in a translation along the dislocation line. The Burgers vector is perpendicular to the dislocation plane. (b) Screw dislocation. The Burgers vector is parallel to the dislocation line. }
\label{fig_disloc}
\end{figure}
Torsion  appears naturally associated to  dislocations \cite{KatVol92,Kleinert89}, line defects that play a very important role in the plastic deformation in crystals.  They are classified in two types. The first type are edge dislocations, shown in Fig. \ref{fig_disloc}(a), which can be 
visualized as the addition of an extra half plane of atoms midway through the crystal. As a consequence, discrete closed paths encircling the dislocation line in the lattice do not close. The vector joining the end with the starting point is called Burgers vector ${\vec b}$ which in this case is perpendicular to the dislocation line. The second type are screw dislocations, which are more difficult to visualize. A schematic one is shown in Fig. \ref{fig_disloc}(b). Again discrete closed paths do not close and the Burgers vector in this case is parallel to the dislocation line. The failure to close of discrete paths can also be seen as the non-commuting of parallel displacements paths going from A to B in different directions of the lattice. This operation, for the parallel transport of a vector, is given by the commutator of two covariant derivatives acting on the vector. The result, going back to the formalism of curved spaces in the continuum (see appendix \ref{sec_curved}) is
\begin{equation}
    [\nabla_\mu,\nabla_\nu]V^\rho=R\,^\rho\,_{\sigma\mu\nu}V^\sigma + \theta^\sigma_{\mu\nu}\nabla_\sigma V^\rho.
\end{equation}
While curvature gives a rotation of the vector, torsion (represented by the torsion tensor $\theta^\sigma_{\mu\nu}$) results in the non-closure of the four successive translations by an extra translation. In a more compact wording, torsion describes the failure of a parallelogram spanned by two vectors to form a closed curve.

Since the emergence of materials with relativistic quasiparticles, like Dirac and Weyl semimetals, establishing a consistent treatment of the coupling of torsion to relativistic fermions, with the associated possible influence of torsion in the electric responses, has become an important task in the context of condensed matter \cite{JCV10,Hughes:2012vg,Parrikar:2014usa,FKBB19,HLZ19,Huang:2019haq,Nissinen:2019wmh,Nissinen:2019mkw,N20,HH20,LN20,HHS20,LO20,KZ18,IY19,IQ20}. Moreover the coupling of electronic excitations to the new geometric objects could potentially induce additional anomalous terms in the effective action which may give rise to novel anomaly--induced transport phenomena. In what follows we will give a  different perspective on torsional anomalies as compared to some of the works in the literature, and address the presence (or more precisely absence) of chiral responses to torsion.

\subsection{Nieh-Yan current}
 Recently, there has been an increasing interest in condensed matter in what has been called torsional (or Nieh-Yan) anomaly \cite{O182,NY282,O283,Y88,Y96,CZ97,Parrikar:2014usa,Hughes:2012vg,FKBB19,HLZ19,Huang:2019haq,Nissinen:2019wmh,Nissinen:2019mkw,N20,HH20,LN20,HHS20}. The conservation equation (or Ward identity) for the axial current in the presence of torsion ($\theta^\sigma_{\mu\nu}$) reads
 \begin{equation}
\label{eq:ward_torsion}
     \nabla_\mu J^\mu_{5}-\theta^\lambda_{\lambda\mu} J^\mu_{5}=\mathcal{A}^g_{5},
 \end{equation}
where the functional $\mathcal{A}^g_{5}$ accounts for the presence of gauge anomalies. It has been argued that $\mathcal{A}^g_5$ receives a torsion contribution, beyond the axial and mixed axial-gravitational anomalies, given by the so called Nieh-Yan term \cite{Nieh:2007zz}, resulting in
 \begin{equation}
     \nabla_\mu J^\mu_5-\theta^\lambda_{\lambda\mu} J^\mu_5=c_T\epsilon^{\mu\nu\rho\lambda}(\eta_{ab}\theta^a_{\mu\nu}\theta^b_{\rho\lambda} - R_{a b \mu \nu} e^a_\rho e^b_\lambda).\label{eq:Nieh-Yan_anomaly_1}
 \end{equation}
For the right-hand side of Eq. (\ref{eq:Nieh-Yan_anomaly_1}), the Nieh-Yan term, to have the correct dimension, the coefficient $c_T$ must have dimensions of inverse square length $c_T\sim l^{-2}$. In vacuum, the only candidate scale is the cut-off $\Lambda$, so that the value of $c_T$ must depend on regularization. This means that the contribution in Eq. (\ref{eq:Nieh-Yan_anomaly_1}) can be removed by a suitable local term in the background fields, usually called counterterm in the language of quantum field theory\footnote{Different regularizations in quantum field theory differ by local counterterms in the background fields.}. 

That the Nieh-Yan term can be removed is further reinforced by noticing that it can be written as a total derivative of well defined fields $e^a_\mu$ and $\theta^a_{\mu\nu}$: $\nabla_\mu (\epsilon^{\mu\nu\rho\lambda}\eta_{ab} e^a_\nu\theta^b_{\rho\lambda})$. The vielbein and torsion tensor are covariant objects under diffeomorphisms and local Lorentz rotations, see Appendix \ref{sec_curved}, and thus the product inside the derivative is well defined. This is unlike the second Chern class $\epsilon^{\mu\nu\rho\lambda} F_{\mu\nu} F_{\rho\lambda}$ or the Pontryagin class $\epsilon^{\mu\nu\rho\lambda} R^a_{\,b\,\mu\nu} R^b_{\,a\,\rho\lambda}$, which determine the chiral and axial-gravitational anomalies, respectively.
The latter two cannot be written as derivatives of well defined tensors. Rather, the corresponding Chern-Simons terms explicitly depend on the local value of $A_\mu$ or $\omega_\mu\,^a\,_b$, which are not tensors but transform as connections. Being a total derivative of well defined fields, the Nieh-Yan contribution to the non-conservation of the axial current in Eq. (\ref{eq:Nieh-Yan_anomaly_1}) must therefore come from a local term in the effective action of the form
\begin{equation}
   S_{eff}^{NY}= c_T\int d^4x\, e\, \epsilon^{\mu\nu\rho\lambda} \eta_{ab}A^5_\mu e^a_\nu \theta^b_{\rho\lambda}.
\end{equation}

The above arguments show that the torsional "anomaly" is not an anomaly in the formal quantum field theory sense, since the corresponding Nieh-Yan term can be removed by choosing a suitably regularization. It is more precise to understand it as a local, regularization dependent, torsion contribution to the axial current
\begin{equation}
\label{eq:Nieh-Yan_current}
    J_5^\mu=c_T\epsilon^{\mu\nu\rho\lambda} \eta_{ab} e^a_\nu \theta^b_{\rho\lambda}.
\end{equation}
This still leaves the problem of the cut-off scale ($\Lambda$) dependence of the coefficient $c_T$, which therefore seems highly ambiguous from the standpoint of relativistic quantum field theory. However in a condensed matter context in which there is a natural cutoff, such currents could appear, as it has been first argued in \cite{Hughes:2012vg,Parrikar:2014usa}. This argument has been subsequently exploited in \cite{FKBB19,HLZ19,Huang:2019haq,Nissinen:2019wmh,Nissinen:2019mkw,N20,HH20,LN20,HHS20}, where the Nieh-Yan term has been discussed in the context of Dirac/Weyl semimetals and superfluids.

\subsection{Absence of universal chiral torsional transport}

We have discussed in depth how the chiral anomaly is responsible for chiral universal transport, in the form of the CME and CVE, as a response to magnetic and vorticity fields. A pertinent question then is wether there exists such kind of chiral transport as a response to geometric torsion. We have argued that the possible existence of an additional Nieh-Yan contribution to the conservation equation of the axial current, is better understood directly as a torsional response in the current (Eq. \eqref{eq:Nieh-Yan_current}). The coefficient $c_T$ of such a response must have units of inverse squared length, and therefore must depend on regularization through the only such possible scale in vacuum, the cut-off. However, the presence of matter could source two additional IR scales, temperature and chemical potential, that could do the trick. It has been argued that, in fact, finite temperature effects could provide a universal coefficient $c_T\propto T^2$ \cite{Huang:2019haq,Nissinen:2019wmh,Nissinen:2019mkw,LO20}.

One can attempt to do a systematic study, as has exactly been done recently in \cite{FL20}, and write down all possible dissipationless chiral transport terms, as a response to the torsion tensor (to linear order), derive the corresponding Kubo formulas, and compute them. This basically searches for parity even coefficients, analogues of the CME and CVE for torsion. To do this, we follow appendix \ref{sec_velocities} and take into account that an equilibrium state of a quantum statistical system is defined not only by the temperature and the chemical potential for the electric charge, but also by the chemical potentials for the rest of conserved charges. Therefore the state is characterized by $T$, $\mu$, and the four components of the velocity $u^a$, which are the chemical potentials for the momenta. $u^a$ is a normalized four velocity, $u_au^a = 1$, so only three components are independent corresponding to the three momenta. The approach in \cite{FL20} was to write down all possible parity even terms in the (axial) current at lowest order in torsion and vorticity $\Omega^\mu=\frac{1}{2}\epsilon^{\mu\nu\rho\lambda} u_\nu\partial_\rho u_\lambda$, with $u_\mu=u_ae^a_\mu$
\begin{equation}
    J^\mu_{(5)}=\rho_{(5)}u^\mu+\sigma^\Omega_{(5)}\Omega^\mu+\epsilon^{\mu\nu\rho\lambda}\Big(\frac{c_{T,(5)}^{||}}{2}P^{||}_{ab}e^a_\mu\theta^b_{\nu\rho}+\frac{c_{T,(5)}^{\perp}}{2}P^{\perp}_{ab}e^a_\mu\theta^b_{\nu\rho}\Big).
    \label{eq:NYanomal}
\end{equation}
$P^\parallel_{ab} = u_a u_b$ and
$P^\perp_{ab} = \eta_{ab} - u_a u_b$ are the projectors parallel and orthogonal to the velocity field, and $\rho_{(5)}$ is the (axial) charge density. The second term is the CVE, and the two last terms represent the response to torsion. In \cite{FL20} it was found that
\begin{equation}
    c_{T,(5)}^{||}=c_{T,(5)}^\perp=0.
\end{equation}
This result means that there is no analogue of the CME and CVE for torsion. In other words, there are no universal chiral torsional responses (at least at lowest order), which may be related to the fact that there is no torsional quantum field theory anomaly as discussed earlier, and as such no anomaly related torsion transport coefficients. This might seem at odds with some recent works \cite{KZ18,IY19,IQ20}, but the apparent contradiction is solved by realizing that those results can be understood just as a response to vorticity in a general background, through the CVE
\begin{equation}
    J^\mu_{(5)}=\sigma^\Omega_{(5)}\Omega^\mu=\frac{\sigma^\Omega_{(5)}}{2}\epsilon^{\mu\nu\rho\lambda}u_au_b e^a_\nu \partial_\rho e^b_\lambda,
\end{equation}
with $u^a=(1,0,0,0)$ in the rest frame of the fluid. By taking small perturbations around flat space-time $e^a_\mu=\delta^a_\mu+h^a_\mu$, the vorticity can be written in terms of the gravitomagnetic field (see Appendix \ref{sec_GEM}) $2\vec{\Omega}=\vec{ B}_g=\vec{\nabla}\times\vec{ A}^g$, with $A^g_i=h^t_i$, and the CVE can be non-zero even if the fluid is at rest
\begin{equation}
    \vec{ J}_{(5)}=\sigma^\Omega_{(5)}\vec{\Omega}=\frac{\sigma^\Omega_{(5)}}{2}\vec{\nabla}\times\vec{ A}^g.
\end{equation}
Note that there is no torsion here. One should not confuse the gravitomagnetic field, arising as a consequence of having vorticity in a general curved background, with the torsion tensor.

\begin{tcolorbox}
\begin{itemize}
\item{\bf The torsional ``anomaly" is not a quantum anomaly: it can be cancelled by a suitable counterterm.}
\item {\bf There is no chiral transport as a response to geometric torsion.}
\item{\bf The chiral vortical effect can explain currents often interpreted as currents resulting from torsion.}
\end{itemize}
\end{tcolorbox}


\section{Experimental advances}
\label{sec_exp}

Throughout this review we have gathered how different conductivity tensors have contributions from various anomalies. Several groups have attempted to measure these contributions experimentally in condensed matter systems that host Weyl semimetals~\cite{AMV18}.

\begin{figure}[!thb]
\includegraphics[width=\textwidth]{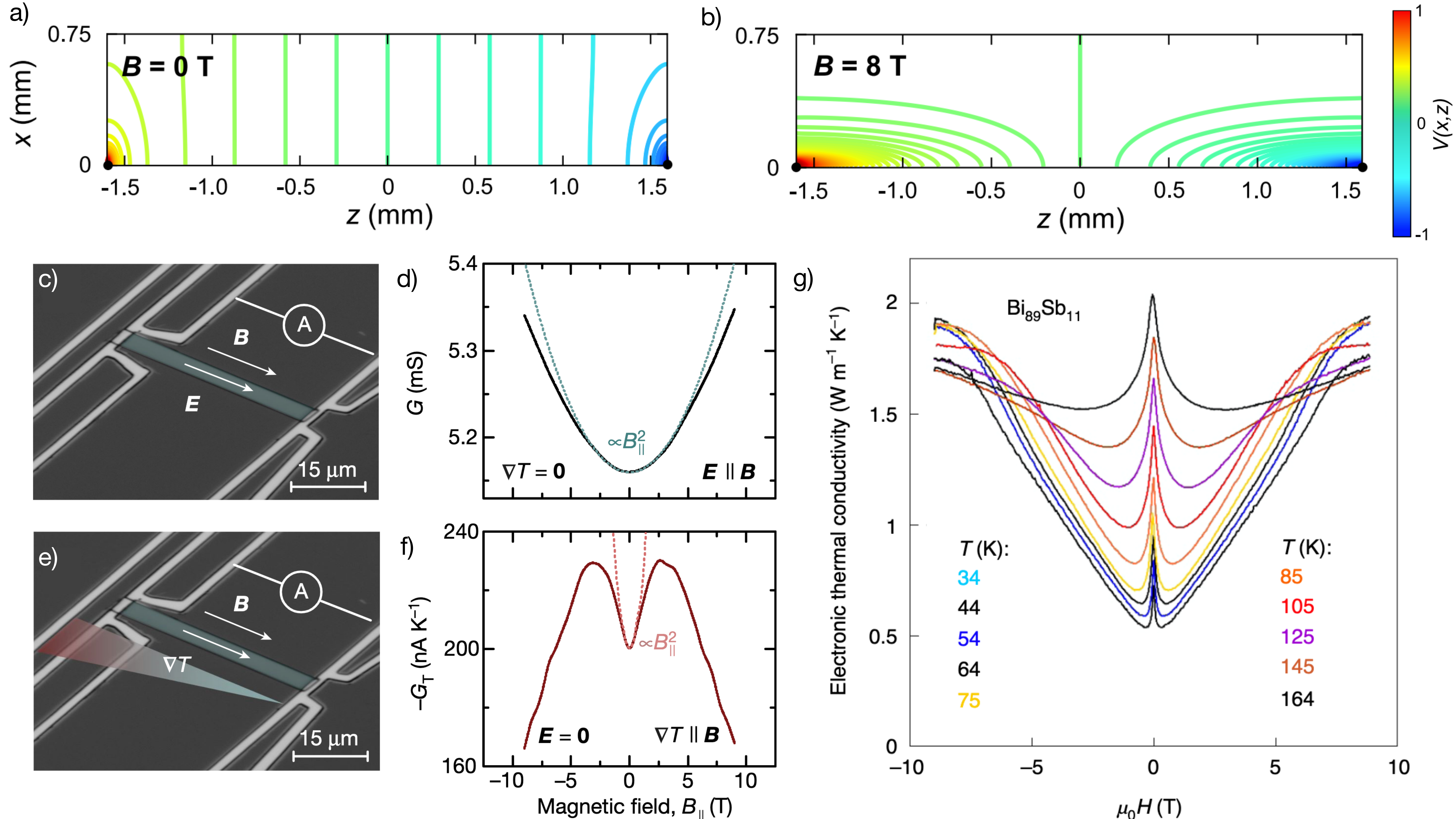}
\caption{a) and b) show calculated equipotential lines in the presence and absence of a magnetic field, respectively. The bending of equipotential lines results in a spurious measurement of intrinsic magneto-conductance known as current jetting (adapted from \cite{Ajeesh:vq}). c) Experimental setup to measure the chiral anomaly enhanced magnetotransport in NbP (adapted from \cite{Gooth2017}). d) The magneto-conductivity is quadratic with field at low magnetic fields and linear at high magnetic fields. e) Experimental setup to measure the mixed-axial gravitational anomaly enhanced magneto-thermal transport in NbP (adapted from \cite{Gooth2017})). f) The anomalous contribution to magneto-thermal conductivity is quadratic with field at low magnetic fields and vanishes at high magnetic fields (adapted from \cite{Gooth2017}). g) Anomalous electronic contribution to the thermal conductivity measured in Ref.~\cite{Heremans2019}.}
\label{fig_experiments}
\end{figure}

In magneto-electric transport it is the longitudinal conductivity that is enhanced by the chiral anomaly~\cite{Son:2012bg}. This phenomenon is referred to as a positive magneto-conductance or, equivalently, a negative magnetoresistance due to the chiral anomaly. The anomalous contribution to conductance in lattice models can be traced back to work by Nielsen and Ninomiya~\cite{Nielsen:1983rb}, which considered a collinear electric and magnetic field, activating the chiral anomaly. Through our arguments in \eqref{sec_magtrans}, the effect of these is to pump right movers into left movers (or vice versa), creating a chiral chemical potential $\mu_5$, which then drives a current through the chiral magnetic effect \eqref{eq:cmecveJ}. At small magnetic fields this results in a $B^2$ dependence of the magneto-conductivity, as dictated by \eqref{eq:lowBLij}, while at high magnetic fields the dependence on $B$ is linear, as dictated by \eqref{eq:highBLij}. 

Numerous theoretical and experimental works have observed negative magnetoresistence consistent with that predicted by the chiral anomaly~\cite{AMV18}. In some materials, typically with low mobility, the intrinsic contribution can be isolated~\cite{Ong2021}.
However, spurious effects complicate the interpretation of an enhanced conductivity based solely on the chiral anomaly. Among them, current jetting is specifically problematic (see Figs.~\ref{fig_experiments}a) and b)).
Current jetting occurs when electrons enter a material with an anisotropic electrical conductivity $\sigma$ and high mobilities through a point contact, bending the otherwise equidistant equipotential lines~\cite{Ajeesh:vq,Reis2016,Ong2021}. Because of this bending, the voltage drop in the sample becomes inhomogeneous, and the value measured becomes sensitive on where the contacts are placed. If sizable, current jetting can jeopardize measuring intrinsic effects, such as the anomalous enhancement of transport. Current jetting is enhanced in a magnetic field because it increases the anisotropies between the different components of the conductivity tensor, especially in high mobility samples. While the strength of this effect can be tested~\cite{Ong2021}, and experimental precautions can be taken against this detrimental effect, e.g. choosing materials with reduced carrier mobility, it severely reduces the pool of materials to test anomalous behaviour.

In Sec. \eqref{sec_magtrans} we saw how anomalous transport enhancement can occur also in the longitudinal magneto-thermal, and magneto-thermoelectric transport due to the mixed axial-gravitational anomaly (see \eqref{eq:lowBLij} and \eqref{eq:highBLij})~\cite{Landsteiner:2011cp}. The first experimental observation of the enhancement of magneto-thermoelectric transport in a Weyl semimetal was reported by Gooth et al in NbP samples~\cite{Gooth2017}.
These samples displayed a longitudinal magneto-conductance ($\sigma_{ii}(B)$) that increased with magnetic field $B$, which was maximal when $B$ was parallel to the applied electric field.
They observed that $\sigma_{ii}(B)\propto B^2$ at low magnetic fields and $\sigma_{ii}(B)\propto |B|$ at high-magnetic fields, in accordance to what is expected from the chiral anomaly (see \eqref{eq:lowBLij} and \eqref{eq:highBLij} and Figs.~\ref{fig_experiments}(c) and (d)). Additionally, they observed that the longitudinal magneto-thermoelectric conductance  increased as $B^2$ at low magnetic fields, and was then suppressed at high fields, again consistent with \eqref{eq:lowBLij} and \eqref{eq:highBLij} (see Figs.~\ref{fig_experiments}e) and f)). These observations, together with the fact that the longitudinal magnetothermal conductance was maximal when $B$ was parallel to the applied thermal gradient, motivated the authors to conclude that their observations were consistent with the signatures of the mixed-axial gravitational anomaly (see section  \ref{sec_magtrans}). 

The mixed-axial gravitational anomaly also predicts the enhancement of the longitudinal magneto-thermal conductivity $\kappa_{ii}(B)$ through \eqref{eq:lowBLij} and \eqref{eq:highBLij}.  Such an  enhancement is consistent with observations in GdPtBi~\cite{Schindler2020} and a Weyl state induced in the alloy Bi$_{1-x}$Sb$_{x}$~\cite{Heremans2019} (see Fig.~\ref{fig_experiments}g)). The benefit of measuring $\kappa_{ii}(B)$ as a probe of anomalous thermal transport is that it does not suffer from the extrinsic effect known as current jetting. The reason is that this measurement does not involve electrical contacts. The electronic thermal transport is achieved through thermalization of thermal contacts mainly by scattering with phonons, which does not occur at a point contact. The measurement of $\kappa_{ii}$ also has the benefit of circumventing other effects, like contact misalignment~\cite{Heremans2019}.

The magnetic field dependence of the different transport coefficients can be derived semiclassically~\cite{Lundgren:2014gy} without an explicit connection to anomalies.
However, Kim and Stone observed in \cite{Stone:2018zel} that the derivations that do take into account the dependence on the different anomalous coefficients ($b_a$ and $d_{abc}$, see \eqref{eq:acoeffgauge}) explicitly offer a chance to understand the physical origin of these anomalies. These authors proposed a physical picture that explicitly connected the gravitational anomaly to heat flow and provided some physical intuition, which we discussed briefly in \ref{sec_transport} and in detail in Appendix \ref{app_bh}.

\begin{tcolorbox}
\begin{itemize}
\item{\bf The chiral anomaly coefficient has been measured through the longitudinal magneto-conductivity, even though some materials suffer from extrinsic effects, notably current jetting.}
\item {\bf Longitudinal magneto-thermal transport does not suffer from current jetting and has been used to probe the mixed-axial gravitational anomaly coefficient.}
\end{itemize}
\end{tcolorbox}


\section{Conclusions}

The seminal work of Luttinger \cite{Luttinger1964} established a deep relation between geometry and thermal transport phenomena of condensed matter systems near equilibrium. 
While this relation allows to understand how Kubo formulas for thermal transport can be derived it is a classical relation at its heart. On the other hand many realms of modern condensed matter physics have to deal with intrinsically quantum mechanical aspects of the  collective behaviour of electronic quasiparticles in crystals. This is especially apparent in the field of Weyl and Dirac matter in which the intrinsically quantum mechanical properties  play a decisive role. This has lead to the extraordinarily interesting situation that concepts originally home to high energy physics such as anomalies of quantum field theories are nowadays at the center of research in condensed matter physics. In view of this development it is necessary  to uplift Luttinger's original ideas into this new quantum era of condensed matter physics. In this review we collected the basic ideas behind Luttinger's original approach but we also have reviewed some of its substantial generalizations relevant this new quantum era. In particular we highlighted the role played by various anomalies in the theory of transport phenomena and its application to Weyl and Dirac matter.

An intriguing aspect of anomaly induced transport is that at first it seems unrelated to the formalism put forward by Luttinger. Both the dependencies of the conductivities \eqref{eq:cmeJ}-\eqref{eq:cveT} temperature and chemical potential can be derived from a microscopic perspective without any reference to anomalies. They simply arise from the standard Sommerfeld integrals involving the Fermi-Dirac distribution. 
However, as first noticed in \cite{Landsteiner:2011cp} the general formulas immediately reveal this relations because they are directly proportional to the anomaly coefficients $d_{abc}$ and $b_a$. The approach in Ref. \cite{Stone:2018zel} relies on a rather interesting marriage of Luttinger's ideas with Hawking's ideas on thermal radiation of black holes. While the original Luttinger formalism is based on classical ideas about gravity it seems in hindsight appropriate that quantum transport, such as the chiral magnetic and chiral vortical effects, calls for the inclusion of ideas from the realm of quantum gravity as is Hawking radiation.
The so far anecdotal similarity between the Sommerfeld integrals that determine quantum transport and the anomaly generating functional highlighted in Ref.~\cite{Loganayagam:2012pz,Stone:2018zel} act as a motivation to develop further the link between chiral anomalies and thermal transport. Another aspect is the role played by the discrete (or global) anomaly pointed out in \cite{Golkar:2015oxw}. It correctly predicts 
the Chiral Vortical coefficient even for exotic theories such as relativistic spin $3/2$ particles. From these considerations it is clear that there is still room for deepened theoretical investigation (see e.g. the recent works \cite{Prokhorov:2020npf, Prokhorov:2021bbv}). For applications in condensed matter physics it will also be important to understand if various forms of ``new" fermions \cite{newfermions} give rise to new forms of anomalies (see \cite{Yee:2019rot} for a partial positive result).

From the quantum field theoretical point of view it is highly surprising that such subtle quantum properties as anomalies can be tested for by the rather mundane operations of applying electric and magnetic fields or subjecting a chunk of matter to a temperature gradient. This is why the developments in condensed matter physics related to Weyl and Dirac metals are so interesting from a fundamental physics perspective. 
What makes it even more exciting is that anomalies give rise to qualitatively new transport effects which actually can be measured in desktop laboratory experiments. Some of these effects might even lead to new applications of technological relevance \cite{Kharzeev:2019ceh}.

In this review we have also discussed the scale anomaly, a new player in the anomaly induced transport phenomena in condensed matter. Scale anomalies, less known than their chiral and gravitational counterparts even in the high-energy context, are realized in condensed matter systems in the limit of  zero-temperature and at the neutrality point of Dirac semimetals.
We reviewed how they lead to a number of new transport and electrostatic quantum effects both in the bulk and at the boundary of semimetals. Contrary to the anomaly coefficients associated with the axial and the mixed axial-gravitational anomalies, the scale anomaly leads to non-quantized anomaly coefficients which are renormalized at all orders of perturbation theory in the electric coupling $e$. The non-quantization of the anomalous conductivities is a distinct feature of the scale anomaly which may help in its experimental detection in Dirac semimetals. Condensed matter experiments can also lead to an alternative observation of the running coupling constants of fundamental quantum field theories. 

We have also reviewed the role of torsion in quantum field theory anomalies, showing that there is no anomaly due to torsion in the strict quantum field theory sense. We further showed that there is no chiral transport as a response to geometric torsion, and that the chiral vortical effect can explain currents often interpreted as currents resulting from torsion.

A recurrent question in the condensed matter community is whether we need to know about anomalies in order to explain the experimental behavior, or whether they are rather a fancy, sophisticated formalism that does not add any new insight into the subject. Similarly to how Newton's theory of gravitation added an additional layer of understanding to the practical calculations of Ptolemy an Kepler, quantum-field theory anomalies seem to reveal a unified perspective to transport phenomena that can be calculated using other, at times more practical techniques.
We view quantum anomalies as the overarching theme of transport phenomena, unifying our understanding of transport in different physical disciplines. In this way, we may conclude with the same sentence that opened our review: {\it In fact, if the gravitational field didn't exist, one could invent one for the purposes of this paper.}
 
\section{Acknowledgements}
We thank B. Bermond, D. Carpentier, A. Cortijo, S. Galeski, J. Gooth, and D. Kharzeev for fruitful discussions and collaboration on related topics. This work  has been supported by the  and Spanish Ministry (MICINN) Grant No. PGC2018-099199-B-I00, A.~G.~G. acknowledges financial support by the ANR under the grant ANR-18-CE30-0001-01~(TOPODRIVE) and the European Union Horizon 2020 research and innovation programme under grant agreement No. 829044~(SCHINES). K.~L.~ acknowledges support from  Centro de Excelencia Severo Ochoa SEV-2016-0597, and by the grant PGC2018-095976-B-C21 from MCIU/AEI/FEDER, EU. M.N.C. is partially supported by Grant No. 0657-2020-0015 of the Ministry of Science and Higher Education of Russia. Y.F. acknowledges financial support through the Programa de Atraccion de Talento de la Comunidad de Madrid, Grant No. 2018-T2/IND-11088.


\appendix

\section{Velocities as Lagrange multipliers}
\label{sec_velocities}

The purpose of this section is to briefly motivate the appearance of the velocities as Lagrange multipliers conjugate to
momentum. Thermodynamics can be derived from an entropy maximization principle. To maximise entropy means that all information that can be erased is indeed absent from the physical system under consideration. Then one has to ask if there is some type of information that can not be erased. Such information does exists and it is stored in the values of the conserved charges such as energy, momentum or $U(1)$ charges. Generically one is lead then to
a description of a physical system that contains no information except for the values of the conserved charges. This is 
achieved by maximizing the entropy under the constraints that the values of conserved charges  are preserved \cite{Jaynes:1957zza}. If $p_n$ is the probability with which a certain microstate $n$ is occupied under the constraints that the $p_n$ sum to one and energy and other conserved charges take specified values we define the entropy by 
\begin{align}
S =& - \sum_n p_n \log(p_n) + \nonumber\\
&+\lambda ( \sum_n p_n -1) - \beta (\sum p_n E_n - E) - \gamma_i (\sum_n p_n P^i_n - P^i) + \nu (\sum_n p_n q_n - Q)\,.
\end{align}
Here the first line is the definition of entropy and in the second line the constraints are collected. 
The index $i$ enumerates the spatial directions. 
The most basic constraint is the conservation of probability, then conservation of energy $E$, momenta $P^i$ and eventual $U(1)$ charge $Q$. For simplicity we consider only a single such charge. We assume a basis of states numbered by $n$. A state is characterized by its energy $E_n$, momenta $P^i_n$ and charge $q_n$. We maximize the entropy with respect to the probabilities $p_n$ and the Lagrange multipliers $\lambda$, $\beta$, $\gamma_i$ and $\nu$. The solution is
\begin{equation}\label{eq:fugacities}
p_n = \exp(\lambda-1-\beta E_n - \gamma_i P^i_n + \nu q_n)\,.
\end{equation}
The probability constraint $\sum_n p_n =1$ fixes the value of $\lambda$ through the equation
\begin{equation}\label{eq:partitionfunction}
Z = \sum_n e^{-\beta E_n - \gamma_i P^i_n + \nu q_n} = e^{1-\lambda}\,.
\end{equation}
The other values of the Lagrange multipliers are fixed by 
\begin{align}\label{eq:ensembleE}
E &= -\frac{\partial \log(Z)}{\partial \beta} \,,\\\label{eq:ensembleP}
P^i &= -\frac{\partial \log(Z)}{\partial \gamma_i} Z \,,\\ \label{eq:ensembleQ}
Q &= \frac{\partial \log(Z)}{\partial \nu} \,. 
\end{align}
Now we note that $\beta$ has a dimension inverse to energy. We introduce a parameter $T$ and a unit $k_B$ such that $k_B T$ has dimension of energy as well and set 
\begin{equation}
\beta = 1/(k_B T)\,.
\end{equation}
Thus $T$ is interpreted as temperature and the unit $k_B$ converts temperature to energy and is the Boltzmann constant. Similarly we set 
\begin{equation}
\gamma_i = v_i/(k_B T)
\end{equation}
and note that $v_i P^i$ 
has dimension of energy and thus $v_i$ is a velocity. If the system has Lorentz invariance it is advantageous to assemble the 
Lagrange multiplier for energy and momentum into a covariant vector by writing 
\begin{equation}
 (\beta, \gamma_i) = \beta_a  = u_a/(k_B T)   
 \label{eq:frame}
\end{equation}
 where $a$ runs over time and space dimensions. Covariance demands $u_a$ to transform as a four-vector. 
 We normalize $u_a u^a = 1$ such that in a rest frame in which $\gamma_i=0$ we obtain the usual thermal ensemble. Thus $u_a$ is naturally a time-like unit four vector, a four velocity. 
 At last we also introduce the chemical potential by writing $\nu = \mu/(k_B T)$
such that $\mu Q$ has dimension of energy as well. Note that the equations (\ref{eq:ensembleE})-(\ref{eq:ensembleQ}) fix $T$, $u_a$ and $\mu$ 
in terms of energy, momentum and charge.  The thermodynamic
free energy is now defined by
\begin{equation}
    F = - k_B T \log(Z)\,.
\end{equation}
If the system breaks translation invariance then momentum is not conserved and spatial velocities
can not be introduced in that way. In that case only energy and charge are conserved and can be used to define a thermal ensemble. 

In quantum theory instead of the Boltzmann weights $w_n=\exp[-(u_a P_n^a - \mu q_n)/(k_B T)]$ we can use the statistical operator
\begin{equation}
\rho = e^{-(u_a \hat{P}^a -\mu \hat{Q})/(k_B T)}\,,
\end{equation}
where operators are denoted by a hat and $\hat P^0=\hat H$ is the Hamiltonian. Now we can define the partition function as
\begin{equation}
Z = \mathrm{tr} (\rho)\,.
\end{equation}
As is well-known the trace can be represented by a Euclidean path integral with period boundary conditions around the Euclidean time direction. The  length of this temporal circle is $\beta=\frac{1}{k_B T}$. 

Hydrodynamics relies on the assumption of local thermal equilibrium. The microscopic dynamics is 
supposed to equilibrate the microscopic degrees 
of freedom on a short time scale and the late time long wave-length behavior can be described by
the effective dynamics of the conserved charge densities. 
The thermodynamic quantities
$T$, $u_a$ and $\mu$ are promoted to local functions
of time and space. Constitutive relations are expressions for the conserved currents in terms of these local thermodynamic variables and can be organized in the manner of effective field theory in a derivative expansion \cite{Kovtun2012}. To zero-th order we encounter
ideal hydrodynamics whereas to first order in derivatives transport coefficients such as conductivities and viscosities enter.
As the possibly simplest example consider (the spatial components) of a conserved $U(1)$ current to zeroth and first order in derivatives
\begin{equation}
{\vec J} = n \vec v - \sigma  \vec\nabla\mu \,.
\label{eq:j_1st_order}
\end{equation}
The first term is convection and the second term is the Fick's law of diffusion. The expansion coefficients $n$ and $\sigma$ have physical interpretation as charge density and conductivity.


\section{Matter fields in a curved background}
\label{sec_curved}

The interaction of curved space with standard matter fields is at the heart of this review. Although we do not have
yet a quantum theory for gravity, a well established formalism exists for describing quantum field theory in a curved space 
background \cite{Birrell82,Weinberg72}. The formalism, based on differential geometry, 
substitutes any given magnitude transforming as a tensor in the flat space by the
corresponding magnitude transforming as a tensor under general transformations in the curved manifold. 
The main tool to ease the task is the definition of a tangent space at each point of the (assumed smooth) manifold
where standard flat geometry applies, and a prescription to move from one tangent space to another involving  the 
definition of appropriate connections. 

\subsection{Curved spaces without torsion}
Scalar  fields couple to the background geometry through the metric tensor $g_ {\mu\nu}(x)$ only. Their partial derivatives transform as vectors and scalar quantities are constructed easily.

Vector (tensor fields  in general but we will refer mostly to vector) fields need additional geometric objects. Since the partial derivative of a vector does not transform as a tensor, we need to define a covariant derivative. This  is a generalization of the concept of  parallel
transport of vectors in a curved geometry where the usual derivative  must be accompanied by a  connection which rectifies the direction of the 
vector. In general we  define
\begin{equation}
\nabla_\mu V^\rho=\partial_\mu V^\rho+\Gamma_{\mu\nu}^\rho V^\nu,
\label{eq_covder}
\end{equation}
where the only requirement of the connection $\Gamma_{\mu\nu}^\rho$ (not a tensor) is that $\nabla_\mu V^\rho$ transforms as a tensor under general coordinate transformations.    
The most commonly used  affine (Levy-Civita) connection fulfills two properties: the metric is covariantly constant:  ($\nabla g=0$) and the connection 
is torsion free (in the presence of torsion a general connection can be split as in eq. \eqref{eq:connection2} ). 
Its expression as a function of the metric is given by the Christoffel symbols:
\begin{equation}
\mathring{\Gamma}_{\mu\sigma}^\lambda=\frac{1}{2}g^{\nu\lambda}\left\{\frac{\partial
g_{\sigma\nu}}{\partial x^\mu}+\frac{\partial g_{\mu\nu}}{\partial
x^\sigma}-\frac{\partial g_{\mu\sigma}}{\partial x^\nu}\right\}.
\label{eq_Christoffel_symbols}
\end{equation}
Note that $\mathring{\Gamma}_{\mu\sigma}^\lambda$ is symmetric in its lower two indices.

To intuitively understand the geometric notion of curvature, we can consider the parallel transport of a vector. Let us transport a vector in one direction, and then in another, and compare the result with transporting it in the opposite order. This operation is given by the commutator of two covariant derivatives acting on the vector. The result is \cite{Weinberg72}
\begin{equation}
    [\nabla_\mu,\nabla_\nu]V^\rho=R\,^\rho\,_{\sigma\mu\nu}V^\sigma.
\end{equation}
where the curvature tensor is given by 
\begin{equation}
R_{\mu\nu\rho}^\sigma=\partial_\nu\Gamma_{\mu\rho}^\sigma- \partial_\mu\Gamma_{\nu\rho}^\sigma+
\Gamma_{\nu\lambda}^\sigma \Gamma_{\mu\rho}^\lambda-
\Gamma_{\mu\lambda}^\sigma \Gamma_{\nu\rho}^\lambda.
\label{eq:Riemann}
\end{equation}
We see that the effect of curvature is basically to rotate the vector, such that parallel transport along the two paths differs by a relative rotation of the vector.

The situation is more complicated for spinors because there are no spinor representations 
in the group of general transformations. A standard way to proceed is  
to introduce the tetrad  formalism 
proposed by Weyl in 1929.  At each point $X$ described in 
arbitrary coordinates $\xi_X^a$ in the manifold,  we introduce a set of orthonormal vectors:  
the vielbein fields (tetrads in (3+1) dimensions)  $e_\mu^a(x)$, 
labeled by $a$ that fixes the transformation between the local
(Latin indices) and the general (Greek indices) coordinates:
\begin{equation}
e_\mu^a(X)\equiv\frac{\partial \xi_X^a(x)}{\partial
x^\mu}{\biggl|}_{x=X}.
\label{eq_fielbein}
\end{equation}
With these we can  derive the  various geometric objects needed to describe the
Dirac physics in the curved space.
The metric tensor of the curved manifold $g_{\mu\nu}(x)$ is
related to the flat, constant  metric $\eta_{ab}$ by the equation
\beq
g_{\mu\nu}(x)=e_\mu^a (x) e_\nu^b (x) \eta_{ab};
\label{eq_g}
\eeq
its determinant, needed to define  scalar  densities, is
given by
\begin{equation}
\sqrt{-g}=[\det(g_{\mu\nu})]^{1/2}\;=\;\det[e_\mu^a(x)]\equiv e.
\label{eq_jacobian}
\end{equation}
The vielbein fields are not unique but are subject to  local Lorentz symmetry transformations
\begin{equation}
    (e')^a_\mu(x) = \Lambda^a\,_b(x) e^b_\mu(x) 
\end{equation}
with $\eta_{bd} = \eta_{ab} \Lambda^a\,_b \Lambda^b\,_d$.
The curved space gamma matrices $\gamma^\mu(x)$ satisfying the
commutation relations
\begin{equation}
\{\gamma^\mu(x)\gamma^\nu(x)\}=2g^{\mu\nu}(x),
\label{curvedgammas}
\end{equation}
are related with the constant, flat space matrices $\gamma^a$ by
\begin{equation}
\gamma^\mu(x)=e^\mu_a(x)\gamma^a.
\label{eq_gammas}
\end{equation}
The remained object needed to complete the analysis is the
spin connection $\Gamma_\mu(x)$ which allows to define the spinor covariant
derivative 
${\mathcal D}_{\mu}=\left[\partial_\mu+\Gamma_\mu\right]$ such that $D_\mu\Psi$
has a standard transformation under a change of coordinates
in both the vector and the spinor indices.
To get the structure of the $\Omega_{\mu}(x)$ matrices it suffices
to consider their transformation properties in flat space. For the
spin one half representation the spin connection  takes the form

\begin{equation}
\Gamma_{\mu}=\frac{1}{4}\omega_{\mu}^{\;
ab}  \sigma_{ab}, 
\label{eq_scon}
\end{equation}
where 
\begin{equation}
  \sigma_{ab} = \frac{i}{2}\left[\gamma_a,\gamma_b\right],
\label{eq:lorentzGen}  
\end{equation}
are the generators of Lorentz rotations for the spinors and
$\omega_{\mu}^{\; ab}$ are the spin connection coefficients:
\begin{equation}
\omega_{\mu\phantom\dagger{b}}^{\phantom\dagger a \phantom\dagger}=e^\lambda_{b} \left(\partial _{\mu}e^a_\lambda +
\Gamma^{\rho}_{\mu\lambda}e^a_\rho\right).
\label{eq_scoef}
\end{equation}
On tangent space vectors $V^a$, the covariant derivative acts as $\nabla_\mu V^a=\partial_\mu V^a+\omega_\mu\,{}^a\,{}_b V^b$. Finally, the free action for Dirac spinors is given by
\begin{equation}
S=i\int d^4 x e \;\bar\psi(x)\gamma^{\mu}(x){\overleftrightarrow{D}}_{\mu}\psi(x),
\label{eq_dircurv}
\end{equation}
where 
\begin{equation}
  \bar\psi(x){\overleftrightarrow{D}}_{\mu}\psi(x)
\equiv \frac{1}{2}  \bar\psi(x){D}_\mu\psi(x)-\frac{1}{2}(D_{\mu}\bar\psi)\psi(x).
\end{equation}

\subsection{ Spaces with torsion}

The extension of the previous concepts to space-times with torsion can be done by introducing the torsion tensor, which is basically the anti-symmetric part of the affine connection
\begin{equation}
\theta^\lambda_{\mu\nu} = \Gamma^\lambda_{\mu\nu} - \Gamma^\lambda_{\nu\mu}\,.
\label{eq:torsion}
\end{equation}
Assuming that the vielbein and metric are covariantly constant $\nabla_\mu e^a_\nu=\nabla_\mu g_{\nu\rho}=0$, the torsion tensor is given by
\begin{equation}\label{eq:deftorsion}
\theta^a_{\mu\nu} =\partial_{\mu} e^a_{\nu} - \partial_{\nu} e^a_{\mu} + \omega_{\mu}\,^a\,_b\, e^b_{\nu} - \omega_{\nu}\,^a\,_b\, e^b_{\mu},  
\end{equation}
If the torsion is not zero, the connection $\Gamma$ can be written as a sum of Levi-Civita connection ($\mathring{\Gamma}$) and the contorsion ($K$)
\begin{equation}
\Gamma^\mu_{\nu\lambda} = \mathring{\Gamma}^\mu_{\nu\lambda} + K^\mu_{\nu\lambda}\,,
\label{eq:connection2}
\end{equation}
where the contorsion is 
\begin{equation}
K_{\lambda\mu\nu}=\frac 1 2 \left( \theta_{\lambda\mu\nu} - \theta_{\mu\nu\lambda} + \theta_{\nu\lambda\mu}\right)\,.
\label{eq:contorsion}
\end{equation}
Only the Levi-Civita connection $\mathring{\Gamma}$ and the spin connection $\omega$ are connections. The vielbein, torsion, and contorsion are covariant objects under diffeomorphisms and local Lorentz rotations. From the action \ref{eq_dircurv} one can derive the Dirac equation in the presence of torsion. It reads
\begin{equation}
i e^\mu_a \gamma^a\left(\nabla_\mu \psi - \frac{1}{2} \theta^\lambda_{\lambda\mu} \psi \right) =0\,.
\end{equation}
with the extra term $\theta^\lambda_{\lambda\mu}$ arising upon integration by parts.
\\

\subsection{ The spin current and the energy-momentum tensor of spinor fields.}

According to the Noether theorem, invariance under space-time translations of a system defined by a field $\Phi$ and a Lagrangian $L$, leads to the definition of the canonical energy-momentum tensor
\begin{equation}
T^{\mu\nu}=L\eta^{\mu\nu}-\frac{\partial L}{\partial (\partial_\mu \Phi)}\partial^\nu \Phi.
\label{canonicalT}
\end{equation}
Generically this tensor is neither symmetric nor gauge invariant. On the other hand, general relativity allows to define 
\begin{equation}
T^{\mu\nu}=2\frac{\delta L}{\delta g_{\mu\nu}},
\label{GRT}
\end{equation}
which is symmetric and gauge covariant. In the case of the spinor field,
he canonical  energy--momentum tensor obtained from eq. \eqref{eq_dircurv} is
\begin{equation}
 {\hat T}^{\mu\nu}=\frac{1}{2}\bar\Psi \gamma^\mu  D^\nu \Psi  ,
\label{eq:stresstensor} 
\end{equation}
an expression that is not symmetric, a standard fact for fields with spin \cite{Kleinert08}.

Since conserved currents are arbitrary up to the addition of a
divergence-less field, we can define 
\begin{equation}
 T^{\mu\nu}=  {\hat T}^{\mu\nu} +\partial_\alpha B^{\alpha\mu\nu},
\end{equation}
with $B^{\alpha\mu\nu}=-B^{\mu\alpha\nu}$. This anti-symmetry  condition ensures the conservation of the modified tensor.  We can also ensure that the new tensor defines the same observables as the old one. The tensor $ {\hat T}^{\mu\nu}$ can be made symmetric by adding the "Belinfante tensor":
\begin{equation}
 B^{\alpha\mu\nu}=\frac{1}{8} \bar\Psi[\gamma^\alpha,\sigma^{\mu\nu}]\Psi,
 \label{eq:Bel}
\end{equation}
what gives 
\begin{equation}
   T^{\mu\nu} = \frac{1}{4}\bar\Psi (\gamma^\mu  D^\nu +\gamma^\nu  D^\mu)\Psi.
   \label{eq:Tsym}
\end{equation}
This is the standard form of the symmetric energy-momentum used throughout this work. As we see, the resulting tensor is a
symmetrization of the canonical tensor in \eqref{eq:stresstensor}.

The Belinfante tensor \eqref{eq:Bel} is related to the spin current as follows:
In relativistic matter with spin there are two in-equivalent definitions of the energy momentum tensor.

1. Considering the tetrads \eqref{eq_fielbein} and the  connection \eqref{eq_scoef} as independent geometric  variables we can write: 
\beq
\delta S=\int d^4 x \vert e\vert (-{\tilde T}^\mu_a\delta e_\mu^a+s^{\mu ab}\delta\omega_{\mu ab}),
\label{eq:stress1}
\eeq
what defines the stress-energy tensor ${\tilde T}^\mu_a$ and the spin current $s^{\mu ab}$.
2. Alternatively one can adopt the Levi-Civita connection of eq. \eqref{eq_Christoffel_symbols} which is defined in terms of the tetrads, and consider the tetrads and the torsion  in eq. \eqref{eq:torsion} as independent geometric variables:
\beq
\delta S=\int d^4 x \vert e\vert(- T^\mu_a\delta e_\mu^a+S^{\mu\nu}_a\delta\theta^a_{\mu\nu}).
\label{eq:stress2}
\eeq
The Belinfante construction corresponds to the  second definition. The spin density is defined by varying the action with respect to torsion:
\beq
S^{\mu}_{ab}=\frac{1}{4}{\bar\Psi}(\gamma^\mu\sigma_{ab}+\sigma_{ab}\gamma^\mu)\Psi,
\label{eq:spincurr}
\eeq
where $\sigma_{ab}$ are given in eq. \eqref{eq:lorentzGen}. 
Since the spin is not a good quantum number,  the spin current is not conserved.  The Noether current 
associated to rotational invariance is the total angular momentum
 $J=S+L$ where the orbital angular momentum current is
\beq
L^{\mu}_{ab}=x_a T^\mu_b+x_b T^\mu_a.
\eeq
As we see, the symmetric (Belinfante) tensor is the canonical Noether tensor with the addition of  the spin density.


\section{Details on conformal anomaly}
\label{sec_curvedform}
\subsection{Conformal anomaly in curved background}

The matter fields experience quantum fluctuations which affect the couplings of the theory and lead to their renormalization. In a curved background, the fluctuations depend on the background geometry of the spacetime, thus affecting not only the couplings but also producing a back-reaction on the curved spacetime. This back-reaction can conveniently be described in the form of an effective quantum action $S_{\mathrm{anom}}$ induced by the fluctuations of the matter fields. 

In this Appendix, we consider the torsion-less spacetime with symmetric Christoffel symbols, $\Gamma^\mu_{\alpha\beta} = \Gamma^\mu_{\beta\alpha}$. In this case, the induced quantum action depends on the vielbein fields~\eq{eq_fielbein} only in the combination of the background metric $g_{\mu\nu}$ given in Eq.~\eq{eq_g}. We limit our discussion to models containing massless fields which possess classical scale invariance. In addition to the curved spacetime, we assume that the system is subjected to a classical electromagnetic background described by the electromagnetic stress tensor, $F_{\mu\nu} = \partial_\mu A_\nu - \partial_\nu A_\mu$.

The induced action provides a gravitational self-interaction of the curved spacetime generated by the quantum fluctuations of the matter fields on the curved background. This interaction contains an anomalous part associated with the scale anomaly and a remaining, regular contribution. The anomalous part of the action is associated with the scale anomaly which renders the expectation value of the trace of the energy-momentum tensor to be nonzero. 
The expectation value may, consequently, be divided into two parts: $\avr{T^{\mu\nu}} = \avr{T^{\mu\nu}}_{\mathrm{anom}} + \avr{T^{\mu\nu}}_{\mathrm{reg}}$ with $\avr{T^\mu_{\ \mu}} = \avr{T^\mu_{\ \mu}}_{\mathrm{anom}} \neq 0$ and $\avr{T^\mu_{\ \mu}}_{\mathrm{reg}} \equiv 0$.

The anomalous part of the induced gravitational action is identified according to Eq.~\eq{eq_energy_momentum_tensor}:
\beqn
\avr{T^\mu_{\ \mu} (x)}_{\mathrm{anom}} = - \frac{2 g_{\alpha\beta}(x)}{\sqrt{-g(x)}}  \frac{\delta S_{\mathrm{anom}}}{\delta g_{\alpha\beta}(x)},
\label{eq_Tmunu_anomalous}
\eeqn 
where the anomalous trace of the energy momentum tensor is given by the following expression~\cite{Deser:1976yx}:
\beqn
\avr{T^\mu_{\ \mu}}_{\mathrm{anom}}
 = - \frac{1}{4} \left[ b \, C_{\mu\nu\alpha\beta} C^{\mu\nu\alpha\beta} + b' \left({\mathcal E} - \frac{2}{3} \Box R \right) + c F_{\mu\nu} F^{\mu\nu}\right].
\label{eq_Tmunu_gravity}
\eeqn
The right-hand side of this formula contains the contributions coming from the background metric (the first two terms) and the electromagnetic fields (the last term). All terms are invariant under the influence of diffeomorphisms and gauge transformations. The coefficients $b$, $b'$ and $c$, which determine the strength of the anomaly~\eq{eq_Tmunu_gravity}, are real-valued numbers which depend on the number and spins of the massless fields. In the standard QED with one species of massless fermion these coefficients are as follows (see, for example, Ref.~\cite{Armillis:2009pq}):
\beqn
b = \frac{1}{320 \pi^2}, 
\qquad 
b' = - \frac{11}{5670 \pi^2},
\qquad 
c = - \frac{e^2}{24 \pi^2}. \quad
\label{eq_anomalous_coefficients}
\eeqn
The coefficient $c$ is determined by the TVV vertex shown in Fig.~\ref{fig_scheme}(a), which, in the language of quantum field theory, couples the graviton to the background electromagnetic field~\cite{Armillis:2009pq}. The coefficients $b$ and $b'$ correspond to the triangular diagram TTT~\cite{Coriano:2017mux} illustrated in Fig.~\ref{fig_scheme}(b). General covariance generates also higher-order, T${}^n$ with $n\geqslant 4$, diagrams which involve multiple insertions of the graviton coupling (for a detailed discussion, see Ref.~\cite{Maglio:2021yaq}). In general quantum field theory, these coefficients depend on the number of scalars,  fermions (including neutrinos), and vector particles~\cite{Capper:1973mv,Capper:1974ed,Capper:1974ic,Christensen:1977jc,Bunch:1978yq,Brown:1976wc,Brown:1977pq}.

The parameters $b$ and $b'$ are exact in one loop, in the sense that a perturbative calculation up to any order gives exactly the same value indicated in Eq.~\eq{eq_anomalous_coefficients}. The parameter $c$, however, is not one-loop exact: every order of perturbation theory contributes to the coefficient $c$. The value given in Eq.~\eq{eq_anomalous_coefficients} corresponds to the leading-order contribution generated by the one-loop QED beta function, $c = - \beta_{\mathrm{QED}}^{\mathrm{1loop}}/(2e)$. 

Notice that the curved spacetime is treated as a classical background which does not involve quantum fluctuations of the metric. The gauge fields involve both classical fields, which explicitly appear in the last term of the anomaly relation~\eq{eq_Tmunu_gravity}, and the quantum fluctuations on top of the classical background, which contribute to the coefficients $b$ and $b'$ in the first two terms in the trace anomaly~\eq{eq_Tmunu_gravity}. Finally, the fermionic fields are of purely quantum origin, whose fluctuations contribute both to the gravitational coefficients $b$ and $b'$ as well as to the electromagnetic coefficient $c$.

As we mentioned, the first two terms in the right-hand side of Eq.~\eq{eq_Tmunu_gravity} are coming purely from the gravity side and contain the seeds of most structures that encode the geometry of the Riemann manifolds. It is therefore instructive to discuss these terms in more detail. In the differential geometry, the Riemann curvature tensor determines the degree to which the spacetime is curved with respect to the flat Minkowski spacetime. In terms of the metric tensor, the Riemann tensor is given in Eq.~\eq{eq:Riemann}.
Physically, the tensor~\eq{eq:Riemann} determines the tidal force which acts on a body that moves along a shortest -- generally, curved -- path between two points in the curved spacetime represented by a Riemann manifold. This path is commonly known as a ``geodesic''. In geometric terms, the tensor~\eq{eq:Riemann} allows to calculate the effect of the background gravitational field on the geometric shape and the occupied volume of a physical body.

The Riemann curvature tensor~\eq{eq:Riemann} may conveniently be divided into three parts following the Ricci decomposition:
\beqn
R_{\mu \nu \alpha \beta} = E_{\mu \nu \alpha \beta} + S_{\mu \nu \alpha \beta} + C_{\mu \nu \alpha \beta},
\label{eq_Ricci_decomposition}
\eeqn
where the constituents, in 3+1 spacetime dimensions, 
\beqn
E_{\mu \nu \alpha \beta} & = & \frac{1}{2} \left( Z_{\mu \beta} g_{\nu \alpha } - Z_{\mu \beta} g_{\mu \alpha } - Z_{\mu \alpha} g_{\nu \beta} + Z_{\nu \alpha} g_{\mu \beta} \right),
\label{eq_E_tensor} \\
S_{\mu \nu \alpha \beta} & = & {\frac{R}{12}} \left(g_{\mu \beta} g_{\nu \alpha} - g_{\mu \alpha} g_{\nu \beta} \right), 
\label{eq_S_tensor}\\
C_{\mu \nu \alpha \beta} & = & R_{\mu \nu \alpha \beta} - S_{\mu \nu \alpha \beta} - E_{\mu \nu \alpha \beta},
\label{eq_C_tensor}
\eeqn
are expressed via the Riemann tensor the Ricci tensor $R_{\mu\nu} = R^{\alpha}_{\ \mu\alpha\nu}$, its traceless part, $Z_{\mu\nu} = R_{\mu\nu} - (R/4) g_{\mu\nu}$ with $Z^\mu_{\ \mu} \equiv 0$, and the scalar (Ricci) curvature $R = R^\mu_{\ \mu}$. The structures $E$, $S$ and $C$ are orthogonal to each other in a sense that $\avr{E,S} = \avr{E,C} = \avr{S,C} = 0$ with $\avr{A,B} = A_{\mu \nu \alpha \beta} B^{\mu \nu \alpha \beta}$.

The first term in the anomalous trace of the energy momentum tensor~\eq{eq_Tmunu_gravity} is expressed via the Weyl tensor~\eq{eq_C_tensor}. Mathematically, the Weyl tensor is a traceless part of the Riemann tensor~\eq{eq:Riemann}: a convolution of any of its indices gives zero (for example, $g^{\mu\nu} C_{\mu \nu \alpha \beta} = 0$, etc). Physically, the Weyl tensor determines how the shape of the body is affected by the tidal force of the gravity as the body moves along a geodesic. The tracelessness of the Weyl tensor implies that it omits a part of the information which is contained in the Riemann tensor. The omitted information is about the change in the volume: it is the Ricci curvature, missed in \eq{eq_C_tensor}, which determines how the volume occupied by a body evolves when the body moves along a geodesic. 

Explicitly, the first term in the anomalous trace~\eq{eq_Tmunu_gravity}, given by the square of the Weyl tensor, reads as follows
\beqn
C^2 \equiv C_{\mu\nu\alpha\beta} C^{\mu\nu\alpha\beta} = R_{\mu\nu\alpha\beta} R^{\mu\nu\alpha\beta} - 2 R_{\mu\nu} R^{\mu\nu} + \frac{R^2}{3}.
\label{eq_Weyl_tensor_squared}
\eeqn
The second term in the anomalous trace~\eq{eq_Tmunu_gravity} involves the Euler (topological) density 
\beqn
{\mathcal E} = {}^*R_{\mu\nu\alpha\beta} {}^*R^{\mu\nu\alpha\beta} \equiv 2 C^2 + \frac{1}{2} \left( S^2 - E^2 \right) \equiv R_{\mu\nu\alpha\beta} R^{\mu\nu\alpha\beta} - 4 R_{\mu\nu} R^{\mu\nu} + R^2,
\label{eq_Euler_density}
\eeqn
where
\beqn
{}^*R_{\mu\nu\alpha\beta} = \frac{1}{2} \epsilon_{\mu\nu\gamma\lambda}R^{\gamma\lambda}_{\phantom{\mu'\nu}\alpha\beta}, 
\eeqn
is the (left) dual of the Riemann tensor~\eq{eq:Riemann}. The geometrical significance of the Euler (topological) density~\eq{eq_Euler_density} comes from the fact that its integral over a closed Riemann manifold $M$ determines its Euler (topological) characteristic,
\beqn
\chi(M) = \frac{1}{32 \pi^2} \int_M d^4 x \, \sqrt{-g(x)} \, {\mathcal E}(x).
\label{eq_Euler_characteristic}
\eeqn
hence the name ``topological'' in the density~\eq{eq_Euler_density}. The quantity~\eq{eq_Euler_characteristic} is a topological invariant of the manifold $M$ which describes the global shape of $M$ which is robust to the smooth deformations of $M$. Finally, the second contribution to the second term in Eq.~\eq{eq_Tmunu_gravity} involves the d'Alembertian differential operator $\Box \equiv \nabla^\mu \nabla_\mu$ expressed via the covariant derivative $\nabla_\mu$. Acting on a scalar, it gives $\Box R = (1/\sqrt{-g})\, \partial_\mu \left(\sqrt{-g} \partial^\mu R \right)$.

The explicit form of the trace anomaly~\eq{eq_Tmunu_gravity} allows us to determine the anomalous action $S_{\mathrm{anom}}$ via the condition~\eq{eq_Tmunu_anomalous}. The latter is, however, expressed as a functional variational equation which fixes the anomalous action up to local regular terms. In this Appendix, we follow the prescription of Ref.~\cite{Riegert:1984kt} to fix this ambiguity. The anomalous action reads as follows~\cite{Riegert:1984kt,Mazur:2001aa,Mottola:2006ew,Armillis:2009pq}:
\beqn 
S_{\mathrm{anom}}[g,A] & = & \frac{1}{8} \int d^4 x \sqrt{- g(x)}  \int d^4 y \sqrt{- g(y)} \left[{\mathcal E}(x) - \frac{2}{3} \Box R(x) \right] \nonumber \\
& & \times \Delta_4^{-1}(x,y)  \left\{  2 b \, C^2(y) + b' \left[{\mathcal E}(y) - \frac{2}{3} \Box R(y) \right] + 2 c \, F_{\mu\nu}(y) F^{\mu\nu}(y)\right\}\,,
\label{eq_scale_action}
\eeqn
where $\Delta_4^{-1}(x,y)$ is the Green's function (the Paneitz operator~\cite{Paneitz1983}) corresponding to the fourth-order differential operator:
\beqn
\Delta_4 = \nabla_\mu \left( \nabla^\mu \nabla^\nu + 2 R^{\mu\nu} - \frac{2}{3} R g^{\mu\nu}\right) \nabla_\nu\,.
\label{eq_Delta_4}
\eeqn
The Paneitz operator~\eq{eq_Delta_4} is a unique fourth-order conformally covariant differential operator: the combination $\sqrt{-g} \Delta_4$ stays invariant under the conformal transformation $g_{\mu\nu} \to e^{2 \tau} g_{\mu\nu}$ where $\tau = \tau(x)$ is an arbitrary function. Due to the presence of the Green's function of the Paneitz operator~\eq{eq_Delta_4}, the action~\eq{eq_scale_action} is a nonlocal functional of the gauge field $A_\mu$ and the background metric $g_{\mu\nu}$. 

It's worth mentioning that the $\Box R$ term, which appears with the $b'$ coefficient in the expression for the trace anomaly~\eq{eq_Tmunu_gravity}, can be removed via a finite local counterterm within a renormalization procedure. The same statement is also true for the corresponding $\Box R$ contribution which enters the anomalous  action~\eq{eq_scale_action} with the same $b'$ coefficient. However the first $\Box R$ term in the action~\eq{eq_scale_action} cannot be removed by a local renormalization procedure. This contribution becomes important for thermoelectric effects described in Section~\ref{sec_Scale} which are also discussed below.

We will now consider the theory in a weak gravitational background for which the metric $g_{\mu\nu}$ can be expanded in terms of a small perturbation ($|h_{\mu\nu}| \ll 1$) over the flat metric
\beqn
g_{\mu\nu} = \eta_{\mu\nu} + h_{\mu\nu}\,.
\label{eq_weak_metric}
\eeqn
In the linearized gravity, the inverse metric tensor is $g^{\mu\nu} = \eta^{\mu\nu} - h^{\mu\nu}$. The indices are raised/lowered with the flat metric tensor $h^{\mu\nu} =  \eta^{\mu\alpha} \eta^{\nu\beta} h_{\alpha\beta}$ and $g^{\mu\alpha} g_{\alpha\nu} = \delta^\mu_\nu + O(h^2)$. Since the Riemann tensor is linear in the perturbation, $R \sim O(h)$, the Weyl-squared term ($C^2$) and the topological term ${\mathcal E}$ do not contribute to the anomalous conformal action~\eq{eq_scale_action}. The linear contribution appears only in the Ricci scalar 
\beqn
R = R^{(1)} + O(h^2) 
\qquad 
{\mathrm{with}} \qquad R^{(1)} = \partial_\mu \partial_\nu h^{\mu\nu} - \eta_{\mu\nu} \square_0 h^{\mu\nu}\,, 
\qquad 
\square_{0} \equiv \eta^{\mu\nu} \partial_\mu \partial_\nu\,,
\label{eq:R:linearized}
\eeqn
where $\square_0$ denotes the flat-space d'Alembertian. 

To linear order in the spacetime curvature, the anomalous action~\eq{eq_scale_action} contains only a mixed gravitational-electromagnetic term
\beqn
S_{\mathrm{anom}} = \frac{c}{6} \int d^4 x \int d^4 y \, F_{\alpha\beta} (x) F^{\alpha\beta} (x) \,
P_{\mu\nu}(x,y) \, h^{\mu\nu}(y) + O(h^2)\,,
\label{eq_action_weak}
\eeqn
where we rearranged the terms and introduced the transverse projection projector
\beqn
P_{\mu\nu}(x,y)  = \eta_{\mu\nu} \delta^{(4)}(x-y) - \square^{-1}_0(x-y) \frac{\partial}{\partial y^\mu} \frac{\partial}{\partial y^\nu} ,\quad
\label{eq_projector}
\eeqn
with the Green function $\square^{-1}_{0}(x-y) \equiv \square^{-1}_{0}(x,y)$ of the flat-space d'Alembertian $\square_{0}$. The would-be double-pole of the original action~\eq{eq_scale_action} reduces to a simple pole which appears in the projector~\eq{eq_projector}.

Consider the conformally flat metric $g_{\mu\nu}(x) = e^{2 \tau(x)} \eta_{\mu\nu}$ with the small scale factor $|\tau| \ll 1$. The linearized perturbation in Eq.~\eq{eq_weak_metric}, $h_{\mu\nu} = 2 \tau \eta_{\mu\nu}$, gives us the leading order expression for the Ricci scalar: $R^{(1)} = 6 \,\square_0 \tau$. The d'Alembertian operator in the linearized expression of the first Ricci scalar in the anomalous action~\eq{eq_scale_action} ``cancels'' two powers of derivatives in the fourth-order differential Paneitz operator~\eq{eq_Delta_4}. We assume that the conformal perturbation factor $\tau$ vanishes at a spatial infinity so that the surface terms do not appear after an integration over the coordinate $y$. As a result, we obtain an absolutely local expression for the anomalous action in the weakly conformal background
\beqn
S_{\mathrm{anom}}{\biggl|}_{g_{\mu\nu} = e^{2 \tau} \eta_{\mu\nu}} = \frac{e^2}{24 \pi^2} \int d^4 x\, \tau(x) \, F_{\alpha\beta} (x) F^{\alpha\beta} (x) + O(\tau^2)\,.
\label{eq_action_weak_conformal}
\eeqn

The variation of the anomalous action~\eq{eq_scale_action} with respect to the electromagnetic gauge field $A_\mu$ gives us the anomalous current:
\beqn
J^{\mu}_{\mathrm{anom}}(x) = - \frac{1}{\sqrt{-g(x)}} \frac{\delta S_{\mathrm{anom}}}{\delta A_\mu(x)}\,.
\eeqn
In a weak gravitational field~\eq{eq_action_weak}, the induced anomalous current gets the following form:
\beqn
J^{\mu}(x) = \frac{e^2}{6 \pi^2} F^{\mu\nu}(x)\partial_\nu \varphi(x) 
\,,
\label{eq:J:SME}
\eeqn
where we used the anomalous coefficient $c$ for one-species electrodynamics~\eq{eq_anomalous_coefficients} and defined the effective scalar field
\beqn
\varphi(x) = - \frac{1}{6}  \int d^4 y \, P_{\alpha\beta}(x,y) h^{\alpha\beta}(y)\,,
\label{eq:v}
\eeqn
via the projector~\eq{eq_projector}. For a conformally flat metric~\eq{eq_g_munu}, the metric perturbation $h_{\mu\nu} = 2 \tau \eta_{\mu\nu}$ is a linear function of the scale factor~$\tau$ provided the latter is treated as a small perturbation. Then Eq.~\eq{eq:v} gives us $\varphi = \tau$ and we come back to the expression for the anomalous current~\eq{eq_J_covariant} in electrodynamics with one fermion species, $N_f = 1$, characterized by the beta function~\eq{eq_beta_QED}. On the other hand, for the result of the anomalous Nernst current, we consider a metric with non vanishing component $h_{00}$. In the static limit we obtain $\varphi=-h^{00}/6$, which gives the current
\beqn
J^{\mu}(x) = -\frac{e^2}{36 \pi^2} F^{\mu i}(x)\partial_i h^{00}(x) 
\,.
\label{eq:anomalous_Nernst_app}
\eeqn

\subsection{Scale electric effect}
We are going to briefly discuss here the scale electric effect (SEE), which is generated by the scale anomaly in the presence of a pure electric field background $\vec E$~\cite{Chernodub16}. The SEE has the form of the usual Ohm law~\eq{eq_SEE} characterized by the metric-dependent anomalous electric conductivity~\eq{eq_sigma_anomalous}. While the SME~\eq{eq_SME} has straightforward condensed-matter applications discussed in Section~\ref{sec_conformal}, its electric counterpart~\eq{eq_SEE} involves a time-dependent gravitational background which poses difficulties in its realization in the condensed matter context. Nevertheless, the scale electric effect has certain interesting consequences in fundamental quantum field theory, most notably in cosmology where the non-stationary gravitational background is inherent to the expanding Universe. 

Surprisingly, the SEE~\eq{eq_SEE} turns the vacuum into an electrically conducting ``medium'', with the electric conductivity~\eq{eq_sigma_anomalous} proportional to the beta function of the theory. In QED-like theories with a positive value of the beta function~\eq{eq_beta}, the anomalous electric conductivity can take negative values contrary to its dissipative Ohm counterpart. The negative conductivity produces a negative value of the Joules heat with the power density~\cite{Chernodub16}:
\beqn
W \equiv \frac{\partial {\mathcal E}}{\partial t} = {\vec J} \cdot {\vec E} = - \frac{2 \beta_e(e)}{e} \frac{\partial \tau(t,{\vec x})}{\partial t} E^2.
\label{eq_negative_heat}
\eeqn
In many other circumstances, the relation $W < 0$ would be a dangerous sign of energy non-conservation. However, the time-dependent gravitational background necessarily corresponds to either an open system or to a sub-part of a closed system, to which the requirement of the energy conservation does not apply. 

The negative vacuum conductivity, which may have played a role during the inflationary stage of the evolution of the early Universe, has also been independently obtained in calculations for fermionic~\cite{Hayashinaka:2016qqn} and bosonic~\cite{Kobayashi:2014zza} electrically charged particles in a expanding de Sitter space via the Schwinger pair-production mechanism. The vacuum in the background electric field produces abundant particle-antiparticle pairs which get diluted and transported along the field as the space expands. Thus, the scale electric effect~\eq{eq_SEE} can be understood as a subtle balance between particle creation and rarefaction, which forms a steady background of particle-antiparticle pairs that makes the space electrically conducting~\eq{eq_sigma_anomalous}. The negative conductivity~\eq{eq_sigma_anomalous} can also lead to an instability in the electromagnetic sector of the Standard Model of particle interactions, which can produce potentially detectable long-wave electromagnetic fields in time-dependent gravitational backgrounds during inflation, that can serve as seeds of strong enough galactic magnetic fields observed nowadays~\cite{Dolgov:1993vg}. The first evidence of the scale electric effect has been mentioned in Ref.~\cite{Dolgov:1993vg}.


\section{The formalism of gravitoelectromagnetism}
\label{sec_GEM}

As we learned in the classical theory of fields, the two long range forces of nature, the Coulomb force between two static charges, and the Newton attraction between two static masses, have the same $r^{-2}$ dependence on the distance.  So the respective potentials are the same differing only on the coupling constants. In electromagnetism, moving charges generate magnetic forces and  Maxwell unification of electricity and magnetism  included the magnetic induction in the elegant gauge formulation that we use today.  Gravito-electromagnetism  is a gauge formulation of linearized gravity in the weak field approximation based on the  analogy of Coulomb and Newton's laws\footnote{We will closely follow the classical review on the matter  ref. \cite{Mash03}.}. Perhaps the most interesting implication of the formulation is the prediction that  the gravitational force
has a magnetic component. Gravito-magnetic fields associated to to mass current arise in reference frames that move relative to a source of static gravitational field \cite{Mash84}.
The analogs to Maxwell gravitational equations are a set of four partial differential equations written for the gravito-electric ${\vec E}_g$ and gravito-magnetic ${\vec B}_g$ fields and their sources: mass density $\rho_M$ and mass current density ${\vec j}_M$. In the  gravito-electromagnetic formulation, the torsion field discussed in  sec.  \ref{sec_torsion} and in Appendix \ref{sec_curved}, is  the gravitomagnetic field strength associated to rotating mass distributions.

As we described in the previous section, in the limit of weak gravitational field the metric tensor can be written as  \eqref{eq_weak_metric}:
\begin{equation}
 g_{\mu\nu}(x)=\eta_{\mu\nu}  +h_{\mu\nu}(x)
    \label{eq:weak},
\end{equation}
where $\vert h_{\mu\nu}\vert \ll 1$.
Under a coordinate transformation
of the form $x^\mu\to x'^\mu=x^\mu-\lambda^\mu$ which leaves the Einstein equations invariant, the metric $h_{\mu\nu}$  transforms as a gauge field:
\begin{equation}
 h_{\mu\nu}\to h_{\mu\nu}+\partial_\nu\lambda_{\mu} +\partial_\mu\lambda_\nu.
 \label{eq:mgauge}
\end{equation}
In this approximation the linearized Christoffel symbols in eq. \eqref{eq_Christoffel_symbols} are
\begin{equation}
    \Gamma_{\mu\nu}^\alpha=\frac{1}{2}\eta^{\alpha\rho}
    (\partial_\nu h_{\mu\rho}+\partial_\mu h_{\nu\rho}-\partial_\rho h_{\mu\nu}),
\end{equation}
the Ricci tensor is

\begin{equation}
    R_{\mu\nu} =\frac{1}{2}[\partial_\sigma\partial_\mu h^\sigma_\nu
    +\partial_\sigma\partial_\nu h^\sigma_\mu
    -\partial_\mu\partial_\nu h
    -\Box h_{\mu\nu}],
\end{equation}
where $h=\eta^{\mu\nu}h_{\mu\nu}$. The trace of the linearized Ricci tensor  is given by Eq.~\eq{eq:R:linearized}.
In terms of the modified metric
\begin{equation}
{\bar h}_{\mu\nu}=h_{\mu\nu}-\frac{1}{2}\eta_{\mu\nu}h, 
\label{eq:metric2}
\end{equation}
and in the transverse gauge $\partial_\nu{\bar h}^{\mu\nu}=0$,
the linearized Einstein equations
read
\begin{equation}
 \partial^2   \bar h_{\mu\nu} = -2 \kappa T_{\mu\nu},
\end{equation}
where $\kappa=8\pi G/c^4$.
The retarded solution is
\begin{equation}
  \bar h_{\mu\nu} =  \frac{4G}{c^4}\int\frac{T_{\mu\nu}(ct, {\vert\vec{x}-\vec{x'}\vert,\vec{x'} )}}{\vert\vec{x}-\vec{x'}\vert}
  d^3x'.
  \label{eq:GEMmetric}
\end{equation}
For a generic mass distribution moving with velocity ${\vec v}$, the mass density $\rho_M$ and mass current density ${\vec j}_M$ are defined by 
\begin{equation}
 T^{00}=\rho_M c^2, \quad {\vec J}_M=\rho{\vec v}.  
\label{eq:Mdensity}
\end{equation}
and the  gravito-electromagnetic gauge potential $  A_{g,\mu}=(\phi, {\vec A_g})$ is defined by the equations

\begin{equation}
\phi=\frac{c^2}{2}{\bar h}_{00} \;,   \;
A_{g,i}=-c^2{\bar h}_{0i}.
  \label{eq:GEMpotentials}
\end{equation}
The gravitoelectromagnetic fields are defined by the standard formulas
\begin{equation}
{\vec E_g}=-{\vec \nabla}\phi -\frac{1}{c}  \partial_t{\vec A}_g\quad,\quad   {\vec B_g}={\vec\nabla}\times{\vec A}_g.
\label{eq:GEF}    
\end{equation}
From these definitions, and in the transverse gauge,    the fields fulfill the standard Maxwell equations:
\beqa
&\vec{\nabla}\cdot\vec{E_g}=-4\pi G\rho, \quad
 \vec{\nabla}\cdot\vec{B_g}=0, \nonumber \\
    &\vec{\nabla}\times\vec{E_g}=-\partial_t \vec{B_g},  \quad
 \vec{\nabla}\times\vec{B_g}=\frac{1}{c^2} ( \partial_t{\vec E}-4\pi G \vec{J}_M),
\label{eq:Gmaxwell} 
\eeqa
and the matter matter current density  satisfy the continuity equation 
$\partial_t\rho_M + \vec{\nabla}\cdot\vec{J}_M = 0$.
\\

For a finite distribution of slowly moving matter with $\vert v\vert <<c$,
 the energy-momentum tensor takes the form
 $T_{ij}\sim \rho v_iv_j+p\delta{ij}$, where $p$ is the pressure. 
 In this approximation and neglecting terms of order $0(c^{-4})$, the line element of gravito--electromagnetism takes the form
 \begin{equation}
     ds^2=-c^2 (1-\frac{\phi}{c^2})dt^2-\frac{2}{c}({\vec A}_g\cdot d{\vec x})dt
     +(1+\frac{\phi}{c^2})\delta_{ij}dx^idx^j.
 \end{equation}
In the Newtonian limit $\phi$ is the gravitational potential and ${\vec A}g=0(c^{-1})$. (From eq. \eqref{eq:GEMmetric} we see that the components $ \bar h_{ij} $ are of order $c^{-4}$).
If the source distribution is confined around the origin of spatial coordinates, then far from the source we have
$\phi\sim GM/r$, $ {\vec A_g}\sim (G/c)({\vec J}\times{\vec
x})/r^3$ where $M$ and ${\vec J}$ are the mass and angular momentum of the mass distribution.

The gravitoelectromagnetic formulation is particularly useful when addressing problems in the presence of rotation or torsion (see Secs. \ref{sec_magnet} and \ref{sec_torsion}). The gravito-magnetic potential, \eqref{eq:GEMpotentials} sources energy currents 
\begin{equation}
 T^{0i}\sim\frac{\delta S}{\delta A_{gi}}
\end{equation}
 and it can  be used to generate Kubo formulas for 
new anomaly related transport coefficients  as the “chiral gravitomagnetic” conductivity 
giving rise to a “chiral gravito-magnetic effect" \cite{Landsteiner:2011tg}.


\section{Black Holes, chiral transport and gravitational anomaly}
\label{app_bh}

 In this appendix we briefly review how the energy current of a single Weyl fermion is related to the gravitational anomaly, following Ref.~\cite{Stone:2018zel}.
\begin{figure}[!thb]
\begin{center}
\includegraphics[scale=0.25,clip=true]{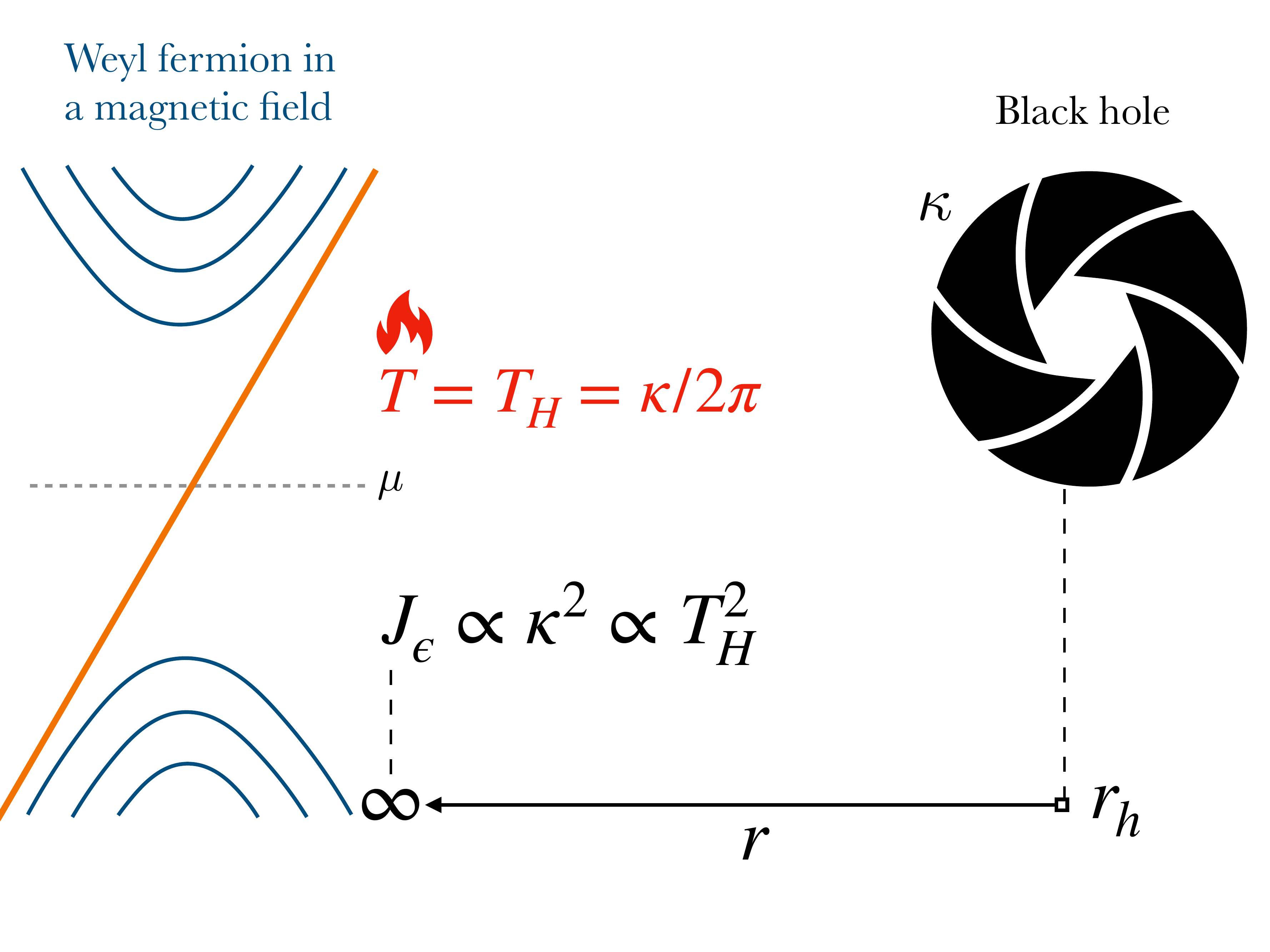}
\end{center}
\caption{Connection between temperature and the axial-gravitational anomaly.  The spectrum of a Weyl fermion in a strong magnetic field $\mathbf{B}$ organizes as Landau levels. The zeroth-Landau level can be regarded as a 1D system with degeneracy $|B|/2\pi$ which can be then placed in a thermal state at temperature $T$. The temperature can be chosen to be the Hawking temperature $T_H$ of a distant black-hole, which is determined by its surface gravity $\kappa$ at the horizon $r=r_h$ as $T_H=\kappa/2\pi$. Integrating the anomaly equation Eq.~\eqref{eq:2Dgravanomaly} for a 1D Weyl fermion between $r=r_h$ and $r=\infty$ results in an energy current proportional to $\kappa^2$ and hence related to the temperature of the thermal bath experienced by the Weyl fermion, see Appendix \ref{app_bh} and Ref. \cite{Stone:2018zel}.}
\label{fig:stone}
\end{figure}
 We start with the gravitational anomaly equation, Eq.~\eqref{eq:covanomalyT}, expressed in 1+1 dimensions:
\begin{equation}
\label{eq:2Dgravanomaly}
    \nabla_\mu T^{\mu\nu}=\frac{c}{96\pi}\frac{1}{\sqrt{-g}}
    \epsilon^{\nu\sigma}\nabla_\sigma R,
\end{equation}
where $c$ is the central charge of the theory ($c=1$ for a Weyl fermion) and $g$ is the determinant of the metric. We choose the background metric
\begin{equation}
\label{eq:metric1+1}
    ds^2 = -f(r) dt^2 + \dfrac{1}{f(r)} dr^2,
\end{equation}
which describes a black hole metric if we require that $f(r)\to 1$ at large $r$ and define the horizon at $r_h$ by $f(r_h) = 0$. For this metric $R= - f^{''}$.
This metric also has a time like Killing vector $\eta^\mu = (1,0)$. Being  a Killing vector is fulfills $\nabla_\mu \eta_\nu + \nabla_\nu \eta_\mu =0$. We use this to define the energy current
\begin{equation}
    J^\mu_\epsilon = T^{\mu\nu}\eta_\nu\,.
\end{equation}

Integrating Eq.~\eqref{eq:2Dgravanomaly} from $r_H$ to infinity in the background metric \eqref{eq:metric1+1} one obtains that~\cite{Stone:2018zel}
\begin{equation}
\label{eq:Jhb}
  J^r_{\epsilon}|^\infty_{r_h} = -\dfrac{c}{96\pi}\left(ff''-\dfrac{1}{2}(f')^2 \right)\biggr\rvert^\infty_{r_h}.
\end{equation}
Choosing the boundary conditions where $J^r_\epsilon$ is zero at the horizon, and recalling that the metric tends to a flat metric when $r\to\infty$, we can infer that the density of energy current is given in this limit as
\begin{equation}
\label{eq:Trt}
   J^r_{\epsilon} = \dfrac{c\kappa^2}{48\pi},
\end{equation}
with $\kappa=f'(r_h)/2$ being the surface gravity of the black hole, which also determines its Hawking temperature as $T_H = \kappa/2\pi$\footnote{We also want to point out that anomalies are central to the derivation of Hawking radiation black holes in the approach of \cite{Robinson:2005pd}.}.

If a Weyl semimetal is placed at a temperature $T_H$, and in a sufficiently strong magnetic field such that only the zeroth Landau level contributes to transport (ultra-quantum limit), its energy current is determined by the gravitational anomaly trough Eq.~\eqref{eq:Trt}.
In fact a stronger statement holds: since the anomaly and the anomaly induced transport stem only form the lowest Landau level the result is still valid for arbitrary magnetic fields. To get the result for the energy current in the full $3+1$ dimensional space time we only need to multiply Eq. (\ref{eq:Trt}) with the degeneracy factor of the Landau level $\frac{B}{2\pi}$. This finally leads to 
\begin{equation}
    J^r_\epsilon = c \frac{T^2_H}{24} B^r\,. 
\end{equation}
For more details we refer to \cite{Stone:2018zel} and also \cite{Jensen:2013vta} which gives a careful treatment based on properties of Euclidean quantum field theory.
Finally we note that the dimensional reduction of the four dimensional anomaly \eqref{eq:covanomalyT} in a strong magnetic field gives precisely the two dimensional anomaly \eqref{eq:2Dgravanomaly}.


\clearpage
\section{Notations}

\begin{table}[!htb]
\centering
\begin{tabular}{lcl}
$A^\mu = (A^0,{\vec A})$ & \qquad\ & 4-vectors \\
$\vec A$   && 3-vectors \\
$\bs A$    && $3 \times 3$ matrices \\
${\vec J}$    && electric current \\
${\vec J}^\epsilon$    && energy current \\
${\vec J}^Q$    && heat current \\
$g_{\mu\nu}$ && metric \\
$h_{\mu\nu}$ or $\delta g_{\mu\nu}$ && metric perturbation\\
$\phi$       && gravitational potential \\
$\vec A_g$                   &&  gravitational (or thermal) vector potential    \\
$h$       && Hamiltonian density \\
$T^{\mu\nu}$ && energy-momentum tensor \\
${\bs L}$           && transport coefficients\footnote{Determined for the transports currents only, Eqs.~\eq{eq transport 1} and \eq{eq transport 2}.}\\
${\bs{\mathcal{L}}}$ && transport coefficients\footnote{Determined from Kubo formulas~\eq{eq_kappa} and \eq{eq_kappa2} which also include magnetization currents~\eq{eq_magnetization_currents}, cf. Eqs.~\eq{eq transport coeff 1} and \eq{eq transport coeff 2}.}\\
${\vec M}$                   && magnetization current \\
                   &&     \\
                   &&     \\
                   &&     \\
                   &&     \\
                   &&     \\
                   &&     \
    \end{tabular}
    \caption{Notations used in the review.}
    \label{tab:notations}
\end{table}

\newpage

\end{document}